\documentclass[12pt]{article}
\usepackage{amsfonts,amsmath}
\usepackage{amssymb}
\usepackage[hypertex] {hyperref}

\makeatletter \@addtoreset{equation}{section} \makeatother

\newcommand{\tu}{\tilde{ g}}

\newcommand{\alM}{\mathbf{M}}

\newcommand{\gD}{\mathfrak{g}}

\newcommand{\Cp}{{\Phi}}
\newcommand{\Cm}{ {\overline{\Phi}}}

\renewcommand{\Re}{\mathop{\rm Re}}
\renewcommand{\Im}{\mathop{\rm Im}}

\newcommand{\pos}{\mathcal{P}}

\newcommand{\JNN}{ \mathrm{J}}

\newcommand{\ppi}{{\mathfrak{p}}}%{{\pi}}
\newcommand{\Os}{{\mathfrak{O}}}%{{\pi}}
\newcommand{\Warpi}{{\Omega}}
\newcommand{\yy}{{Y}}
\newcommand{\tr}{ {tr}}
\newcommand{\zz}{{W{}}}

\newcommand{\oc}{\hat{\mathfrak{c}}}

\newcommand{\za}{{{A}}}
\newcommand{\zb}{{{B}}}

\newcommand{\X}{{  \mathbf{X}}}
\newcommand{\Y}{{\mathbf{Y}}}

\newcommand{\XX}{{X^\prime\,{}}}
 \newcommand{\gY}{{\mathcal{Y}}}%{{\Upsilon}}

\newcommand{\Z}{{{\cal X}}}

\newcommand{\W}{W}%{\mathcal{W}}

\newcommand{ \cZ }{{\overline{\mathcal{X} \rule{0pt}{10pt}}}}
\newcommand{\cY}{{\overline{\gY\rule{0pt}{10pt}}}}

\newcommand{\B}{{\cal B}}

\newcommand{\Gm}{{\cal V}}
\newcommand{\GmX}{{ \mathfrak{ V}}}

\newcommand{\be}{ \begin{equation}}

\newcommand{\bst}{{\star\,}}
\newcommand{\bp}{ {\,\stackrel{ \!\op}{\oprod}}}
\newcommand{\bpd}{ {\,\stackrel{ \opd}{\oprod}}}

\newcommand{\U} {{\mathcal U}}

\newcommand{\NNN}{{ \mathcal{ N}}}

\newcommand{\III}{\mathcal{I}}%

\newcommand{\ee}{\end{equation}}
 \newcommand{\TT}{  {\hat{\mathrm{T}}}}
  \newcommand{\TTT}{  {T}}

\newcommand{\PPP}{  {J} }
\newcommand{\JJJ}{ \mathcal{J}}

\newcommand{\bee}{\begin{eqnarray}}
\newcommand{\beee}{\begin{array}}
\newcommand{\eee}{\end{eqnarray}}
\newcommand{\eeee}{\end{array}}
\newcommand{\hhh}{{}}%{{\hhh}}

\newcommand{\gn}{\nu}
\newcommand{\gm}{\mu}

\newcommand{\gx}{\xi}

\newcommand{\ga}{\alpha}
\newcommand{\pa}{{\ga^\prime}}

\newcommand{\pb}{{\gb^\prime}}
\newcommand{\pga}{{\gamma^\prime}}
\newcommand{\gb}{\beta}

\newcommand{\gga}{\gamma}

\newcommand{\M}{{\cal M}}
\newcommand{\Mi}{{ M^d}}
\newcommand{\E}{{\cal E}}

\newcommand{\G}{{\cal G}}

\newcommand{\T}{\mathcal{T}}
\renewcommand{\S}{{\cal S}}
\newcommand{\Q}{{\cal Q} }
\newcommand{\rhs}{{\it r.h.s.} }

\newcommand{\lhs}{{\it l.h.s.} }

\newcommand{\ie}{{\it i.e.,} }
\newcommand{\ls}{\!\!\!\!\!\!}

\newcommand{\gd}{\delta}

\newcommand{\gl}{\lambda}

\newcommand{\gvep}{\varepsilon}
\newcommand{\gs}{\sigma}
\newcommand{\bz}{{\bar z}}

\newcommand{\go}{\omega}

\newcommand{\by}{{\bar{y}}}

\newcommand{\q}{\,,\qquad}

\newcommand{\da}{{{\ga^\prime}}}
\newcommand{\db}{{{\gb^\prime}}}

\newcommand{\nn}{\nonumber}

\newcommand{\half}{\frac{1}{2}}

\newcommand{\ptl}{\partial}
 \newcommand{\p}{\partial}
\newcommand{\D}{{\cal D}}
\newcommand{\f}{\frac}
\newcommand{\A}{{\cal A}}%%
\newcommand{\C}{{\cal C}}

\newcommand{\bu}{{\bar{u} }}
\newcommand{\bv}{{\bar{v} }}

\newcommand{\bZ}{\overline{{Z}}}

\newcommand{\ZIGM}{\mathfrak{H}_M}
\newcommand{\GZM}{\mathfrak{H}_M\times\mathbb{C}^M}

\newcommand{\starbul}{\star}%{\circledast}

\newcommand{\opd}{\,\triangleright\,}%%{ \circ}
\newcommand{\op }{\,\triangleleft\,}%{\diamond}
\newcommand{\ten}{{ \times }}
\newcommand{\oprod}{{\,\bowtie\,}}
 \newcommand{\oprodeq}{{\,\underline{\bowtie}\,}}
\newcommand{\Sq}{  \mathrm{{D}}  }

 \newcommand{\strop}{ {tr_{\opd}}}
\newcommand{\stropd}{ {tr_{\op}}}
\newcommand{\stropr}{ {tr_{\oprod}}}

\begin{document}

\begin{flushright}
{\small FIAN/TD/03-13\\
      January 2013}\,\,\,
\end{flushright}
\vspace{1.7 cm}

\begin{center}
{\large\bf Operator algebra  of free  conformal currents via
twistors}

\vspace{1 cm}

O.A. Gelfond$^1$ and M.A.~Vasiliev$^2$ \vglue 0.3  true cm

${}^1$Institute of System Research of Russian Academy of Sciences,\\
Nakhimovsky prospect 36-1, 117218, Moscow, Russia

\vglue 0.3  true cm

${}^2$I.E.Tamm Department of Theoretical Physics, Lebedev Physical
Institute,\\
Leninsky prospect 53, 119991, Moscow, Russia

%gel@lpi.ru, vasiliev@lpi.ru \\
\end{center}

\vspace{0.4 cm}

\begin{abstract}
\noindent
Operator algebra of (not necessarily free) higher-spin
conformal conserved currents in generalized matrix spaces, that
include $3d$ Minkowski space-time as a particular case,  is shown to
be determined by an associative algebra $M$ of functions on the
twistor space. For free conserved currents, $M$ is the universal
enveloping algebra  of the higher-spin algebra. Proposed construction
greatly simplifies computation and analysis of correlators of
conserved currents.  Generating function for $n$-point functions of $3d$
(super)currents of all spins, built from ${\mathcal N}$ free constituent massless
scalars and spinors, is obtained in a concise form of certain determinant.
Our results agree with and extend  earlier bulk computations in the  HS
$AdS_4/CFT_3$  framework. Generating function for $n$-point functions of
$4d$ conformal  currents is also presented.

\end{abstract}

\newpage
\tableofcontents

\newpage

\section{Introduction}

We apply methods of unfolded dynamics  to derivation of the operator
algebra of conserved currents in three and higher dimensions. The
problem is analyzed within the formulation in generalized matrix
space $\M_M$ that provides a natural framework for description of
conformal fields in various space-time dimensions
\cite{F,BL,Bandos:1999qf,Vasiliev:2001zy,Mar,Bandos:2005mb}. In the
case of $\M_2$ equivalent to usual $3d$ Minkowski space, our
analysis reproduces operator algebra of free conformal conserved
currents of all spins built from free $3d$ massless scalars and
spinors. We derive full operator algebra of free conserved currents
and, in particular, the explicit form of $n$-point functions,
including relative coefficients. Obtained results agree with
previously available in the literature on two-point, three-point
(see, e.g.
\cite{Osborn:1993cr,Petkou:2003zz,Giombi:2011rz,Costa:2011mg,
Giombi:2012ms,Stanev:2012nq, Zhiboedov:2012bm,Colombo:2012jx} and
references therein) and $n$-point functions \cite{Didenko:2012tv}
for conserved currents of any spin, extending them to supercurrents.

Our construction exhibits  covariance under an infinite-dimensional
symmetry which extends usual higher-spin (HS) symmetries, that act
individually on $n$-point functions with definite $n$, to much
larger multiparticle symmetries deduced in \cite{Vasiliev:2012tv}
from the research of this paper. Mixing states with different number
of particles, multiparticle symmetries relate $n$-point functions
with different $n$. Being represented by the universal enveloping
algebra of the HS algebra, multiparticle algebra $\alM$ can  be realized
\cite{Vasiliev:2012tv} in terms of an infinite set of oscillators
reminiscent of those of String Theory. Recognition of this symmetry
as a promising candidate for the symmetry of a HS generalization of
String Theory  is an important by-product of our analysis.

The key observation is that, when all spins are involved, conserved
currents are described by unrestricted functions $ J(Y_i|X) $ of
the doubled number of spinor (twistor) variables $Y^A_i$, $i=1,2$
($A=1,\ldots M$) \cite{tens2}. In these terms, free $3d$ massless
fields are described by functions of two-component spinors $y^\ga$
and space-time coordinates $x^{\ga\gb}=x^{\gb\ga}$, where
$\ga\,,\,\gb=1,2$ are $3d$ spinor indices, while $3d$ conserved
currents $J(y^\ga_i|x^{\ga\gb})$ depend on a pair of spinors
$y_{1}^\ga$, $y_{ 2}^\ga$. Spin $s$ primary HS currents are
\be
\label{phcur} J^\pm_{\ga_1\ldots \ga_{2s}}(x) = \f{\p^{2s}}{\p
y_\pm^{\ga_1}\ldots \p y_\pm^{\ga_{2s}}} J(y^\ga_i|x)\Big \vert_{y=0}\q
y^\ga_\pm=y^\ga_1\pm y^\ga_2\,. \ee

OPE of $ J(y^\ga_i|x ) J(y'{}^\ga_i|x') $ develops a singularity at
$(x ,y )\to (x',y')$. In particular, for $x =x'$, $ J(y^\ga_i|x )
J(y'{}^\ga_i|x) $ is a distribution with respect to
$y^\ga-y'{}^\ga$.   The form of  OPE at the same $x$ is very simple
in terms of the twistor variables. However, making no sense at  $y
=y'$, such OPE cannot be directly used for the analysis of currents
(\ref{phcur}) of definite spins evaluated at $y=0$. Hence, to
compute OPE for currents of definite spins it is necessary to
separate their space-time coordinates $x$, $x'$ while for the tower
of all spins OPE can be defined at $x=x'$.

In more  detail, introducing for general $M$, \be\label{Jgab}
 \III^2_{  g  }=\int_{\mathbb{R}^{2M}}   d^{2M }  {Y} \,\,  \,\,
g{}( Y_1, Y_2) \,  J(Y_1,\,Y_2 |0)\q
 \ee%
  where $g{}( Y_1, Y_2)$ is an arbitrary (test) function, operator
algebra has the form \bee\label{schem} \III^2_{  g  }\III^2_{  g' }=
:\III^2_{ g }\III^2_{  g'  }: + \III^2_{(  g\op g'+   g'\opd g ) } +
\strop  ({g}\op {g}') Id \,, \eee%
 where $\op$ and $\opd$  denote certain
convolution products related to the star product via
half-Fourier transform and representing particular cases of
 certain ``butterfly" product law
(see Section \ref{Vertex
algebra}). The first term in (\ref{schem}) is regular (terms of this
type are usually skipped). The other two are singular. The second
term implies that currents generate the HS algebra (see Section
\ref{Charges}). The third term is central. The number ${\mathcal N}$
of free fields from which the currents are constructed enters through the
definition of $\strop $ which includes usual trace over color indices.
Extension of this formula to $ \III^2_{  g  }(x)\III^2_{  g'  }(x')$ is
uniquely determined by the unfolded current conservation equations,
reproducing usual formulas.

Current operator algebra is fully determined in terms of the
multiparticle  algebra  $\alM$, leading to the remarkably simple
 generating function for all $n$-point functions, expressed in terms
of certain determinant with respect to butterfly product. This
butterfly formula works both for $3d$ and for $4d$ Minkowski space.
Since all other operators have zero VEV, a $n$-point function
coincides with the coefficient in front of the unit operator $Id$ of
the operator algebra.
As announced in (\ref{schem}),
such coefficients are  expressed in terms of a trace over the
butterfly-product algebra.  Being similar to the Ansatz  of
bulk computation in \cite{Colombo:2010fu,Colombo:2012jx,Didenko:2012tv}, here it
results from direct computation which specifies the Green functions
involved. Namely, we compute Wightman functions. Of course, at
least naively (\ie away from singularities), from  these results it
is easy to obtain $T$-ordered correlators
 by inserting appropriate step-functions in time coordinates.

Our original motivation  was to study further
 $AdS_4/CFT_3$ HS correspondence that acquired considerable
attention after important work of Klebanov and Polyakov
\cite{Klebanov:2002ja}
 where it was argued that the HS
gauge theory of \cite{more} should be dual to the $3d$ $O(\NNN)$ sigma
model in the $\NNN\to \infty$ limit (for fermionic counterpart see
\cite{Sezgin:2003pt}). This conjecture was checked by
Giombi and Yin \cite{Giombi:2012ms} (and references therein) who
were able to show in particular how the bulk computation in HS
gauge theory reproduces conformal correlators in the free $3d$
theory. Recently, Maldacena and Zhiboedov
\cite{Maldacena:2011jn,Maldacena:2012sf} addressed the question on
restrictions imposed on a boundary $3d$ conformal theory by HS
conformal symmetries. Assuming unitarity and locality
 they were able to show that a $3d$
conformal  theory, that possesses a HS conserved current, should be
free. This conclusion seemingly suggests that any $AdS_4$ HS theory
should be equivalent to a free boundary theory at least in the most
symmetric vacuum. However, in \cite{Vasiliev:2012vf} it was shown
that, beyond $\NNN\to \infty$ limit, holographically dual of the
$AdS_4$ HS theory is a  nonlinear $3d$ $HS$ theory of
conformal  currents interactings with $3d$ conformal HS
gauge fields which were not allowed in the analysis of
\cite{Maldacena:2011jn,Maldacena:2012sf}.

Results of this paper further  support  the
 idea  of manifest duality of the bulk $AdS_4$ HS
 theory and that of boundary currents (plus boundary conformal fields)
 via unfolded dynamics approach \cite{Vasiliev:2012vf}.
The appearance of convolution product in the OPE (\ref{schem}) is related via
a half-Fourier transform  to the star product underlying
minimal HS interactions (see Section \ref{Half}). This fits the standard expectation that the nontrivial part
of boundary correlators should result
from interactions in the bulk. Moreover, as explained in Sections
\ref{cch} and \ref{conc}, via replacement of boundary  propagators of currents
by bulk-to-boundary propagators of bulk fields, the boundary computation of the OPE
admits the bulk deformation that does not affect the final result.

From the perspective of a multiparticle HS theory, the computation of $n$-point
 functions is essentially classical. There are two elements that make the result
 quantum from  usual perspective. One is that algebra $\alM$ is itself non-commutative as a result of
 non-commutativity of the HS algebra which is the algebra of oscillators (\ie Weyl
 algebra). Another is the origin of the central term proportional to the
 number ${\mathcal N}$ of free constituent fields in the example of currents
 of this paper. As discussed in \cite{Vasiliev:2012tv}, the
 ambiguity in ${\mathcal N}$ is encoded in the definition of trace on $\alM$,  which enters
 the  definition of $n$-point function. Hence, both of these quantum effects acquire ``classical"
 interpretation in terms of HS-like computations in multiparticle theory.
It is tempting to speculate that such ``classical" multiparticle consideration should be equivalent to
 quantum HS theory with all its multiparticle states involved.

The rest of the paper is organized as follows. In Section
\ref{currents}, we recall  relevant facts of the unfolded
formulation of fields of different ranks, conserved currents and
charges. In Section \ref{Dfunc}, the construction and
properties of  $\D-$functions of the  unfolded massless field
equations are recalled. In Section \ref{Quantization}, quantization prescription
for free fields in the generalized matrix space is recalled and
quantum currents are introduced. In Section \ref{Charges}, algebra
of charges is shown to be isomorphic to the star-product algebra
associated with  conformal HS algebra. In Section \ref{Vertex
algebra}, operator algebra of currents in twistor space is
introduced and  determined for any $M$
in terms of butterfly-product algebra. Derivation of butterfly
formulae form the multiparticle algebra of \cite{Vasiliev:2012tv}
is given in Section \ref{BMult}.
Space-time operator algebra is presented in Section \ref{Contact}.
General formulae for $3d$ and $4d$ $n$-point functions
are obtained in Section \ref{Correlators}. Numerous examples
are considered in Section \ref{Examples}.
Conclusions
and perspectives are discussed in Section \ref{conc}.

\section{Fields,   currents and charges}
\label{currents}
\renewcommand{\W}{W}
\renewcommand{\U}{U}
\subsection{Fields}
\label{Fields} As   shown in \cite{Vasiliev:2001zy}, conformal
fields in various dimensions are conveniently described in terms of
generalized space-time $\M_M$ with symmetric matrix coordinates
$X^{AB}=X^{BA}$ ($ A,B = 1\ldots M$). In the unfolded formulation of
\cite{Vasiliev:2001zy}, conformal fields are formulated in terms of
0-forms $C^\pm (Y|X)$ that depend both on the coordinates $X^{AB}$
and on the spinor (twistor) coordinates $Y^A$.
Rank-one unfolded  equations are
\be \label{dgydgyh+} \left (
\f{\p}{\p X^{ A B}} \pm \,i\,\hhh \f{\p^2}{\p Y^ A \p Y^ B}\right )
C^\pm(Y|X) =0\,. \ee
In the case of $M=2$, these are massless
equations for  scalar and spinor fields in three space-time
dimensions, described, respectively, by even
\big($C^\pm(Y|X)=C^\pm(-Y|X)$\big) and odd
\big($C^\pm(Y|X)=-C^\pm(-Y|X)$\big) functions of $Y$
\cite{Shaynkman:2001ip}.

For $M>2$, equations (\ref{dgydgyh+}) are
 related to conformal equations in higher dimensions as
discussed in
\cite{BL,Bandos:1999qf,Vasiliev:2001zy,Mar,Bandos:2005mb}. In
particular, at $M=4$ they describe a tower of $4d$
conformal (massless) fields of all spins
\cite{BL,Bandos:1999qf,Vasiliev:2001zy}. In this section we consider
the case of arbitrary $M$, keeping in mind that it is related to the
$AdS_4/CFT_3$ HS correspondence at $M=2$.

Unfolded formulation \cite{Ann} is useful in many respects (for more
detail and references see \cite{Bekaert:2005vh}). In particular,
given function $C(Y |0)$ of spinors $Y^A$, it reconstructs a
solution of Eq.~(\ref{dgydgyh+}) by
\be \label{XpoY} C ^\pm(Y|X) = \exp{\left (\mp i \hhh
X^{AB}\f{\p^2}{\p Y^A \p Y^B }\right )} C(Y|0)\,. \ee

As explained in  \cite{gelcur}, $+$ and $-$ in Eq.~(\ref{dgydgyh+})
distinguish between positive and negative frequencies, \ie particles
and antiparticles. Namely,
\be \label{cbfopmX+} C^\pm(Y|X)
=\f{1}{(2\pi)^{\f{M}{2}}} \int d^M\xi\,c^\pm (\xi) \exp \left[\pm
\,i \Big ( \hhh\,\, \xi_{\za}\xi_{\zb} X^{\za\zb}+Y^B \gx_B \Big
)\right]\,
\ee
are complex conjugated to each other \be \label{cc}
c^- (\xi) = \overline{c^+ (\xi)}\q C^- (Y|X) = \overline{C^+
(Y|X)}\,. \ee

It is useful to analytically continue $C^\pm(Y |X )$ to $C^\pm(\gY
|\Z )$ \cite{gelcur} where
 \be\label{Sieg} \Z^{AB}
=X^{\za\zb}+i\,\X^{\za\zb}\q X^{\za\zb}= \Re \Z^{AB}\q \X^{\za\zb}=
\Im \Z^{AB}\,, \ee
 \be
\label{SiegY}\gY^{\za}=Y^{\za}+i\,\Y^{\za}\q Y^{\za}= \Re \gY^{A}\q
\Y^{\za}= \Im \gY^{A}. \ee The real part $X^{AB}$  of $\Z^{\za\zb}$
is identified with the coordinates of the generalized space-time
containing
 Minkowski space as a subspace. The imaginary part
$\X^{\za\zb}$ is required to be positive definite and was treated in
\cite{Mar} as a regulator that makes the Gaussian integrals
well-defined (\ie physical quantities are obtained in the limit
$\X^{\za\zb}\to 0$; note, that the complex coordinates $Z^{\za\zb}$
of \cite{Mar} are related to $\Z^{\za\zb}$ as $ \Z^{AB} =
i\bZ^{AB}$).
 $\Z^{\za\zb}$ belong to the upper Siegel half-space $\ZIGM$ \cite{Siegel,Mumford}.
Evidently,  $-\cZ^{AB}\in \ZIGM $ provided that $\Z^{AB}\in \ZIGM $
and vice versa. The complex variables $\gY^A$ extend  Siegel space
to \emph{ Fock-Siegel space} $\GZM$.

Continuation of the functions $ {C}^\pm$(\ref{cbfopmX+}) to
Fock-Siegel space $\GZM$ is
 \bee \label{Csieg+}  {C}^+(\gY|\Z)
&=&\f{1}{(2\pi)^{\f{M}{2}}}\int d^M \gx\,\, {c}^+(\gx)\exp
\left[i\big(\,\hhh \gx_A\gx_B \Z^{AB}+ \gx_A
\gY^A\big)\right]\,,
\\
\label{Csieg-}  {C}^-(\cY|\cZ) &=&\f{1}{(2\pi)^{\f{M}{2}}}\int d^M
\gx\,{c}^-(\gx)\exp \left[- i\big(\,\hhh  \cZ{}^{AB}\gx_A\gx_B
+ \gx_A \cY{}^A\big)\right]\,\,.
\eee

 Generalization of the conventional classical field
to unfolded dynamics approach is provided by mutually conjugated {\it
full  fields} \be\label{lowpmq}
\Cp_j(Y|X)={C}_j^+(Y|X)+i^{\ppi_j} {C}_j^-(i Y|X)\q \Cm_j(
Y|X)= {C}_j^-(Y|X)+i^{\ppi_j} {C}_j^+(i Y|X)\,, \,\ee
where $j=1,\ldots,\NNN$ is the color index,
 and parity  $\ppi_j=0,1 $ distinguishes
between bosons and fermions
\be\label{simos0}
c_j^\pm(-\xi)=(-1)^{\ppi_j} c_j^\pm(\xi).
\ee

 Note that, in terms of representations of the
$AdS_4$ symmetry $sp(4;\mathbb{R})$, $3d$ conformal fields
correspond to singletons. Hence, adding a singleton
is equivalent to the change ${\mathcal N}\to{\mathcal N}+1$ of the number
${\mathcal N}$ of $3d$ boundary conformal
fields, which fact is in agreement with the results of \cite{Leigh:2012mz}
where, however, singleton was described as a bulk field.

Let $\rho$ be an involutive linear map \be\label{rho}\rho(
{C}_j^{\pm}(\yy|X ))=(i)^{\ppi_j}{{ C}}_j^{\mp}(i\yy|X )\,,
\ee \be\label{rhoroC} \rho^2( {C}_j^{\pm}(\yy|X ))=(-1)^{\ppi_j}
{C}_j^{\pm}(-\yy|X )= {C}_j^{\pm}(\yy|X )\quad:\quad \rho^2
= Id\,. \ee
Then \be\label{lowpmqrho} \Cp_j(Y|X)=(1+
\rho)({C}_j^{+}(\yy|X )) \q \Cm_j(Y|X)=(1+
\rho)({C}_j^{-}(\yy|X )) \,. \ee Hence \be\label{rhoi}\rho\left(
\Cp_j (\yy|X )\right) =  \Cp_j(\yy|X ) \q \rho\left( {\Cm}_j{}
(\yy|X )\right) =  \Cm_j (\yy|X ) . \ee Note that \be\label{rhoii}
\Cp_j (\yy|X ) = i^{\ppi_j} \Cm_j(i\yy|X ) \q {\Cm}_j{} (\yy|X ) =
i^{\ppi_j} \Cp_j (i\yy|X )\,. \ee

The factors of $i$ in (\ref{lowpmq}) are introduced in such a way
that, containing positive and negative frequencies, $\Cp_j(Y|X)$ and
${\Cm}_j{} (\yy|X )$ obey unfolded equations
 with definite signs of the second terms
\be \label{dgydgyhq} \left ( \f{\p}{\p X^{ A B}} + \,i\,\hhh
\f{\p^2}{\p Y^ A \p Y^ B}\right ) \Cp_j{} (Y|X) =0\q \left (
\f{\p}{\p X^{ A B}} - \,i\,\hhh \f{\p^2}{\p Y^ A \p Y^ B}\right )
\Cm_j{} (Y|X) =0 \,. \ee

\subsection{Currents and charges}
\label{cch} Let us recall, following \cite{gelcur}, main properties
of the unfolded formulation of conserved currents. Rank-two field
equation reads as \cite{tens2} \be \label{rank2eq} \left (\f{\p}{\p
X^{ A B}} - \,i\,\hhh \f{\p^2}{\p U^{( A}\p V^{B)}}\right ) \,
\PPP{}(U,V|X) =0\,.\quad \ee $ \PPP{}(U,V|X)$, that satisfies
Eq.~(\ref{rank2eq}), will be called rank-two current field (or,
simply, current). Clearly, rank-two fields can be interpreted as bi-local
 fields in the twistor space. In this respect they are somewhat
analogous to space-time bi-local fields sometimes used for the description
of currents (see e.g \cite{Jevicki:2012fh,Nikolov:2001iz} and references
therein). The important difference between these two approaches is however that
rank-two fields  are unconstrained in the twistor space while bi-local space-time fields should obey additional differential
conditions with respect to the doubled coordinates. It is this property that
makes  twistor description  efficient.

In unfolded formulation, fundamental dynamical
fields, which are primaries in the conformal setup, are described by
cohomology of the operator $\sigma_-$. For Eq.~(\ref{rank2eq}), this
is the cohomology group $H^0(\gs_-)$ where
\be\label{sigma-}\sigma_-=dX^{ A B}\f{\p^2}{\p U^{ A}
\p V^{B}}\,.
\ee
As shown in \cite{tens2}, the primary currents at any $M$ are
represented by the following components $j$ of $\PPP{}(U,V|X)$
\be\label{primcurUV}
  j_1(U|X)=\PPP{}(U,0|X)\q j_2(V|X)=\PPP{}(0,V|X)\q j_{1,2\,} =C_{AB}(X)U^A V^B\q
  \ee
  where
$C_{AB}(X)=-C_{BA}(X)$. All other components of $\PPP{}(U,V|X)$ are
expressed via derivatives of the primary currents by
unfolded equations (\ref{rank2eq}).

Any current $\PPP (U,V|X)$ can be decomposed into a sum of currents
$\PPP^h{}(U,V|X) $ of different ``helicities" $h$, that satisfy \be
H \PPP^h{}(U,V|X) = h \PPP^h{}(U,V|X)\,, \ee where the helicity
operator \be\label{helicitytoka} H=\half\left(V^A\f{\p}{\p
V^A}-U^A\f{\p}{\p U^A}\right) \ee commutes to the operator on the
\lhs of Eq.~(\ref{rank2eq}), and $h$ is some integer or
half-integer.

Since  $\PPP( U,V |X)$ obeys (\ref{rank2eq}), one can use  the
evolution formula analogous to (\ref{XpoY}) \be\label{Xsdvigtoka}
\PPP( U,V |X ) = \exp{\left (  i \hhh X^{AB}\f{\p^2}{\p V^A \p
U^B}\right )}\PPP  (U,V |0 )\,. \ee

A rank-two current field gives rise to the $M$-forms \cite{gelcur}
\be\label{Warpi} \Warpi(\PPP)\, =\, \half\, \Big(i\hhh\, d\,
X{}^{AB}\f{\p}{\p U^B} +    d\, V{}^A
\Big)^{\,M}\,\,\PPP(U,\,V\,|X)\Big|_{ \,U=0}\,\qquad%\q
\ee and \be \label{Warpi'} \widetilde{\Warpi} {}(\PPP)\,= \,\half \,
\Big(i\hhh\, d\, X{}^{AB}\f{\p}{\p V^B} +    d\, U{}^A
\Big)^{\,M}\,\,\PPP(U,\,V\,|X)\Big|_{V=0}\,\q \ee that are closed by
virtue of (\ref{rank2eq}). As a result, on solutions of
(\ref{rank2eq}), the charges \be \label{Q}   Q (\PPP)=\int_{\Sigma }
\Warpi (\PPP)\q \widetilde{Q} (\PPP)= \int_{\Sigma }
\widetilde{\Warpi} (\PPP) \ee are independent of local variations of
a $M$-dimensional integration surface $\Sigma \subset \M_M\otimes
\mathbb{R}^M$. For currents that decrease fast enough at space
infinity of $\M_M$, the charges (\ref{Q}) turn out to be independent
of the time parameter in $\M_M$, thus being conserved.

An important realization of a rank-two current field is provided by
the generalized stress tensor constructed from bilinears of free
rank-one fields $C_i^\pm$ in $\M_M\otimes \mathbb{R}^{M}$, that
satisfy (\ref{dgydgyh+}) and carry   color indices $i,j=1,2,\ldots \NNN$
\be\label{Tbilin} T_{ij}( U,\,V\,|X) =C_i^+( V-U|X)\,C_j^-(
{U}+V|X)\,. \ee

 Eq. (\ref{rank2eq}) is equivalent to
\be \label{rank2eqY} \left (\f{\p}{\p X^{ A B}} + \,i\,\hhh
\f{\p^2}{\p \yy_1^{  A}\p \yy_1^{B}} -\,i\,\hhh \f{\p^2}{\p \yy_2^{
A}\p \yy_2^{B}}\right ) \, \T_{ij}(Y_1,\,Y_2\,|X)=0\,\q \ee where
\be\label{UVYY} %\yy_1=V-U\q \yy_2=V+U\,.
\T_{ij}(Y_1,Y_2|X) = T_{ij} (U(Y),V(Y)|X)\q V(Y)=\half(\yy_1+\yy_2)\q
U(Y)= \half(\yy_2-\yy_1)\,. \ee

The first-order differential operators $\A(\yy|X)$ \be\label{param12}
\A_a{}^B(\yy|X)=2i\hhh X^{AB}\f{\p}{\p \yy^A}-a\yy^B\,\q\,
\A_a{}_C(\yy|X)= \f{\p}{\p \yy^C}\,\q a=+,- \,, \ee which have the
property \be\label{param12i}
\A_{\pm}{}(-i\yy|X)=i\A_{\mp}{}(\yy|X)\,, \ee obey
 \be \label{dgydgyh1} \left [ \f{\p}{\p X^{ A B}}
\pm \,i\,\hhh \f{\p^2}{\p \yy^A \p \yy^
B}\,,\,\A_{\pm}{}(\yy|X)\right ]=0\,. \ee
Hence, any polynomial of
these operators $\eta(\A_{+}{}(\yy_1|X),\A_{-}{}(\yy_2|X))$ obeys
\be \left[ \left (\f{\p}{\p X^{ A B}} + \,i\,\hhh \f{\p^2}{\p
\yy_1^{  A}\p \yy_1^{B}} -\,i\,\hhh \f{\p^2}{\p \yy_2^{  A}\p
\yy_2^{B}}\right)\,,
\eta(\A_{+}{}(\yy_1|X),\A_{_-}{}(\yy_2|X))\right ] =0\,. \ee
Therefore, \be J_\eta(Y|X) =
\eta^{ij}(\A_{+}{}(\yy_1|X),\A_{-}{}(\yy_2|X)) \T_{ij}(
\yy_1,\,\yy_2\,|X) \ee
obeys the current equation (\ref{rank2eq}) with the identification (\ref{UVYY}).

Most of polynomials $ \eta^{ij}(\A )$  lead  to  exact forms
(\ref{Warpi}) and (\ref{Warpi'}) and, hence, zero charges (\ref{Q}).
Indeed, consider the following operators \bee \label{parameters}
\B_A(U,V|X) &=& \f{\p}{\p U^A} =-\A_+{}_A(\yy_{1
}|X)+\A_-{}_A(\yy_{2 }|X)\,\q
\\ \nn
\B{}^B(U,V|X)  &=& i\hhh X^{AB}\f{\p}{\p
U^A}+V^B=\half(-\A_+{}^B(\yy_{1 }|X)+ \A_-{}^B(\yy_{2 }|X))\q \eee
\bee\label{exparam} \widetilde{\B}_A(U,V|X) &=& \f{\p}{\p
V^A}=\A_+{}_A(\yy_{1 }|X)+\A_-{}_A(\yy_{2 }|X)\,\q
\\ \nn  \widetilde{\B}{}^B (U,V|X) &=&\, i\hhh X^{AB}
\f{\p}{\p V^A}+U^B =\half(\A_+{}^B(\yy_{1 }|X)+\A_-{}^B(\yy_{2 }|X))
\,, \qquad \eee which  obey  Heisenberg commutation relations
\be\label{compar} [\widetilde{\B}{}_B(U,V|X),
\B{}^C(U,V|X)]=\gd_B^C\q[\B{}_B(U,V|X),
\widetilde{\B}{}^C(U,V|X)]=\gd_B^C\,. \ee

As  shown in   \cite{BRST2010}, the form $\Warpi(\PPP_\eta)$
(\ref{Warpi}) is exact provided that
\be\label{zeropar}
\eta(\B\,,\widetilde{\B} )= \widetilde{\B}{}^B(U,V|X) \eta_B(\B\,,\widetilde{\B}   ) +
\widetilde{\B}_A(U,V|X) \eta^A(\B  \,,\widetilde{\B} ) \ee for some $\eta_A$ and
$\eta^A$. Analogously, $\eta(\B  \,,\widetilde{\B})$ of the form
\be\label{zeropar'} \eta(\B\,,\widetilde{\B} )= \B{}^B(U,V|X)
\eta_B(\B \,,\widetilde{\B}  ) + \B_A(U,V|X) \eta^A(\B  \,,\widetilde{\B} )
\ee
leads to the exact form
$\widetilde{\Warpi}(\PPP_\eta)$  (\ref{Warpi'}). As a result,
nonzero charges $Q $ and $\widetilde{Q}$ are represented,
respectively, by $\widetilde{\B}$-independent ${\eta}(\B(U,V|X))$
and  $\B$-independent  ${\eta}(\widetilde{\B}(U,V|X))$
which can be interpreted as
 parameters of global HS symmetry transformations generated by the charges
\bee\nn Q(\PPP_{\eta })= Q(\PPP_{{\eta(\B) }}) \q
\widetilde{Q}(\PPP_{\eta })= \widetilde{Q} (\PPP_{
\eta(\widetilde{\B})}) \,. \eee

Since
\be\label{heliB} H (\B )= -\half \B \q H (\widetilde{\B} )=
\half \widetilde{\B}\,,
\ee where $H$ is the helicity operator
(\ref{helicitytoka}),  the charges $\widetilde{Q} $ and $Q $ are
supported by the parameters of non-negative and non-positive
helicities, respectively. As shown in Section \ref{Charges}, the
charges $\widetilde{Q} $ and $Q $  generate the $N=2$
supersymmetric HS algebra where the number of charges of a given
spin is doubled.

 Consider involutive maps  $ P_{1 },P_2 $ ($(P_i)^2=1$)
\be  \label{hiYY} P_1 \big(A(\yy_1,\yy_2)\big)=A(-\yy_1,\yy_2)\,\q
P_2 \big(A(\yy_1,\yy_2)\big)=A( \yy_1,-\yy_2)\,. \ee Equivalently,
in terms of  $ U,V $,
\be \label{hiUV} P_1\big(\B(V,U
)\big)=\B(U,V)\,\q P_2\big(\B(V,U )\big)=\B(-U,-V)\,. \ee
 Since from
Eqs.~(\ref{parameters}), (\ref{exparam}) it follows that
$\widetilde{\B} \Leftrightarrow\B$ when $U  \Leftrightarrow V$, we
observe that \bee  \label{mapB+-}&  P_1\big(\B
(U,V|X)\big)=\widetilde{\B}(U,V|X)\q&
  P_1\big(\widetilde{\B} (U,V|X)\big)= {\B} (U,V|X)\q
\\ \nn&
   P_2\big(\B (U,V|X)\big)=\widetilde{\B}(-U,-V|X)\q &
 P_2\big(\widetilde{\B} (U,V|X)\big)= {\B} (-U,-V|X)\,.\eee
For the bilinear stress tensor $T_{ij}$  (\ref{Tbilin}) this gives
\be  \label{TVU} \widetilde{T}_{ij}( U,\,V\,|X):=
P_1\big(T_{ij}(U,\,V\,|X)\big) =C_i^+(-  V+U |X)\,C_j^-( {U}+V|X)\,.
\ee Since $C_i^+(-Y|X)$  solves  (\ref{dgydgyh+}) provided that
$C_i^+(Y|X)$ does, $\widetilde{T}_{ij}( U,\,V\,|X)$ is also a
bilinear stress tensor (\ref{Tbilin}).
 Hence
\be\label{chiWarpi}  P_1\left(\Warpi (\PPP_{\eta(\B)})\right)\,=\,
\widetilde{\Warpi}(\widetilde{\PPP}_{\eta(\widetilde{\B})})\q
  \widetilde{\PPP}_{\eta(\widetilde{\B})}=
\eta^{ij} (\widetilde{\B})\widetilde{T}_{ij}( U,\,V\,|X) \ee and
 \be \label{Q+-} Q (
\PPP_{\eta(\B)})=\widetilde{Q}(\widetilde{\PPP}_{\eta(\widetilde{\B})}).
\ee

The construction of currents admits several generalizations. In
particular, rank-$2r$ currents considered in \cite{gelcur} obey
\be\label{rank2reqX} \left (\f{\p}{\p X {}^{ A B}} -
\,i\,\hhh\sum_j^r \f{\p^2}{\p U_j{}^ A \p V_j{}^ B}\right ) \,
\PPP^{2r}{}(U,V|X) =0\,. \ee
A multilinear ``stress tensor" is
\be \label{stress2r}
T^{2r}(U,\,V\,|X)=\prod_{j=1}^r\,C^+_{i_j}(V_j-U_j|X)\,C^-_{k_j}(U_j+V_j|X),
\ee where $C^\pm_m(Y|X) $ are solutions of positive- or
negative-frequency rank-one   equations. Then
\be\label{prodtensX}
\PPP^{2r}{}(U,V|X)=\eta(\B,  \widetilde{\B} ) T^{2r}(U,\,V\,|X), \ee
where $\eta(\B  \,,\widetilde{\B} )$ are polynomial parameters that depend
on   $ \B_j$  (\ref{parameters}) and $\widetilde{\B_j}$(\ref{exparam})
with $U\to U_{j}$ and $V\to V_{{j}}$, solves (\ref{rank2reqX}).  The
multilinear stress tensor (\ref{stress2r}) is nothing but the
bilinear stress tensor (\ref{Tbilin})  built from the rank-one
solutions in $\M_{rM}$, that result from rank-$r$ solutions in
$\M_M$ realized by the $r$-linear products of  rank-one   solutions in
$\M_M$.

Alternatively, one can consider multilinear currents of the form
\be\label{Ntok} \JJJ^{2r}_{\eta_1\ldots\eta_n }(\yy_1, \ldots,
\yy_{2r}|X_1,\ldots,X_r) =\prod_j \JJJ _{\eta_j }(\yy_{2j-1},
\yy_{2j}|X_j)\q \ee where $\eta_j=\eta_j(\A(\yy_{2j-1},\yy_{2j
}|X_j) )$ with $\A$ (\ref{param12}).

It is important to note that unfolded equation (\ref{rank2reqX}) as
well as its particular cases of ranks one (\ref{dgydgyh+}) and two
(\ref{rank2eq}) have a form of covariant constancy conditions \be
\label{D0} D_0 \C =0\q D_0 = d+\omega_0 \,, \ee where $\C$ is any of
the fields $C$, $\Phi$ or $J$ and $D_0$ is the covariant derivative
of the conformal algebra $sp(4)$ for $M=2$ or its extension $sp(2M)$
for higher $M$ (see e.g. \cite{Vasiliev:2001zy}). To describe a
conformally flat space-time, the connection $\go_0$ is required to
be flat \be D_0^2 =0\,. \ee The  equations considered in this paper
represent (\ref{D0}) in the particular case of   $\go_0$
corresponding to Cartesian coordinate system where the only nonzero
component is vielbein associated with the
 translations in the conformal group. However, one can equally well
describe the same systems in any other conformally invariant
coordinates associated with different $\go_0$.

Moreover, in accordance with the general analysis of
\cite{Vasiliev:2012tv}, this  allows one immediately extend
consideration to a larger space exhibiting the same symmetry simply
by adding additional coordinates. The particularly important case is
the extension of $d$-dimensional Minkowski space to $AdS_{d+1}$.
(For the constructive definition of the respective connections see
\cite{Vasiliev:2012tv}.) As discussed in some more detail in the
preamble of  Section \ref{Vertex algebra}, this simple  observation
makes the  correspondence between boundary and bulk computations
nearly tautological.

 \section{$\D-$functions}

\label{Dfunc}
  $\D-$functions  of the massless
Klein-Gordon-like equation \be \label{oscal} \Big ( \f{\p^2}{\p X^{
A B} \p X^{ C D}} - \f{\p^2}{\p X^{ A C} \p X^{ B D}}\Big ) b(X)=0\,
\ee and the Dirac-like equation \be \label{ofer} \f{\p}{\p X^{ A B}}
f_ C(X) - \f{\p}{\p X^{ A C}} f_ B(X) =0\,  \ee in $\M_M$ were
introduced in \cite{Mar} as their singular solutions resulting from
the integral representation (\ref{cbfopmX+})
\be \label{cD}
\D^\pm(X) =\mp  i(2\pi)^{-M} \int  d^M \gx\,\,
\exp\left[  \pm i\hhh ( \gx_A\gx_B X^{AB})\right]\,. \ee
 The $Y$ dependence of
$\D-$functions reconstructed in \cite{gelcur} via the unfolded
equations  (\ref{dgydgyh+}) is
\be \label{DXY} \D^\pm (\yy|X) = \mp
i(2\pi)^{-M} \int  d^M \gx\,\, \exp\left[  \pm
i(\hhh \gx_A\gx_B X^{AB}+\gx_A\yy^A)\,\right] \,. \ee Note that
normalization  of  $\D^\pm$ (\ref{cD}), (\ref{DXY}) differs from
that of \cite{Mar,gelcur} by the  factor of $2^{-M}$ to achieve
 \be\label{Ddelta}\D^\pm (\yy|X)\Big|_{X\to 0}= \mp i
\delta^M(\yy)\, .\quad
\ee \!Hence,  $\D^\pm (\yy|X)$ (\ref{DXY})
belong to the class of solutions of (\ref{dgydgyh+}), that includes
distributions in $\yy$.

Recall that $\D$-functions have to be distinguished from Green
functions. The former solve homogeneous field equations with
$\delta$-functional initial data while the latter solve
inhomogeneous field equations with the $\delta$-functional
right-hand-side. Nevertheless they have similar form away from
singularity. As discussed in \cite{Mar}, the difference is
essentially due to  step-functions in time. As a result, away from
singularity,  computation of chronologically ordered functions with
causal propagators gives the same result as  Wightman functions
 with appropriately ordered arguments. In this paper we compute Wightman
 functions based on $\D$-functions.

Continuation of $\D^+$ to Fock--Siegel space
\be  \label{D^+} \D^+(\gY|\Z)
=  {-i}{ (2\pi)^{-M}} \int d^M\xi\,  \exp\left[ i (\hhh \Z^{\za\zb}\,
\xi_{\za}\xi_{\zb}+\gY^A\xi_A )\right] , \ee gives \bee \label{dy} \D^+
(\gY|\Z) =  \f{-i}{2^M \pi^{M/2}}
\,\,s^{-1} \,\exp ( - \f{ i}{4\hhh} \Z^{-1}_{AB}\gY^A \gY^B )\q
\Z^{-1}_{AB}\Z^{BC} =\delta_{A}^{C}\,, \eee where
\bee\label{dete}
s^2(\Z)={\det \Big(-i\hhh\Z   \Big)} \,.
\eee Evidently, \be
\label{D^-}
\D^-(\gY|\Z)=\overline{\D}^+(\cY|\cZ)\,.
\ee
 As shown in \cite{Mar}, for non-degenerate $X^{AB}$
\be \label{Dsol} \D^\pm (X )
=\mp\f{i}{2^M\pi^{\f{M}{2}}} \exp \left(\pm\f{i\pi I_X}{4}\right)
\f{1}{\sqrt{\hhh|\det (\Z)|}}\Big |_{\Im\Z\to 0}\,, \ee
 where \be \label{iner} I_X = n_+ - n_-  \ee is the inertia index of
the matrix $X^{A\,B}$ with $n_+$ and $n_-$ being, respectively, the
numbers of positive and negative eigenvalues of $X^{AB}$.   Integral
representation (\ref{D^+}) provides  definition of the regularized
expression in the complex space.

 {}From (\ref{dy}) and (\ref{Dsol})  it follows that
\bee\label{DInert+}\D^\pm (\yy |X)=\D^\pm (X )\exp \left( \mp \f{ i}{4\hhh}
X^{-1}_{AB}\yy^A \yy^B \right)\,=\qquad\qquad \\ \nn
\mp\f{i }{2^M  \pi^{\f{M}{2}}}
\exp
\left(\pm\f{i\pi I_X}{4}\right)
\f{1}{\sqrt{\hhh|\det (\Z)|}}
\exp
\left( \mp \f{ i}{4\hhh} \Z^{-1}_{AB}\yy^A \yy^B \right)\Big
|_{\Im\Z\to 0}      \,. \eee

  $\D$-functions provide  integral representation for
solutions of  field equations. The simplest way to see this is to
observe that $\D^\pm(Y'-Y| X'- X )$ solves (\ref{dgydgyh+})
with respect to both $\XX, Y'$ and $X,Y$. Hence, with respect to
primed variables,
 \be
\label{DCDC}  \D^\pm(V'\mp U'-Y| X'- X) C^\mp(V'\pm U'|X')  \, \ee
is a bilinear stress tensor of the form (\ref{Tbilin}) that depends
on $Y$ and $X$ as parameters. The respective (spin one)
conserved charge gives a solution  $C^\mp(Y|X)$ of (\ref{dgydgyh+})
independent of the choice of integration surface and reproducing
$C^\mp(Y'|X')$ at $X=X'$, $Y=Y'$ due to (\ref{Ddelta}).

In this paper, we will use the evolution formulae in the space
of  $\yy$ variables \cite{gelcur},  which  in the limit $\Im \Z\to0$
can be conveniently written  as
\be\label{DCYlow} {C}^a( \yy|X) =
i\int d^M  \yy^\prime\,\, \D_a(\yy^\prime - \yy|X^\prime-X)
C^a(\yy^\prime |X^\prime)\qquad \mbox{no summation over $a=+,-$}\,,
\ee
 where
 \be\label{lowindD} \D_a( \yy|
X)=\gvep_{ab}\D^b( \yy| X) \q \gvep_{-+}=-\gvep_{+-}=1\,. \ee
 The following simple consequence of
 (\ref{DCYlow}) plays the key
  role in the analysis of   Section \ref{Contact}
\be\label{intDD} \int      d^M p \,\D_{-}(p - \yy |X) \D_{+}(p
-\yy' |X')=
 - i  \D_{-}( \yy -\yy' |X-X')=-i  \D_{+}( \yy -\yy' |  X'-X)
   \,.
\ee

Away from singularities, \ie at $\det X\ne0$, from
Eqs.~(\ref{DInert+})  it follows that \be
\label{intDDpmM}
     \D_{\pm}( \yy| X  )=
   \D_{\pm}(  X  ) \exp
\left( \pm \f{ i}{4\hhh} X^{-1}_{AB}\yy^A \yy^B \right)
 \q \D_{+}( \yy| X  )=
\D_{-}( \yy|- X  )
 \,, \ee
 \be\label{intDDinM}
 \D_{-}( i Y| -X  )=
 \D_{+}( i Y| X  )=\exp\left( \f{ i\pi I_{ X  }}{2}\right)\D_{-}(  Y|  X )
    =\exp\left( \f{ i\pi I_{ X  }}{2}\right)   \D_{+}(  Y| - X )
   \,.
 \ee

Formula (\ref{DCYlow}), providing at fixed $X^\prime$
the map from the twistor space to (boundary) $X$ space, allows us
to call $\D_a$ twistor-to-boundary $\D$-functions. In accordance with
the discussion in the end of Section \ref{cch}, their extension to
twistor-to-bulk $\D$-functions  with the same initial data (\ref{Ddelta})
provides an extension of the whole setup to the bulk.

 \section{Quantization}
\label{Quantization}

 As usual, the decomposition into positive-- and
negative--frequency parts (\ref{cbfopmX+})  gives rise to the
quantum  creation and annihilation operators \cite{Mar} $\oc_j^\pm
(\xi)$ ($j$ is color index).
The
(anti)commutation relations are \cite{Mar} \bee\label{comrelYCbf}
[\oc_j^- (\xi_1) , \oc_k^+ (\xi_2  )]_\pm =\gd_{jk}
\half\Big(\delta^M(\xi_1-\xi_2)+(-1)^{\ppi_j\ppi_k}
\delta^M(\xi_1+\xi_2)\Big)\q
 [\oc_j^\pm (\xi_1) , \oc_k^\pm (\xi_2  )]_\pm =0\,,
 \eee
where   $\ppi_j=0,1 $ is the boson-fermion parity (\ref{simos0}).
For quantum fields
\be \label{Chat+}
\hat{C}_j^\pm(\yy|X) = {(2\pi)^{-\f{M}{2}}}\int d^M
\gx\,\,\oc_j^\pm(\gx)\exp\left[ \pm i \big(\,\hhh  \gx_A\gx_B X^{AB}+\,
 \gx_A \yy^A\big)\right]\,
\quad
\ee
this gives
\be\label{comCCXX} \ls
[\hat{C}_j^-(\yy|X)\,,\,\hat{C}_k^+(\yy'|X')]_\pm=\f{1}{2i}\gd_{jk}
\left\{\D^-\left(  \yy-\yy'  |X-X'\right) +(-1)^{\ppi_j
\ppi_k}\D^-\left( \yy+\yy' |X-X'\right)\right\},\quad \ee with
$\D^-$   (\ref{DXY}). By virtue of (\ref{Ddelta}), at $X=X'$ this
gives \be\label{comCCXX0}
[\hat{C}_j^-(\yy|X)\,,\,\hat{C}_k^+(\yy'|X )]_\pm= \half
\gd_{jk} \big(\gd(\yy- \yy')+(-1)^{\ppi_j \ppi_k}\gd(\yy+ \yy')\big)
\,.\quad \ee

\renewcommand{\Cp}{\hat{\Phi}}
\renewcommand{\Cm}{ \hat{\overline{\Phi}}}
\renewcommand{\T}{ \hat{\cal{T}}}

Generalized  quantum HS stress tensor is
\be\label{Tq} \T_{j\,k}(\yy_1,\yy_2|X)=: { \Cp}_j (  \yy_1|X)
\Cm_k (  \yy_2|X ) :\, \ee with
the quantized full fields
(\ref{lowpmq}) ${ \Cp}_j$,  ${ \Cm}_j$. As usual, normal ordering  sends $\hat{C}^+$ to the
left and $\hat{C}^-$ to the right, \ie \be\label{normord}
:\hat{C}^-_k(  \yy_2|X) \hat{C}^+_j(  \yy_1|X):=
 (-1)^{\ppi_j\ppi_k}\hat{C}^+_j(  \yy_1|X) \hat{C}^-_k(  \yy_2|X)\,.
\ee Substitution  of (\ref{lowpmq})   into (\ref{Tq}) gives
\be\label{rhoT}
\T_{j\,k}(\yy_1,\yy_2|X)%\Big|_{\yy_1 =(U-V)  \,, \yy_2=(U+V)}
=\sum_{a,b=+,-} (\kappa^a_1)^{\ppi_j} (\kappa^b_2)^{\ppi_k}
\TT^{ab}_{j\,k}(\kappa^a_1\yy_1,\kappa^b_2\yy_2|X)\,,\qquad \ee where
\be \label{Tab} \TT^{a\,b}_{j\,k}( \zz_1, \zz_2|X)=
\,:\hat{C}_j^a(\zz_1 |X) \hat{C}_k^b(  \zz_2|X): \, \, \ee and
\be\label{kappa}\kappa_1^+=\kappa_2^-=1\q\kappa_2^+=\kappa_1^-=i\,.
\ee
 Note that
 \bee \label{parTab}&& \TT^{a\,b}_{j\,k}( \zz_1, \zz_2|X)=
(-1)^{\ppi_j\ppi_k}\TT^{b\,a}_{k\,j}( \zz_2, \zz_1|X)
 \q\\ \nn&&\TT^{a\,b}_{j\,k}( \zz_1, \zz_2|X)=
(-1)^{\ppi_j} \TT^{a\,b}_{j\,k}( -\zz_1,  \zz_2|X)=
 (- 1)^{\ppi_k}\TT^{a\,b}_{j\,k}( \zz_1,- \zz_2|X)  \,
\, \qquad \eee and
  \be\label{kappaprop}(\kappa_1^a)^{-1}=a\kappa_1^a
\q(\kappa_2^a)^{-1}=-a\kappa_2^a\q i\kappa_2^a=-a\kappa_1^a \,. \ee
\vskip5pt

Let us extend the action of $\rho$ (\ref{rho}) to
$\T{}_{j\,k}( \yy_1\,,\yy_2|X)$. Assuming that $\rho$ is an
antiautomorphism and using (\ref{rhoi}) we obtain
 \be\label{Tant}\rho\left(
\T_{j\,k}(\yy_1,\yy_2|X)\right)=\,: \rho\left(\Cm_k (  \yy_2|X )\right)  \rho\left(
\Cp_j ( \yy_1|X)\right): \,=\, (-1)^{ \ppi_k\ppi_j}\T_{j\,k}(\yy_1,\yy_2|X) \,.
\ee
Also, by virtue of (\ref{rhoii})  along with
 \be\label{ppijk} i^{\ppi_{j,\,k }} =
i^{\ppi_j+ \ppi_k}  (-1)^{ \ppi_k\ppi_j}\,\q \ppi_{j,\,k }=
(\ppi_j+\ppi_k\,) {\mod 2}\qquad ( \{p\mod 2\} =0\,\, \mbox{or} \,\, 1)\,,
\ee
where $\ppi_{j,\,k } $ is the boson-fermion parity of
$\T{}_{j\,k}( \yy_1\,,\yy_2|X)$,
\be\label{rhoTab}  \T{}_{j\,k}(
\yy_1\,,\yy_2|X)   =   i^{\ppi_{j,\,k }} \T{}_{k\,j}
(i\yy_2\,,i\yy_1|X)\,. \ee

{}From
\be
 { \Cp}^\dagger_j (  \yy|X) = \Cm_j (  \yy|X )\q
\ee
where $\dagger$ denotes Hermitian conjugation, it follows that
\be
\label{tdag}
\T^\dagger_{j\,k}(\yy_1,\yy_2|X)=\T_{k\,j}(\yy_2,\yy_1|X)
\ee
and, hence,
\be
\label{rdag}
\T^\dagger_{j\,k}(\yy_1,\yy_2|X)= i^{\ppi_{j,\,k }+2s}   \T_{j\,k}(\yy_1,\yy_2|X) \,,
\ee
since spin $s$ of a current is defined as the homogeneity degree in $Y$
\be \label{spin}
 \T{}_{j\,k}(\lambda\yy_1\,,\lambda \yy_2|X)=\lambda^{2s} \T{}_{j\,k}(\yy_1\,,\yy_2|X)\,.
\ee
 In particular, this implies that   $\T_{jk}$ is (anti)Hermitian for
(odd)even spins.

Consider the full current
\be\label{quanJeta} \JJJ_\eta(\yy_1,\yy_2|X) =
\eta^{j\,k} (\A  )\T_{j\,k}(\yy_1,\yy_2|X)\,,
\ee
called later on $F$-current,
with parameters $\eta^{j\,k} (\A )$   depending on $\A_\pm(\yy|X)$ (\ref{param12}). Since $\Cp_j$, $\Cm_j$
obey the unfolded   equations  (\ref{dgydgyhq}),
$\JJJ_\eta(\yy_1,\yy_2|X)$ obeys Eq.~(\ref{rank2eqY}). Hence,
$\Warpi (\JJJ_{\eta})$ (\ref{Warpi}) is a closed $M-$form, defining
a conserved charge (\ref{Q}).  $F$-currents
are analogues of classical currents in unfolded dynamics.
{\it Space-time currents} are
\be\label{quanJetaspace}
\JJJ_\eta(X)=\JJJ_\eta(\yy_1,\yy_2|X)\big|_{\yy=0}\,. \ee

To relate statistics of $\eta^{jk}$ to that of  $\T_{jk}$, \ie
$\ppi_{j,\,k }$ (\ref{ppijk}), we require
\be\label{Jot-Y}
\JJJ_\eta(-\yy_1,-\yy_2|X) =\JJJ_\eta(\yy_1,\yy_2|X)\,:\qquad
\eta^{j\,k} (-\A)=(-1)^{\ppi_{j,\,k }}  \eta^{j\,k} (\A)\ee with
Grassmann (odd)even $\eta^{jk}$ for $\ppi_{j,\,k }=(1)0$\,, \ie
\be\label{:etaeta':}
  \eta^{j\,k}(\A(\yy|X))\eta'{}^{m\,n}(\A(\yy'|X'))= (-1)^{\ppi_{j,\,k }\ppi_{m,\,n }}
\eta'{}^{m\,n}(\A(\yy'|X'))  \eta^{j\,k}(\A(\yy|X))\,\,.
 \ee
In addition,   symmetry parameters  are required to have the same
$Y$-parities as
  elementary fields
  \be\label{Comreleta} \eta^{k\,m}
(\A_+(\yy_{1}|X),\A_-(\yy_{2}|X) )
=(-1)^{\ppi_k}\eta^{k\,m}(\A_+(-\yy_{1}|X),\A_-(\yy_{2}|X) )
=(-1)^{\ppi_m}\eta^{k\,m} (\A_+(\yy_{1}|X),\A_-(-\yy_{2}|X) )\,.
\ee

 Taking into account properties (\ref{param12i}) of $\A$, let $\rho$
act   on   parameters  $\eta^{j\,k}(\A ) $ as
\be\label{rhoeta}
\rho\big(\eta^{j\,k} (\A_+(\yy_1|X)  ,\A_- (\yy_2|X) )\big)=
i^{\ppi_{j,\,k }} \eta^{k\,j} ( i\A_-(\yy_2|X) ,i \A_+ (\yy_1|X) )\,.\quad
\ee
Hence, by virtue of   (\ref{Jot-Y}), (\ref{Comreleta}) and  (\ref{rhoeta}),
the action of $\rho$ remains involutive on $\eta$. As a result,
\be\label{etaoe} \eta=\eta^+ +\eta^-\q \eta^\pm
=\half\big(\eta\pm\rho  (\eta)\big) \,. \ee
The
terms with $ \eta^-$ do not contribute to the space-time currents $
\JJJ_{\eta (X)}$. Indeed,
due to (\ref{param12i})
\bee\label{quanJetaspacerho}
\JJJ_{\eta^\pm }(\yy|X)\big|_{\yy=0}&=&
 \eta^\pm {}^{j\,k} (\A_+(\yy_{1}|X ),\A_-(\yy_{2}|X) )\T_{j\,k}(\yy_1,\yy_2|X)\big|_{\yy=0}
\\ \nn&=&
\pm\eta^\pm {}^{j\,k} (\A_+(i\yy_{2}|X ),\A_-(i\yy_{1}|X) )\T_{j\,k}(i\yy_2,i\yy_1|X)\big|_{\yy=0}
=\pm\JJJ_{\eta^\pm }(\yy|X)\big|_{\yy=0}\eee
since the substitution
$\yy\Rightarrow i\yy$  along with the exchange $ 1\Leftrightarrow
2$ has no effect at $\yy_{1,2}=0$.

More generally, we  introduce $A$-currents that contain
different frequency parts and can be used for
computation of amplitudes
 \be\label{rhoJ} \JNN_\gga(\yy_1,\yy_2|X  ) =  \sum_{a,b
}\JNN^{a\,b}_\gga(\yy_1,\yy_2|X)\,,
\ee
where
 \bee\label{rhoJX12}
 \JNN^{a\,b}_\gga(\yy_1,\yy_2|X )
=  (\kappa^a_1)^{\ppi_j} (\kappa^b_2)^{\ppi_k}\gamma_{a\,b}^{j\,k}
\left( \rule{0pt}{14pt} \A_+ \left( \yy_1|X \right),
 \A_-\left( \yy_2|X \right)\right)
\TT^{a\,b}_{j\,k}(\kappa^a_1\yy_1,\kappa^b_2\yy_2|X ) \eee with
arbitrary $\gamma_{a\,b}^{j \,k} \left( \rule{0pt}{14pt} \A_+ \left(
\yy_1|X\right),
 \A_-\left( \yy_2|X\right)\right)$.

$F$-current $\JJJ _{\eta{} }$ is a
particular case of $\JNN_{\gamma{} }$ at $
\gamma_{ab}^{k\,j}=\eta^{k\,j}$ $\forall\,\, a,b$, \ie
\be\label{JNNab}
 \JJJ_{\eta{} }=\sum_{a,b }\JJJ^{ab} _{\eta{} }\q
\JJJ^{ab} _{\eta{} }(\yy_1,\yy_2|X)=\JNN^{ab}_{\gamma{}
}(\yy_1,\yy_2|X) \Big|_{\gamma_{ab}=\eta\,}\,. \ee

The action  of $\rho$    on   parameters
$\gamma_{ab}^{j\,k}(\A_+(\yy_{1}|X ),\A_-(\yy_{2}|X ) ) $ is
defined analogously
\be\label{rhogga}%
\rho\big(\gamma_{ab}^{j\,k} (\A_+(\yy_1|X)  ,\A_- (\yy_2|X)
)\big)= i^{\ppi_{j,\,k }}\gamma_{ba}^{k\,j} (  i\A_-(\yy_2|X) , i
\A_+ (\yy_1|X) )\,.\quad \ee Again,
  \be\label{ggaoe} \gga=\gga^+ +\gga^-\q \gga^\pm
=\half\big(\gga\pm\rho  (\gga)\big) \,. \ee Analogously to
(\ref{:etaeta':}), we require \be\label{:ggagga':}
  \gga_{a\,b}^{j\,k}(\A(\yy|X))\gga'{}_{c\,d}^{m\,n}(\A(\yy'|X))=
  (-1)^{\ppi_{j,\,k }\ppi_{m,\,n }}
\gga'{}_{c\,d}^{m\,n}(\A(\yy'|X))
\gga_{a\,b}^{j\,k}(\A(\yy|X))\,.
 \ee

Now we are in a position to consider algebra of quantum charges
 and OPE of currents. The current operator algebra contains all quantum  $2r$-linear currents (\ref{Ntok})
which are normal-ordered products of  the bilinear ones
\be \label{cancurunfolr}
\JJJ^{2r}_{\eta_1\ldots\eta_r}(\yy_1,\ldots,\yy_{2r}|X_1,\ldots, X_r) =
 \eta^{j_{1}\,k_{1};\ldots ; j_{r}\,k_{r}}(\A_1,\ldots,\A_{ r} )
\T_{j_{1}\,k_{1};\ldots ; j_{r}\,k_{r}}
(\yy_1,\ldots,\yy_{2r}|X_1,\ldots,X_r)\,,
\ee where
\be\nn \eta^{j_{1}\,k_{1};\ldots ;
j_{r}\,k_{r}}(\A_1,\ldots,\A_{ r} ) = \prod_{\ga=1}^r\Big(
\eta_\ga^{j_{\ga}\,k_{\ga}} (\A_+(\yy_{2\ga-1}|X_\ga
),\A_-(\yy_{2\ga}|X_\ga )  )\Big)\,, \ee \be \label{stress2rY}
\T_{j_{1}\,k_{1};\ldots ; j_{r}\,k_{r}}(\yy_1,\ldots,\yy_{2r}|X_1,\ldots,
X_r) =\,\, :\Big(\prod_{\ga=1}^r
\T_{j_{\ga}\,k_{\ga}}(\yy_{2\ga-1},\yy_{2\ga}|X_\ga )\,\Big):\,.
\quad\ee

\section{Algebra  of  charges}
 \label{Charges}
\subsection{Charges}
\label{Manifestcharge}

As shown in Section \ref{currents}, nonzero charges $Q$ and
$\widetilde{Q}$ (\ref{Q}) are supported by currents with parameters
$\eta (\B )$  and $\widetilde\eta (\widetilde{\B} )$ with $\B$
(\ref{parameters}) and $\widetilde{\B}{}$ (\ref{exparam}),
respectively. The decomposition  (\ref{rhoJ}) of currents induces a
similar decomposition  of the differential forms (\ref{Warpi})  and
charges (\ref{Q})
\be\label{rhoQ} Q_{\eta }=\sum_{ab}Q_{\eta }^{ab}
\q Q^{ab}_{\eta }=\half \int_{\Sigma^{ab}} \Warpi(\PPP^{ab}_{\eta
})\, \ee
and similarly for $\widetilde Q$.

Let us first consider the charges $Q$ with parameters $\eta(\B)$
independent of $\widetilde{\B}{}$. In this case,
\be\label{rhoWarpiab} \Warpi(\PPP^{ab}_ {\eta})\,=\,\f{1}{2M!} \,
dW ^{A_1}\wedge\ldots\wedge dW ^{A_{M }} \gvep_{A_1\ldots A_M}
\,\,\PPP^{ab}_{\eta}(U,\,V\,|X)\Big|_{ \,U=0}\q
dW^A=
i\hhh\, d\, X{}^{AB}\f{\p}{\p U^B} +    d\, V{}^A\,,
\ee
where
$\gvep_{A_1\ldots A_M}$ is the fully antisymmetric tensor
 ($\gvep_{ 1\ldots M}=1$)
and
\be
\PPP^{ab}_{\eta}( U,\,V \,|X)=\JJJ^{ab}_ {\eta}(V-U,V+U\,|X)
\ee
with $\JJJ^{ab}_ {\eta}$
(\ref{JNNab}). As a first step,  we show   that \be\label{Qres}
Q^{++}_{\eta } =Q^{--}_{\eta }=0 \q
 Q_{\eta }= Q ^{+-}_{\eta} +   Q^{-+}_{\eta}= Q_{\eta^a } \q
Q ^{-+}_{\eta^a}  =   Q^{+-}_{\eta^a} \q  a=\mathrm{sign}(i^{M})\,
\ee with $\eta^a $ (\ref{etaoe})   ($a=+,-\,,\,\,\,M$ is even).

In particular, for $M=2$, a bosonic  parameter
$\eta^{\mathrm{sign}(i^2)}   $ is   antisymmetric, while for $M=4$
$\eta^{\mathrm{sign}(i^4)}  $ is symmetric. Note that, in the case of
$M=4$, the fields and currents depend on coordinates $X^{AB}$ of the
ten-dimensional matrix space. As shown in \cite{gelcur},  global
symmetry parameters of the current in the matrix space should be
singular to reproduce a nonzero current in Minkowski space. This
singularity is $\rho$-odd. Hence genuine
symmetry parameters of the currents in Minkowski space are also
$\rho$-odd. This phenomenon is expected to occur for
higher $M$ as well.

Let us   show that $\Warpi(\PPP^{++}_{\eta})\,$ (\ref{rhoWarpiab})
with  ${\eta}={\eta}(\B{}_A,\B^A)$ is   exact. Indeed,
{  by virtue of (\ref{Chat+}) and (\ref{kappaprop}),
for any parameter $\eta^{jk}=\eta^{jk}(\A )$
with $\A$ (\ref{param12}) $\JJJ{}^{ab}_\eta$ (\ref{JNNab}) acquires the form
 \bee\label{JetaF}
\JJJ{}^{ab}_\eta( \yy_1, \yy_2|X)\, =\f{(\kappa^a_1)^{\ppi_j}
(\kappa^b_2)^{\ppi_k}}{(2\pi)^{M}}
 \int_{\mathbb{R}^{2M}} d^M\xi_1 d^M\xi_2\,:\hat{c}_j^a (\xi_1)\hat{c}_k^b (\xi_2):
 \quad\\ \nn
\eta^{jk}\,  \exp  \Big (a  \,i \big (
\hhh\,\, \xi_1{}_{\za}\xi_1{}_{\zb} X^{\za\zb}+ \kappa^a_1\,\yy_1^B
\gx_1{}_B \big )+
  b  \,i \big (
\hhh\,\, \xi_2{}_{\za}\xi_2{}_{\zb} X^{\za\zb}+\kappa^b_2\yy_2^B
\gx_2{}_B\big )\Big ) \,,
 \eee
where
\be
 \eta^{jk}(\A_+{}^A(\yy_1|X),\A_+
 {}_A(\yy_1|X),\A_-{}^A(\yy_2|X),\A_-{}_A(\yy_2|X))\to
 \eta^{jk}(i\kappa_1^a\f{\p}{\p \gx_1 {}_A},
a  i\kappa_1^a \gx_1 {}_A ,
i\kappa_2^b\f{\p}{\p \gx_2 {}_A},
bi\kappa_2^b \gx_2 {}_A)\,.
\ee
Hence, by virtue of (\ref{parameters}) and  partial integration,
(\ref{rhoWarpiab}) with  ${\eta}=(\B{}_A,\B^A)$  gives
\bee
\label{go++} \Warpi(\PPP^{++}_{\eta})= (i)^{\ppi_k }
\f{1}{2(2\pi)^{ M}M!}
\int_{\mathbb{R}^{2M}}\,d^M \gx\,d^M \gl\,\,
d W ^{A_1}\wedge\ldots\wedge dW ^{A_{M }}  \,\,\,\gvep_{A_1\ldots A_M}
\qquad\qquad \\ \nn
 \exp \Big (i\hhh X^{AB}(\gx_A\gx_B+ \gl_A\gl_B)
 +iV^C(i\gl_C+\gx_C)
\Big)
\eta^{j\,k} \left(-i(\gx-i\gl)\,,  i \f{\p}{\p (\gx +
i\gl)}  \right)
\oc_j^+(\gx) \, \oc_k^+(\gl)
\,,
\eee where }$dW ^A=\Big(i \hhh\,  d\, X{}^{AB}(\gx-i\gl)_B   +     \,
d\,V{}^A\Big)$.
 It is easy  to see that $
\Warpi(\PPP^{++}_\eta)=d\go_{M-1}^{++}$  where
\bee \label{M-1form}
\go_{M-1}^{++} = \f{(i)^{\ppi_k+1 }}{2M!(2\pi)^M }
\int_{\mathbb{R}^{2M}} d^M\,
\gx\, d^M\gl\, \,
\,\,
dW ^{A_1}\wedge\ldots\wedge dW ^{A_{M-1}} \gvep_{A_1\ldots A_M}
\f{\pos^{{A_M}\,B}
(\gx_{B}-i\gl_{B})}{\pos^{C\,D}(\gx_{C}\gx_{D}+\gl_{C}\gl_{D})}\\\nn
\exp \Big (i\big(\hhh X^{AB}(\gx_B-i\gl_B) + V^A\big)(\gx_A+i\gl_A)
\Big)%\\ \nn
  \eta^{j\,k}\left(-i(\gx-i\gl),i  \f{\p}{\p (\gx +i \gl)}
\right)
\oc_j^+(\gx)  \oc_k^+(\gl)
\eee
with any positive definite matrix $ \pos^{{C}\,D}$. Here
it is important that  the singularity in (\ref{M-1form})
is integrable for $M\ge2$. Analogously, $\Warpi(\PPP^{--}_\eta)$
(\ref{rhoWarpiab}) is also exact. Hence, \be Q^{++}_{\eta }
=Q^{--}_{\eta } =0. \ee

To prove the remaining relations in Eq.~(\ref{Qres}), consider $Q
^{+-}_{\eta } $. Following \cite{cur} (see also \cite{gelcur}), we
choose the integration surface $ \Sigma^{+-}$ to be a $M-$dimensional plane
in $\M_M\times \mathbb{C}^M$ parametrized by coordinates $x^i$ with
$i=1\ldots M$, \ie
\be \label{plane}\hhh X^{AB} =  \U^{AB}_i x^i\q
V^{A } =U^{A }= 0\, \ee for some $\U^{AB}_i$.  Using (\ref{JetaF}) and (\ref{plane}) along
with (\ref{UVYY}) and (\ref{parameters}), we obtain from
(\ref{rhoWarpiab})
\bee \nn Q ^{+-}_{\eta }   &=& \half (2\pi)^{-M}
\int_{\mathbb{R}^{M}} d^M  x \int_{\mathbb{R}^{2M}} d^M  (\gx+\gl) d^M (\gx-\gl)\,\,
 \det\Big | \Os  \big\{\gx + \gl\big\}\Big|
\, c_j^+ (\gx ) c_k^- (\gl) \quad
\\ \label{Qin} &&
\eta^{j\,k}\left(-i(\gx+\gl)\,,- i\Big(\f{\p}{\p \gx}- \f{\p}{\p \gl}\Big)
\right)\exp \Big (i{\Os }^A_n\big\{\gx + \gl\big\}\,(\gx_A -
\gl_A{})x^n\Big ) \,,\qquad
 \eee where
 \be\label{Os} \Os^{A}_n \big\{\gx + \gl\big\}= \U^{AB}_n
(\gx_B + \gl_B)\,.  \quad \ee
Let     $\U^{AB}_n$   in
(\ref{plane}) be chosen in such a way that the roots of
$ \det\Big | \Os  (\xi  )\Big|$ are independent  functions of
$\xi $ (that is generically true). Observing that the measure
in (\ref{Qin}) is equivalent to $\delta^M (\gx - \gl)$, we obtain
 \bee \label{Qgintpmres} Q^{+-}_{\eta}  &=&\half
     \int d^M  \gx\,\eta^{j\,k}\left(-2i\gx \,,
     \f{i}{2}  \f{\p}{\p \gn}
\right)
 \oc_j^+( \gx+\gn )  \oc_k^-(\gx-\gn)
\Big|_{\gn=0}   \,.\quad\eee

Analogously \bee \label{Qin-+} Q^{-+}_{\eta } = \half
i^{\ppi(\eta)+M} \int d^M \gx\, \, \eta^{k\,j}\left( -2\gx \,,
\half\f{\p}{\p \gn}\right)
 \oc_j^+( \gx+\gn )  \oc_k^-(\gx-\gn)
\Big|_{\gn=0}
   \,.
 \eee
Here the factor of $i^{ M}$ results from $U$-derivatives
 in  the operator $\Big(d\,X{}^{AB}\f{\p}{\p U^B}
\Big)^{\,M} $ in (\ref{rhoWarpiab}).

To complete the proof of  (\ref{Qres}), by virtue of
(\ref{parameters}) and (\ref{rhoeta})
 we obtain
 \be \label{QRES2} Q_{\eta} = \half \int d^M
\gx\, \left(1+  i^M \rho
  \right)
 \,\eta^{j\,k}\left(-2i\gx \,,  \f{i}{2}  \f{\p}{\p \gn}
\right)
 \oc_j^+( \gx+\gn )  \oc_k^-(\gx-\gn)
\Big|_{\gn=0}   \,.
   \qquad \ee

Integration of  charges over the twistor space with coordinates $V$
using \be\label{eta0char} \eta^{m\,n}
(\B{}_A(U,V|X),\B{}^A(U,V|X))\Big|_{X=0}=\eta^{m\,n} ( \p_U\,,V )\,
\ee
 gives equivalent results. However,  there is a
subtlety that, to keep integrals well-defined, one has to integrate
different components $Q^{ab}$   over different cycles in the complex
twistor space, which is possible because every $Q^{ab}$ represents a
conserved charge. To give precise meaning to integrals it is useful
to use the formulation in Fock-Siegel space
 which makes the $\xi$ integration well-defined.
Skipping details, the proper definition of the charges (\ref{rhoQ})
is \bee \label{proQ+-} Q_{\eta}^{+-}&=& \half\int_{\Sigma^{+-}} d^M
v \quad \eta^{jk}(  \p_u,\,v)\T{}^{+-}_{j\,k}(v-u,v+u|0)|_{u=0}
\q\qquad\qquad\,
\\
\label{proQ++} Q_{\eta}^{++}&=& \half {i}^\ppi  \int_{\Sigma^{++}}
d^M z\quad \eta^{j\,k}( \p_\bz,\,z)\T{}^{++}_{j\,k}( z-\bz,
i(z+\bz) |0)|_{\bz=0} \q \qquad \qquad
\\
\label{proQ--} Q_{\eta}^{--}&=& \half{i}^\ppi    \int_{\Sigma^{--}}
 d^M   z \quad
\eta^{jk}( \p_\bz,\,z)\T{}^{--}_{j\,k}(i(z-\bz) ,   z+\bz |0)|_{\bz
=0} \q\qquad\qquad
\\
\label{proQ-+}\qquad Q_{\eta}^{-+}&=&\half{i}^M
 \int_{\Sigma^{-+}}
 d^M  v  \quad
\eta^{j\,k}(      i\p_u,\,iv) \T{}^{-+}_{j\,k}(   -v-u,u-v|0)|_{u=0}\q
 \quad\, \eee where
$\Sigma^{-+}=\Sigma^{+-}=\mathbb{R}^M$ while
$\Sigma^{--}=\Sigma^{++}=\gga^M$ where $\gga^M$ is any
$M$-dimensional closed cycle in $\mathbb{C}^M(z,\bz)$. With this
definition, the charges $Q_{\eta}^{++}=Q_{\eta}^{--}=0$ as being
represented by integrals of analytic functions. (Using formulation
in Fock-Siegel space, the corresponding contours can be deformed to
infinity where the integrals are zero.) For $Q^{+-}_{\eta}$ and  $
Q^{-+}_{\eta}$ this prescription again gives (\ref{Qres}).

By (\ref{Q+-}), the analysis of charges $\widetilde Q$ is analogous,
giving \bee \label{Qgintpmres+}
\widetilde{Q}^{+-}_{\widetilde\eta(\widetilde{\B})}&=&
      \half\int d^M \gx\,   \widetilde \eta^{j\,k} \left(2i\gx \,,  -\f{i}{2}  \f{\p}{\p
\gn}\right)  \oc_j^+( \gx+\gn )  \oc_k^-(\gx-\gn) \Big|_{\gn=0}
\,\q
\\
   \label{Qgintpmrespm+} \widetilde{Q}^{-+}_{\widetilde\eta(\widetilde{\B})}&=&
      \half i^{M +\ppi({\eta})}
\int d^M \gx\,  \widetilde \eta^{j\,k} \left( -2\gx \,,    \half
\f{\p}{\p \gn} \right)  \oc_j^+( \gx+\gn )  \oc_k^-(\gx-\gn)
\Big|_{\gn=0} \,.\quad \eee
By virtue of (\ref{rhoeta}) and
(\ref{exparam}), the counterpart of (\ref{QRES2}) is
\be
\label{QRES2+} \widetilde{Q}_{\widetilde\eta(\widetilde{\B})}    =
\half \int d^M \gx\,
 \left(1+  i^M \rho \right)  \widetilde\eta^{j\,k} \Big(
2i\gx \,,  - \f{i}{2}  \f{\p}{\p \gn}\Big) \oc_j^+( \gx+\gn )
\oc_k^-(\gx-\gn) \Big|_{\gn=0} \, \,. \qquad \ee

\subsection{Higher-spin algebra}
\label{Commutatorchargecharge}

Let us show that, for $\rho$-odd $\eta$ and $\eta'$,
\be\label{QQ} [Q _{\eta}, Q_{\eta'  }]=
   Q_{[\eta\,,\,\eta']_ * }\q
\ee where
 \be \nn[Q _{\eta}, Q_{\eta'   }] =
Q _{\eta}\, Q_{\eta'}-  Q _{\eta'}\, Q_{\eta}\,\q
 [a\,,b]_ * = a * \,b-    b * \,a\,
\ee  and
\be  \label{stars}
 (a  * \,b)^{k\,n}(u,v)=
\gd_{jm} a^{k\,j} \left(u ,\,    v
\right)\exp\left(\f{\overleftarrow{\p}}{\p  u_C
}\f{\overrightarrow{\p}}{\p v^C} -\f{\overleftarrow{\p}}{\p
v^C}\f{\overrightarrow{\p}}{ \p u_C }\right) b^{m\,n}\left(u ,\,
v \right)\,. \ee An important property of the star product  is that
it possesses the antiautomorphism $\rho$ \cite{Vasiliev:1986qx}
 \be
\rho(a^{mn}(u,v)) :=  (i)^{\ppi(a)} a^{nm}(iu,iv)\q a^{mn}(-u,-v)=
(-1)^{\ppi(a)} a^{mn}(u,v)\,, \ee \ie assuming that the coefficients
of $a(u,v)$ are endowed with the
 Grassmann parity $\ppi(a)$,
\be \label{anti}
\rho(a * b) = \rho(b) * \rho (a)\,.
 \ee
 Hence,  for $\rho$-odd    $a$ and $a'$, $[a\,,\,a']_ * $ is also $\rho$-odd.

The proof of (\ref{QQ}) is straightforward. By virtue of
(\ref{Qres}),
 \bee \label{Qgintpm2}
[Q_{\eta},Q_{ \eta'} ] &=&
\int    d^M   v     d^M   v'
\,\,
\eta^{j\,k}\left( {\p_ u},    v \right)
{\eta'}^{m\,n}\left( {\p_{ u'}},\,  v'{} \right) \qquad
\\\nn&&
\left[\widehat{C}_j^+(v-u|0)  \widehat{C}_k^-(v+u|0)\,,
\,\widehat{C}_m^+(v'-u'|0)  \widehat{C}_n^-(v'+u'|0)\right]
 \Big|_{u=u'=0}\,.\quad \eee
Using (\ref{comCCXX0}) along with (\ref{:etaeta':}),  we obtain
  \bee\label{commQQ}\nn [Q _{\eta}, Q _{\eta'}   ]\!\!&=&\!\!
 \gd_{jm}\int     d^M   v\left ( \eta^{k j}\left( {\p_ u} ,
v-u' \right) {\eta' } { }^{m\,n}\left( {\p_{ u'}},
v+u  \right)-
{\eta' } {}^{k\,j}\left(\p_{ u'},    v-u{}\right )
\eta^{m\,n}\left( {\p_ u},   v+u' \right)\right)
\\ &&
\widehat{C}_k^+\left(v'-(u+u') |0\right)
 \widehat{C}_n^-\left(v'+u+u' |0\right)  \Big|_{u=u'=0}\,.
  \eee
Hence
\bee\nn [Q_{{\eta}}, Q_{{\eta'}} ] =
   \Big [ \int     d^M   v \Big(
(\eta  * {\eta'})^{m\,j}\left( {\p_ u } ,\, v
\right)
 -
  ({\eta'}  * \eta )^{mj}\left( {\p_ u },\, v  \right)
\Big)\widehat{C}_m^+(v-u  |0) \widehat{C}_j^-(v+u  |0)\Big ]
\Big |_{u =0}\,,\quad \eee
which implies (\ref{QQ}) by virtue of (\ref{Qres}), (\ref{proQ+-}), (\ref{proQ-+}).

Using that $Q$ is related to $\widetilde{Q}$ via (\ref{Q+-}) it can
be shown that \bee\label{tilQtilQ} [\widetilde{Q} _{{\tilde \eta}},
\widetilde{Q}_{{\tilde \eta'}} ] &=&  {Q}_{{[K \tilde \eta\,, K
\tilde \eta']_ * }}\q
 [\widetilde{Q}_{{\tilde \eta}},  {Q}_{{\eta'}} ]=
\widetilde{Q}_{K{[K\tilde \eta\,,  \eta']_ * }} \, \eee with the
convention
 \be \label{klein} K^2=1\q K\eta  = (-1)^{\ppi
(\eta)} \eta\,K. \ee
These commutation relations coincide with
  those of the HS subalgebra of the algebra of
\cite{Fradkin:1987ah} described by the symmetry parameters
$\eta(u,v;K)$ that, in addition to spinor variables,  depend on the
Klein operator $K$
 possessing properties (\ref{klein}). In other words, combining
 $Q$ and $\widetilde Q$ into $\Q$ with the parameter
\be \varepsilon  =\eta +K \tilde \eta \,, \ee the commutation
relations (\ref{QQ}) and (\ref{tilQtilQ}) combine to
\be\label{varvar} [\Q _{\varepsilon}, \Q_{\varepsilon'  }]=
   \Q_{[\varepsilon\,,\,\varepsilon']_ * {}}\,.
\ee More precisely, in the case of $M=2$, the resulting algebra of
charges $\Q$ is isomorphic to the HS algebra $ho(\NNN,\NNN|4)$ in
notations of \cite{kon}. In this paper, it appears as the $3d$
conformal HS algebra acting on $3d$ conformal fields realized as
rank-one $M=2$ fields of Section \ref{currents}. In accordance with
consideration of \cite{kon}, the doubling of fields is because the
$3d$ conformal HS symmetry acts both on bosons and  fermions. In
particular, its even subalgebra decomposes into direct sum of two
isomorphic algebras acting on $3d$ massless scalar and spinor,
respectively. All other types of HS algebras $hu(n,m|4)$,
$ho(n,m|4)$ and $husp(n,m|4)$ can be obtained along the lines of
\cite{kon} as proper subalgebras of $ho(\NNN,\NNN|4)$ preserving
appropriate additional structures (bilinear forms or gradings) with
respect to color indices.

Note that multiparticle algebra $\alM$ of \cite{Vasiliev:2012tv} can be realized as the algebra
of all polynomials of $Q_\varepsilon$ with various $\varepsilon$, that includes
unity. By virtue of (\ref{varvar}) it can be realized as the algebra
of symmetrized polynomials of $\varepsilon$, or, equivalently, as
the algebra of polynomials of a single $\varepsilon$. This is the
realization considered in \cite{Vasiliev:2012tv}.

\section{Twistor current operator algebra}
\label{Vertex algebra}

Free bilinear twistor currents   are most conveniently described as
\be\label{Jg}
  \III^2_{  g  } =
\sum_{a,b=+,-}\int   d^{2M }  {\zz} \,\,  \,\,
g{}^{m\,n}_{a\,b} ( \zz_1, \zz_2) \,  \TT^{ab}_{m\,n}(\zz_1,\,\zz_2
|0) \ee with $\TT^{ab}_{k\,j}(\zz_1,\,\zz_2 |X)$ (\ref{Tab}).
 Symmetry properties (\ref{parTab}) of  $\TT^{ab}_{m\,n}(\zz_1,\,\zz_2 |0)$
 imply that
  \bee\nn\label{Comrelg}
g{}^{m\,n}_{a\,\,b} ( \zz_1, \zz_2) &=&(-1)^{\ppi_m}
 g{}^{m\,n}_{a\,\,b}(-  \zz_1, \zz_2)=(-1)^{\ppi_n}
 g{}^{m\,n}_{a\,\,b}(  \zz_1,- \zz_2) \q\\ \mu(g)&=&g \,, \eee
where
\be \label{mug} \mu(g){}^{m\,n}_{a\,\,b}( \zz_1,\zz_2):=
(-1)^{\ppi_m\ppi_n}g{}^{n\,m}_{b\,\,a} ( \zz_2, \zz_1)\,. \ee
Also,
 we require
 \be\label{:gg':}
  g{}^{n\,m}_{b\, a}\, g'{}^{j\,k}_{c\,d} = (-1)^{\ppi_{n,\,m}\ppi_{j,\,k}}
   g'{}^{j\,k}_{c\,d}\, g{}^{n\,m}_{b\, a}\,
 \ee
 with $
 \ppi_{j,\,k }$ (\ref{ppijk}).
Otherwise, the  function $g{} ( \zz_1, \zz_2) $ remains
arbitrary. Note that due to (\ref{:gg':})
 \be\label{mutuallyJ} \, :\,  \III^2_{  f  }\, \III^2_{  g  }\,:\,=  \,:\,
\III^2_{ g  }\, \III^2_{  f  }\,: \quad \forall \,\, g\,,f\,. \ee

In Section \ref{Contact}, using  evolution formula (\ref{DCYlow}),
the space-time dependent current will be represented as the twistor
  current with the function $g{} ( \zz_1, \zz_2) $ that depends
in a specific way on space-time coordinates as   parameters. This
trick  fully reduces  construction of space-time operator
product algebra to the much simpler problem in the twistor space.
More generally, via uplifting of twistor-to-boundary $\D$-functions to
twistor-to-bulk ones as discussed in the end of Section \ref{Dfunc},
it can be straightforwardly extended  to the bulk where the
 boundary conformal algebra  acts.

\subsection{OPE of twistor currents}
\label{Fusionat0}

A basis of current operator algebra consists of
 $2n$ twistor currents with various $n$
 \be\label{Jg2N}  \,\,:
\prod_{j=1}^n\, \III^2_{  g_j  }\, :\,  \q \III^{0} = Id\,. \ee
Since currents (\ref{Jg2N}) are symmetric multilinear functionals of
$g_j $, the twistor operator algebra can be  represented by
nonlinear functionals of a single function $g( \zz_1, \zz_2)$
(skipping color indices). This leads to the construction of algebra $\alM$ described
in \cite{Vasiliev:2012tv}. To determine full current operator algebra
it suffices to compute $ \III^{2n} \III^2$ for any $n$. Though the
final results can  be derived by
 methods of \cite{Vasiliev:2012tv}, here we will use a slightly different
setup based on interesting algebraic
structures most appropriate for   description of $n$-point  functions.

The cases of $A$-currents  with the parameters $g{}^{m\,n}_{a\,b} (
\zz_1, \zz_2)$, that are independent for different values of indices
$a,b$ modulo symmetry properties (\ref{Comrelg}), and
 $F$-currents with  $g{}^{m\,n}_{a\,b} ( \zz_1,
\zz_2)$ at different $a,b$  related to each other in accordance with
 (\ref{Tq}), are different. We start with the more general case of $A$-currents.

From the commutation relations (\ref{comCCXX0})   along with
 (\ref{Comrelg}) and (\ref{:gg':}) it follows that
\be\label{PAIRINGint}\III^2_{  g  } \III^2_{g'} = :\III^2_{  g
} \III^2_{g' } : +   \III{}^2 {}_{ (g  \op g'   + g'\opd  g )
}  + \strop ( g  \op g')\III^0\,,
  \ee where
   \be\label{PAIRintop}
 ( g  \op g'{})^{mn}_{ab}(\zz_1,\zz_2) = 2 \gd_{kj}\tau^{dc}\int
      d    {p}\, \,g^{mk}_{a\,c}(\zz_1,p) \, g'{}{}^{jn}_{d\,b}(p,\zz_2)\,,
\ee
   \be\label{PAIRintdi2}
 ( g  \opd g'{})^{mn}_{ab}(\zz_1,\zz_2) =   2
        \gd_{kj}\tau {}^{c\,d}\int
      d    {p}\, \,g^{mk}_{a\,c}(\zz_1,-p) \, g'{}{}^{jn}_{d\,b}(p,\zz_2)\,,
\ee  \be \label{tau} \tau^{ab}=\delta^a_+\delta^b_- \, \ee and
  \be\label{PAIRint00}
\!\! \stropd(g)=
  \gd_{mn}\tau^{ab}\!\!\int\!\!
      d    {p} \, g^{mn}_{a\,b}(-p,p)
      \q  \strop  (g   )=
  \gd_{mn}\tau^{ba}\!\!\int\!\!
      d    {p} \, g^{mn}_{a\,b}( p,p)
     \,. \ee

  The product laws $\op$ and $\opd$ are
both associative similarly to the  product of matrices $ (h h')_{ab}
= h_{ac}t^{cd} h'_{db}$ with some $t^{cd}$.
 Moreover, they are mutually associative, \ie
   \bee\label{opdopass}
( ( g  \opd g'{})\op f)^{mn}_{ab}(\zz_1,\zz_2) =
  (g\opd ( g'  \op
f{}))^{mn}_{ab}(\zz_1,\zz_2)\,,\\\nn
 (g\op  ( g'  \opd f{}))^{mn}_{ab}(\zz_1,\zz_2)  =
  ( ( g  \op  g'{})\opd f)^{mn}_{ab}(\zz_1,\zz_2)\,,
\eee hence providing an example of a biassociative algebra discussed
in \cite{Vasiliev:2012tv}.
 One can see that
  \be
  \label{stopg} \strop (g   )=
 \stropd (\mu(g)   )\,,
\ee and
\be%
\label{sstrtr} \stropd ( g  \op f)=  \stropd ( f  \op g)\,\q
  \strop  ( g  \opd f)=    \strop ( f  \opd g)\,,
\ee%
 \ie  $ \stropd $ and $\strop\,$ are supertraces on the algebras
$A_{\op }$ and $A_{\opd}$ with the respective products. In addition,
the following relation holds
\be  \label{opopdst} %
\strop ( g \op f)= \stropd ( f  \opd g)\,.
\ee%

By virtue of (\ref{mug}) and (\ref{:gg':}),
    \be\label{PAIRintdi}
      g  \opd g'{}      = \mu\Big(  \mu(g'{}) \op \mu (g)  \Big)\,
        .\ee
From (\ref{PAIRintdi}) it follows in particular that if $g$ and $g'$
obey (\ref{Comrelg}), $(g  \op g'   + g'\opd  g )$
   also does, which, in fact, is the reason for appearance of this combination
   in Eq.~(\ref{PAIRINGint}). Also note that, by virtue of
   (\ref{stopg}) and (\ref{PAIRintdi}),
   \be
 \stropd ( g  \op f)=\strop  ( g  \opd f)\q \forall g=\mu(g),\, f=\mu(f)\,.
   \ee

Consider the  commutator \be\nn [\III^2_{  g  }\,, \III^2_{g'}] =
\III^2_{  g  } \III^2_{g'}-
 \III^2_{  g'  } \III^2_{g}\,. \ee
Because of Eq.~(\ref{PAIRINGint}), $\III^4$   does not contribute.
Taking into account   (\ref{opopdst}) we obtain \be\label{bullcom}
[\III^2_{  g  }\,, \III^2_{g'}] = \III^2_{[g\,,g']{}_\starbul{}} +
\half(\strop+\stropd)\big([g\,,g']_\starbul{} \big )\III^0 \,, \ee
 where
\be\nn [g\,,g']{}_\starbul{}  = g\starbul g' -  g'\starbul g\,, \ee
  \be\label{bull}
 ( g  \starbul g'{})^{mn}_{ab}(\zz_1,\zz_2) =( g  \op g'{})^{mn}_{ab}(\zz_1,\zz_2)-
  ( g  \opd g'{})^{mn}_{ab}(\zz_1,\zz_2)
 \,.\quad
\ee

The product law $\starbul$  is  associative.
More generally, from (\ref{opdopass}) it follows that all linear
combinations of $\op$ and $\opd$ are mutually associative,
forming a particular case of
 {\it multi-associative} algebra   by which we
mean a linear space $A$ over the field ${\mathbb K}$ endowed with a set of
such mutually associative products $*_1 , \ldots  , *_k$ that any their
linear combination $\ga^1 *_1+\ldots +\ga^k *_k$, $\ga^i\in$ ${\mathbb K}$
 is associative. In particular, it is useful to introduce
  \be\label{dot}
 ( g  \cdot g'{})^{mn}_{ab}(\zz_1,\zz_2) =( g  \op g'{})^{mn}_{ab}(\zz_1,\zz_2)+
  ( g  \opd g'{})^{mn}_{ab}(\zz_1,\zz_2)
 \,.\quad
\ee

Analogously to the product (\ref{PAIRINGint}) of two bilinear
currents, it is not difficult to obtain the product   of a $2n-$linear
current
(\ref{Jg2N}) and  a bilinear  current using (\ref{comCCXX0}),
(\ref{Comrelg}), (\ref{:gg':}) and (\ref{opdopass})
 \bee\label{JnJ2}
:\,\III^2_{  g_1  } \ldots\III^2_{g_n}: \,%\III^{2n}_{   g_1\ten\ldots.\ten g_n    }
  \III^2_f\, =\,
:\,\III^2_{  g_1  } \ldots\III^2_{g_n}\,\III^2_{  f  }\,:\,%\III^{2n+2} _{   g_1\ten\ldots.\ten g_n\ten \,f }
  + \sum_{k  }
  :\,\III^2_{  g_1  } \ldots  \widehat{\III^2_{g_k}}\ldots\III^2_{g_n}
 \, \III^2_{(g_k  \op f   + f\opd  g_k ) }\,:
%\III^{2n } _{   g_1\ten\ldots\widehat{g}_{k} \ldots\ten g_n\ten (g_k\op\,f+f\opd\,g_k  ) }
     \qquad\qquad\\ \nn  +\sum_{k}
          \strop(g_k\op \,f):\,\III^2_{  g_1  } \ldots  \widehat{\III^2_{g_k}}\ldots\III^2_{g_n}\,:\,+
            \sum_{k>p  }
 :\,\III^2_{  g_1  } \ldots  \widehat{\III^2_{g_p}} \ldots  \widehat{\III^2_{g_k}}
 \ldots\III^2_{g_n} \,\III^2_{(g_k\op f \opd g_p +g_p\op f \opd g_k  ) } \,:\,, \qquad\eee
where $\widehat{\III^2_{g_k}}  ,\,\,\widehat{\III^2_{g_p}}   $ denote omitted terms.

Using (\ref{opdopass}), (\ref{sstrtr}),
(\ref{opopdst}) and (\ref{JnJ2}), one can obtain, in particular,
  \bee\label{PAIRINGint3}  \III^2_{  g_1  } \III^2_{g_2}\III^2_{g_3} &=&
:\,\III^2_{  g_1  } \III^2_{g_2}\III^2_{g_3}\,: +
 :\,\III^2_{  g_1  } \III^2{}_{   \,  g_2 \op g_3  }\,:+   :\,\III^2_{  g_1  }\III^2{}_{   \,  g_3 \opd g_2  }\,: \\ \nn&+&
  :\,\III^2_{  g_2  } \III^2{}_{  \, g_1  \op g_3}\,:
  + :\,\III^2_{  g_2  } \III^2{}_{   \, g_3  \opd g_1}\,:
 +:\,\III^2_{  g_3  } \III^2{}_{   g_1 \op\, g_2     }\,:+
 :\,\III^2_{  g_3  } \III^2{}_{   g_2 \opd\, g_1    } \,: \\ \nn&+&
  \III{}^2 {}_{    g_1\op g_3 \opd g_2 }\,+\III{}^2 {}_{    g_3\opd g_1  \op g_2 }\, +
  \III{}^2 {}_{    g_2\opd g_1   \op g_3 } \\ \nn&+& \III{}^2 {}_{    g_2 \op g_3  \opd g_1  }+
\III{}^2 {}_{   g_1\op g_2 \op g_3}+\III{}^2 {}_{   g_3\opd g_2 \opd g_1}
\\ \nn&+&
  \strop(g_1\op g_2)\III^2_{g_3} +    \strop(g_2\op g_3)\III^2_{g_1
}+
 \strop(g_1\op g_3)\III^2_{g_2 }
 \\ \nn&+&
    \strop(g_1\op g_2\op g_3  )\III^0 + \stropd( g_3\opd g_2 \opd  g_1)\III^0\,. \eee
Analogously, Eq.~(\ref{JnJ2}) allows us to derive in Section \ref{NprodG}
the explicit formula for
the product of any number of bilinear currents $\III^2_{g_1}\ldots \III^2_{g_n}$.

 \subsection{Star-product interpretation}
\label{Half} Current equation (\ref{Xsdvigtoka}) is mapped to a
first-order PDE system by the half-Fourier transform \cite{gelcur}.
A similar transformation is known to map the convolution product to
Weyl-Moyal star product  (see, e.g., \cite{BerezShub}).

Indeed,    half-Fourier transform can
be introduced as follows
 \bee\label{halfFourier}
 \widetilde{g} ( W,V) &=&(2\pi)^{-M /2}\int d^M\, {U} \exp (i W{}_C
{U}{}^C) g(V-U,V+U)
 \q \\\nn
 g(V-U,V+U)&=& (2\pi)^{-M /2}\int d^M\,W
 \exp (-i W{}_C U {}^C) \widetilde{g}  (W,V)
\,,\qquad \eee or, alternatively,
 \bee
 \label{halfFourier2} \ddot{g}  ( U,W )
&=&(2\pi)^{-M /2}\int d^M\, {V} \exp (i W{}_C {V}{}^C) g(V-U,V+U)
 \q \\\nn
 g(V-U,V+U)&=& (2\pi)^{-M /2}\int d^M\,W{}
 \exp (-i W{}_C V {}^C) \ddot{g}  (U,W)
\,.\qquad\eee
Note that, for functions of definite symmetry $ G (U,V)=(-1)^{\pi }G
(V,U)$, \bee\label{halfFourierfBF1} \widetilde{G }  ( W ,V) =
(-1)^\pi\ddot{G }  (   W,V )
 \,.\qquad\eee

   Direct substitution of (\ref{halfFourier}) into (\ref{PAIRintop}) gives
\be\label{starprod}
 \widetilde{( g  \op g')}{}^{mn}_{ab}({W,\,V})=  2^{1+3 M/2}   \pi^{  M/{2}}
 \gd_{kj}  (\widetilde{g}_{a\,-}^{m,\,k} * \widetilde{g}'{}_{+\,b}^{j,\,n})     (W,\,  V) \q \ee
   where
\be\label{defstar} F  * G\,(x,y) = (2\pi)^{ - 2M }\int d s\,dt\,du\,dv\, \exp
 \left[ i\left( u_C\,v^C-t_C\,s^C   \right) \right]
 \,F  (x+t   ,  y+v    )
  G   (x+u    ,  y+s    )
  \q \ee
which is equivalent to the star product (\ref{stars}). In
 these terms,  (\ref{PAIRint00}) reads as
\be
 \stropd ( g)  =
  (2\pi)^{M/2 }  \delta_{nm}\tilde g^{nm}_{+-} (0,0)\q
 \strop ( g ) =   (2\pi)^{M/2 }  \delta_{nm}\ddot g^{nm}_{-+} (0,0)
\,.
 \ee

Thus, the half-Fourier transform relates the formalism of convolution
product convenient for the analysis of boundary currents to the star-product
formalism familiar for the bulk analysis of HS theories \cite{more}. In fact,
in the framework of bulk HS gauge theory, the half-Fourier transform also plays
crucial role,  relating adjoint and twisted adjoint representations
of HS algebra. In the bulk formalism, various (half-)Fourier transforms are
naturally represented by the Klein operators
(see, e.g., \cite{Didenko:2009td}).

\subsection{Butterfly algebras}
\label{Butterfly formula}

For any associative algebra $A_\star$,
butterfly algebra $A_{\stackrel{B}{\oprod}}$ is the unital
associative algebra spanned by elements   $ g_{  j_{ 1} , \ldots
,j_{ k}  }\in A_\star$   endowed with a set of
integers $j_{ 1}, \ldots  ,j_{ k} $.
 Given three elements $B_+,\,\,B_-,\,\,B_0 \in A_\star$,
the   butterfly product law $\stackrel{B}{\oprod}$ is defined as follows
\be  g_{
j_{ 1} , \ldots  ,j_{ k}  }\stackrel{B}{\oprod}  f_{  i_{ 1} , \ldots , i_{ m}  }=
\left\{ \beee{ll}\label{kmprodnueq0}
 g_{  j_{ 1} , \ldots  ,j_{ k}  }\star B_-\star    f_{  i_{ 1} , \ldots , i_{ m}  } &\mbox{if } j_{ k}< i_{ 1}\,,\\
 g_{  j_{ 1} , \ldots  ,j_{ k}  }\star B_+\star   f_{  i_{ 1} , \ldots,  i_{ m}  } &\mbox{if } j_{ k}> i_{ 1} \,,\\
  g_{  j_{ 1} , \ldots  ,j_{ k}  }\star B_0\star   f_{  i_{ 1} , \ldots,  i_{ m}  } &\mbox{if } j_{ k}= i_{ 1} \,. \eeee\right. \ee
$\stackrel{B}{\oprod}  $ is associative by virtue of associativity
of $\star$.
Unit  element $e_{\stackrel{B}{\oprod}}$ of $A_{\stackrel{B}{\oprod}}$
is identified with $g_{  j_{ 1} , \ldots  ,j_{ k}  }$
  with the empty set of indices, \ie $k=0$.
Trace $tr_\star$ of $A_\star$ induces a trace of $A_{\stackrel{B}{\oprod}}$
\be\label{troprodenueq} \tr_{\stackrel{B}{\oprod}}\big(  g_{
j_{ 1} , \ldots , j_{ k}}\big) =\left\{ \beee{ll}
\tr_\star\big(  B_+\star    g_{  j_{ 1} , \ldots , j_{ k}}\big) &\mbox{if } j_{ 1}< j_{ k}
\,,\\
\tr_\star\big(  B_-\star  g_{  j_{ 1} , \ldots , j_{ k}}\big) &\mbox{if } j_{ 1}>j_{ k}\,,
\\
\tr_\star\big(  B_0\star  g_{  j_{ 1} , \ldots , j_{ k}}\big) &\mbox{if } j_{ 1}=j_{ k}\,\, \mbox{or}\,\, { k}\le1\,,   \eeee\right.
 \ee
\be\label{troprodciknueq}
\tr_{\stackrel{B}{\oprod}}\big(     g_{  j_{ 1} , \ldots , j_{ k}}\stackrel{B}{\oprod} f_{  i_{ 1},, \ldots , i_{ m}}\big) =
\tr_{\stackrel{B}{\oprod}}\big( f_{ i_{ 1}, \ldots , i_{ m}}\stackrel{B}{\oprod}   g_{     j_{ 1} ,
\ldots , j_{ k}} \big)\,.
\ee

 Clearly, products $\stackrel{B_1}{\oprod}$ and $\stackrel{B_2}{\oprod}$
with any two sets $B_1=(B_{1+}, B_{1-},B_{1\,0}) $ and
$B_2=(B_{2+}, B_{2-},B_{2\,0} )$ of elements of $A_\star$ are mutually associative and
\be\label{trab}
tr_{\stackrel{B_1}{\oprod}}(g \stackrel{B_2}{\oprod} f)=
tr_{\stackrel{B_2}{\oprod}}(f \stackrel{B_1}{\oprod} g)\,.
\ee

As shown below, multiple products of
$J^{2n}$ currents can be described by the butterfly algebra  $A_\oprodeq $
   with
   \be\label{oprodeq}
 {\oprodeq} =\,\stackrel{B}{\oprod} \q
   B_0= B_+=-\Pi_+ ,\,
 \,  B_- = \Pi_-
\q
\ee
 where $\Pi_\pm$ are defined so that
\be\label{star+-}
\Pi_+ +\Pi_- = e_\star\q f\opd g = - f\star \Pi_+ \star g\q f \op g =
f\star \Pi_- \star g\,.
\ee
Note that, as shown in \cite{Vasiliev:2012tv}, $\Pi_\pm$ are projectors,
\ie  $\Pi_\pm\star \Pi_\pm=\Pi_\pm$.

We will also use the following products
\be
\label{opordeqnu}
\bp=\, \stackrel{B }{\oprod}\,: \,\,\, B_0=B_\pm=\Pi_-\q
\bpd= \,\stackrel{B }{\oprod} \,:\,\,\, B_0=B_\pm=-\Pi_+\q
\stackrel{-}{\oprod}=  \,:\,\,\, B_+=B_0=0\,,  B_-=e_\star\,,
\ee
where $e_\star$ is the unit element of $A_\star$.
Instead of the product
$\stackrel{B}{\oprod}$ with $ B_0=B_\pm=e_\star$ we will use
notation $\star$. Note that, by (\ref{star+-}),
 \be\label{star+-b}
\bp -\bpd =  \star \q \stackrel{-}{\oprod} = {\oprodeq}- \bpd
\,.
\ee

For any $B$,   $A_{\stackrel{B}{\oprod}}$ contains  ideals
  $I^k_{\stackrel{B}{\oprod}}\subset A_{\stackrel{B}{\oprod}}$ spanned by such its elements, that contain at least $k$ coinciding
  successive indices $j_n=j_{n+1}=j_{n+2}=\ldots = j_{n+k-1}$. Algebras
  $A^k_{\stackrel{B}{\oprod}}=A_{\stackrel{B}{\oprod}}/I^{k+1}_{\stackrel{B}{\oprod}}$
are spanned by elements that do not contain more than $k$ coinciding
  successive indices.

  It turns out that multiple  current products are most compactly  described by the
  butterfly algebra
  $A_\oprod \sim A_{ {\oprodeq}}/I^2_{\oprodeq} $  spanned by
$g^{nm}_{ab}{}_{  j_{ 1} , \ldots  ,j_{ k}  }(W_1,W_2)$ endowed with
ordered sets of integers $j_{ 1} , \ldots  ,j_{ k} $ such that
$j_n\neq j_{n+1}$ $\forall n$. The butterfly product law is \be
\label{kmprod} g_{j_{ 1} , \ldots  ,j_{ k}  }\oprod g_{  i_{ 1} , \ldots , i_{ m}  }=
\theta(i_1-j_k)
g_{  j_{ 1} , \ldots  ,j_{ k}  }\op  g_{  i_{ 1} , \ldots , i_{ m}  }
+ \theta(j_k- i_1)
g_{  j_{ 1} , \ldots  ,j_{ k}  }\opd g_{  i_{ 1} , \ldots,  i_{ m}  }\,,
\ee
where
\be
\label{theta}
\theta (n) = 1 \quad \mbox{at $n>0$}\q
\theta (n) = 0 \quad \mbox{at $n\leq 0$}\,.
\ee
$A_\oprod$ possesses a trace
\be\label{troprod} \stropr\big( g_{
j_{ 1} , \ldots , j_{ k}}\big) =
\theta(j_k - j_1)
\strop\big( g_{  j_{ 1} , \ldots , j_{ k}}\big) + \theta(j_1 -j_k)
\stropd\big(g_{  j_{ 1} , \ldots , j_{ k}}\big).
\ee
 A useful property  is that, for $   {{j_{m}}}=\max(j_{ 1} ,
\ldots,{j_{ k}}) $,
\be\label{trmax} \stropr\big( g_{  j_{ 1}}\oprod   \ldots
\oprod g_{j_{ k}}\big) =
     \strop\big(
       g_{  j_{m+1 }}\oprod \ldots \oprod g_{j_{k}}\oprod
        g_{j_{ 1 }} \oprod  \ldots \oprod g_{j_{m-1}}
     \op{g}_{j_{ m}} \big)\, \,.
 \ee

\subsection{Butterfly formulae for multiple products and $n$-point functions}
\label{NprodG}

In terms of butterfly algebra $A_{\oprod}$, operator products
 (\ref{PAIRINGint}) and (\ref{PAIRINGint3}) can be rewritten as
\bee\label{oprod2} \III^2_{g_{1}} \III^2_{g_{2}} &=&\!\!
\sum_{i_j=1,2;\,i_1\ne i_2}\left\{\f{1}{2!}\,
  \, :
 \III^2_{ g_{      i_1}  }   \III^2_{ g_{ i_2}} \, :
+
 \III^2_{ g_{    i_1} \oprod  g_{i_2  }}+\f{1}{2}\stropr\big( g_{  i_1} \oprod  g_{ i_2}\big)\III^0
 \, \right\}   \q \\ \label{oprod3} \III^2_{g_{1}}
\III^2_{g_{2}}\III^2_{g_{3}}&=&\ls\sum_{i_j=1,2,3\atop\,i_1\ne i_2\ne i_3\ne i_1}
%\ls&&\ls
\left\{\f{1}{3!}\,
  \, :
 \III^2_{ g_{      i_1}  }   \III^2_{ g_{ i_2}}   \III^2_{ g_{ i_3}} \,:
 +  :\left(
 \III^2_{ g_{    i_1} \oprod  g_{i_2  }}+\f{1}{2}\stropr\big( g_{  i_1} \oprod  g_{i_2}\big)\III^0
\right)\,
 \III^2_{ g_{   i_3  }}
 \,:
\right.\\ \nn&+&\left.
 \III^2_{ g_{     i_1 } \oprod  g_{i_2} \oprod  g_{i_3 }}
 +\f{1}{3}\stropr\big( g_{  i_1 } \oprod  g_{i_2} \oprod  g_{i_3}\big)\III^0
  \right\} \,.\qquad  \eee

\newcommand{\Gf}{{ \widetilde{\cal{ V}}}}
\newcommand{\Gif}{\mathbf{ G}}
As shown below, these formulae admit a natural generalization to the operator product
of any number of multilinear currents
$
 \III^{2n}_{g }=:\underbrace{\III^{2 }_{g }\cdots \III^{2 }_{g }}_n:\,$
\be\label{oprodnmul}
\III^{2n_{1}}_{g_{1}}\,\ldots \III^{2n_{k}}_{g_{k}} = \left(
\f{\p^{ n_{1}}}{ (\p
 \gm^{ {1} })^{n_{1}}}\ldots \f{\p^{ n_{k}}}{ (\p
 \gm^{ {k} })^{n_{k}}} {E}(\Gm(\gm)
)\right)\Big |_{\gm=0}\,,   \ee
where
\be\label{Veq}
 \Gm(\gm)=\sum_{j=1}^\infty \gm^j g_j\,
\,,
\ee
\be \label{PTGB} {E}(\Gm(\mu)
)=\exp \Big ( -\stropr\big(   ln_\oprod (  Id_\oprod-\Gm(\mu)   )   \big)\Big )
\exp_{\ten}
\Big(
         \G(\Gm(\mu))    \oprod
     \big(
          Id_\oprod-     \G(\Gm(\mu))
     \big)^{-1}_\oprod
\Big)\,,
\ee%
      \be \label{Gkrugloe}
  \G (g) = \III^2_{  g} \q \G\big(g\big) \oprod  \G\big(g'\big) = \G\big(g\oprod g' \big)\,,
\ee
\be
ln_\oprod(Id_\oprod-\Gm(\mu)) = - \sum_{k=1 }^\infty\f{1}{k}
\underbrace{\Gm(\mu)\oprod \ldots \oprod \Gm(\mu)}_k\,,
\ee
\be
 \exp_{\ten}(A)=\sum_{n=0}^\infty\f{1}{n!}
\underbrace{A\ten\ldots\ten A}_n \q\underbrace{A\ten\ldots\ten A}_0 = Id\,,
\ee
where the commutative product $\ten$  encodes the normal
ordering
\be \III^{2n}_g\ten \III^{2m}_f\sim\, :\III^{2n}_g\,
\III^{2m}_f:\,. \ee
Using that $g_i\oprod g_i=0$ and $tr_\oprod (g_i)=0$ by virtue of (\ref{kmprod}) and
(\ref{troprod}), Eq.~(\ref{oprodnmul}) at $k=1$ implies that
$\exp_\times\big(\G(\mu g )\big)$ is the generating function for $2n$-linear currents via
\be\label{defI2n}
\III^{2n}_{g} = \f{\p^{ n}}{ (\p
 \gm^{  })^{n}}\exp_\times\big(\G(\mu g )\big)\Big |_{\mu=0}\,.\ee
Hence, Eqs.~(\ref{oprodnmul}), (\ref{PTGB}) express the operator product of any number of
multilinear currents via a linear combination of  multilinear currents
with composite arguments like in
Eqs.~(\ref{oprod2}), (\ref{oprod3}).

Since $\ten$  is associative and
commutative, it is sometimes convenient to
identify  $\ten$ with ``usual" product. In this notation, formula (\ref{PTGB})
takes the form
 \be \label{PTGBdet} {E}(\Gm(\mu)
)=\det{}^{-1}_\oprod(Id_\oprod-\Gm(\mu))
\exp
\Big(
\G(\Gm(\mu))    \oprod \big(       Id_\oprod-\G(\Gm(\mu))      \big)^{-1}_\oprod
\Big)\,,
\ee%
where
 \be
\label{det}
\det{}_\oprod(Id_\oprod-\Gm(\mu))=
\exp  tr_\oprod (ln_\oprod(Id_\oprod-\Gm(\mu)))\,.
\ee

 By (\ref{PTGB}), the correlator of  multilinear currents
\be
 \Phi^{n_1\ldots n_k} (g_1,\ldots, g_k)
 = \langle
\III^{2n_{1}}_{g_{1}}\,\ldots \III^{2n_{k}}_{g_{k}}
  \rangle\,
\ee
is represented by the coefficient in
front of the central element  in the multiple product of multilinear
currents, \ie
 \be\label{oprodn0}
 \Phi^{n_1\ldots n_k} (g_1,\ldots g_k) = \left(
\f{\p^{ n_{1}}}{ (\p
 \gm^{ {1} })^{n_{1}}}\ldots \f{\p^{ n_{k}}}{ (\p
 \gm^{ {k} })^{n_{k}}} det^{-1}_\oprod
\big |Id_\oprod-\Gm(\mu)\big |\right)\Big |_{\gm=0}
\,. \ee
We call Eqs.~(\ref{oprodnmul})  and (\ref{oprodn0})
butterfly formulae. To obtain operator product and $n$-point equations
for bilinear currents it remains to set ${n_1=\ldots= n_k}=1$.

Note that, as expected for Wightman functions, the result is not
symmetric with respect to permutations of indices $1,2,\ldots k$
because butterfly product $ \oprod$ is sensitive to the ordering.

 The direct proof of Eq.~(\ref{oprodnmul}) is quite simple.
As an illustration let us prove (\ref{oprodnmul}) for the
product of bilinear currents.
From (\ref{JnJ2}) it follows
\be\label{expI}%
\exp_{\ten}(\III^2_g)\, \III^2_f=
  \exp_{\ten}(\III^2_g)\ten \left(\III^2_f+\III^2_{f\opd g+ g\op f}+
\III^2_{g\op f\opd g}+\strop(g\op f)\III^0\right)\,.
\ee%
Substitution of $ g=\sum_{k=1}^\infty \Gm_k(\mu)$ with $\Gm_k(\mu)={\underbrace{ \Gm(\mu) \oprod\ldots
\oprod  \Gm(\mu)}_k}$
 into (\ref{expI}) gives
 \bee\nn
E(\Gm ) \III^2_f=
 E(\Gm) \ten \left(\III^2_f+\III^2_{f\opd \sum_{k=1}^\infty \Gm_k+ \sum_{k=1}^\infty
 \Gm_k\op f}+
\III^2_{\sum_{k=1}^\infty \Gm_k\op f\opd \sum_{n=1}^\infty \Gm_n}+
\strop(\sum_{k=1}^\infty \Gm_k\op f)\III^0\right) \qquad
\\\label{EI}=
 E(\Gm  ) \ten \left(\III^2_f +
 \sum_{n\ge1}\sum_{k=0}^n
\III^2_{ \Gm _k\op f\opd  \Gm_{n-k}}+\strop(\sum_{k=1}^\infty  \Gm_k\op f)\III^0 \right)\,.\qquad\qquad\qquad\eee

On the other hand,   direct
differentiation gives \be \label{mu2} \f{\p}{ \p {\gm^{i }}}
E(\Gm(\gm) )\Big|_{\gm^i=0}= E(\Gm(\gm) ) \ten\left( \III^2_{g_{i
}}+\sum_{k>1}  \stropr(\Gm_k (\gm)\oprod g_i)\III^0 + \sum_{n\ge1}
\sum_{k=0}^n \III^2_{\Gm_k(\gm) \oprod {g_{i }}\,\oprod
   \Gm_{n-k}(\gm)}
 \right) \Big|_{\gm^i=0}\,
.
\ee
Let  $i_{max}$ be the maximal label among nonzero  $\gm^j$.
Comparison of (\ref{EI}) with $f=g_{i_{max}}$
 and (\ref{mu2}) at $i= {i_{max}}$  gives
\be  \label{mu2EI} \f{\p}{  \p {\gm^{i_{max} }}} E(\Gm(\gm)
)|_{\gm^{i_{max}}=0}= E(  \breve{\Gm}(\gm) ) \III^2_{g_{i_{max}}}\q
\breve \Gm(\gm) =  \Gm(\gm)\big |_{\gm^{i_{max}}=0}\q
\breve\G(\Gm(\gm)) = \G(\Gm(\gm))\big |_{\gm^{i_{max}}=0}\,
\ee
because in
this case $\breve{\Gm} _k \op f\opd \breve{\Gm}_{n-k}
= \breve{ \Gm}_k \oprod f\oprod \breve{ \Gm}_{n-k} $  and
$\strop(\sum_{k=1}^\infty  \breve{\Gm}_k\op f)=\stropr(\breve{\Gm}_k (\gm)\oprod f)$.
The inductional proof of (\ref{oprodnmul})  is completed by observing that
$
\p_{\gm_{i
}} E(\Gm(\gm) )|_{\gm=0 }= \III^2_{g_{i }}\,.$
Since bilinear currents generate the full operator algebra this proves
butterfly formula. An alternative derivation
from the multiparticle algebra is presented in the next section.

Butterfly formulae encode information on all $n$-point functions and current
operator algebra  in an amazingly compact and efficient way as
 demonstrated in the sequel by practical computations.
Concise and elegant form of butterfly  formulae suggests that they should
have some simple origin in a more general setup which
is believed to result from a multiparticle HS theory. Appearance of the
butterfly determinant indicates that it may admit some Gaussian
(one-loop) representation.  It is also worth to mention that the
sum of all butterfly powers in the exponential in (\ref{PTGB})
describes all connected operator products.

\section{Butterfly formulae from multiparticle algebra}
  \label{BMult}

\subsection{Multiple operator products from multiparticle algebra}

In \cite{Vasiliev:2012tv} it was shown that the generating function  for
  $2n$-currents
$\III^{2n}_g$ (\ref{defI2n}), which are  order-$n$ homogeneous functionals of $g$,
has the form
\be
\widetilde G_g = det^{-1}_\star\left | e_\star-\Pi_+\star g\right |
\exp \left(g\star (e_\star-\Pi_+\star g)^{-1}_\star \right)\,,
\ee
\be\label{minusop}
(e_\star -A)_{\star}^{-1}=e_\star+A+ \ldots+ \underbrace{A\star A\star\ldots \star A}_n+\ldots\,.
\ee

Note that in \cite{Vasiliev:2012tv}  notation $\nu$ was used instead of
$g$ and the preexponential factor had power ${\mathcal N}$. The difference
is that star product of \cite{Vasiliev:2012tv} does not contain color indices,
hence corresponding to the case of ${\mathcal N}=1$, while in this paper
star product includes matrix multiplication over color indices. As a result,
the trace of this
paper is ${\mathcal N}$ times larger than that of \cite{Vasiliev:2012tv}, that,
in accordance with (\ref{det}),  brings a power of  ${\mathcal N}$ to all determinants. Taking this into account,
the associative product, that reproduces the operator algebra of $A$-currents,
is \cite{Vasiliev:2012tv}
\be
\label{A}
\widetilde G_{g_1} \diamond \widetilde G_{g_2}=
\f{1}{det_\star|  \left(e_\star +\Pi_+\star\Upsilon_2\right) |
 det_\star | e_\star-\Pi_+ \star g_1 |\, det_\star | e_\star-\Pi_+\star g_2 |}
\widetilde G_{\mathbf{g} }
 \,,
\ee
where
\be\label{resg}
{\mathbf{g}  }= \left(
\Upsilon_2\star\left(e_\star +\Pi_+\star\Upsilon_2\right)_\star^{-1}
\right)\,,
\ee
\be\label{Ups2}
\Upsilon_2 =\left(g_1\star(e_\star -\Pi_+\star g_1)_\star^{-1}+e_\star\right)
\star\left(g_2\star(e_\star -\Pi_+\star g_2)_\star^{-1}+e_\star\right)-e_\star\,.
\ee
 By virtue of
Eqs.~(\ref{star+-}),   (\ref{kmprod}),
in   terms of  butterfly product   $\oprod$, ${\mathbf{g}  }$ (\ref{resg}) reads as
\be\label{resgMV}
{\mathbf{g}  }=
 (e_\oprod +g_1  )\oprod
(e_\oprod-g_2 \oprod g_1  )^{-1}_\oprod  \oprod g_2  +
(e_\oprod +g_2 )\oprod(e_\oprod-g_1 \oprod g_2 )^{-1}_\oprod \oprod  g_1 \,.
\ee

Using that the maps $f(g)=g\star (e_\star -\Pi_+ g)^{-1}_\star$ and
$f^{-1}(g)=g\star (e_\star +\Pi_+ g)^{-1}_\star$ are mutually inverse, \ie
$f^{-1}(f(g)) = g$, it is easy to obtain the composition law for
the product of any number $n$ of generating functions that encodes the
operator product of any number of operators
\be
\label{F}
\widetilde G_{g_1} \diamond \ldots \diamond   \widetilde G_{g_n} =
\f{\tilde{\eta}(g_1)\ldots\tilde{\eta}(g_n)}
{\tilde{\eta}(\Upsilon_n\star\left(e_\star +\Pi_+\star\Upsilon_n\right)_\star^{-1})}
\widetilde G_{\mathbf{g} }  \,,
\ee
where
 \be\label{etamult}
\tilde{\eta}(A)=\exp\left[- tr_\star\left( \log_\star(e_\star-\Pi_+\star A)\right)\right] \q\,\,\,
%\ee\be\label{resg}
{\mathbf{g} }= \left(
\Upsilon_n\star\left(e_\star +\Pi_+\star\Upsilon_n\right)_\star^{-1}
\right)\,,
\ee
\be\label{Upstil}\ls\!\!
\Upsilon_n=\left(\tilde{ g}_1 +e_\star\right)
\star\ldots\star\left(\tilde{ g}_n +e_\star\right)-e_\star\,
\q  \tilde{ g_i}= { g}_i\star(e_\star -\Pi_+\star g_i)_\star^{-1}\,.
\ee
In these terms, the operator product
$
\III^2_{g_1}\ldots\III^2_{g_n}$ is described by the part
of (\ref{F}) polylinear in   ${g_1}$, \ldots, ${g_n}$.
OPE of multilinear currents $\III^{2n_i}_{g_i}$ are associated with the
terms of order $n_i$ in $g_i$.

To obtain from here butterfly formulae it is convenient to
use notations (\ref{opordeqnu}), rewriting (\ref{F}) in the  form
\be
\label{Fb}
\widetilde G_{\tilde{ g}_1}\diamond \ldots \diamond  \widetilde G_{\tilde{ g}_n} =
 {\exp\Big[\sum_{m\le n}
 tr_\bpd\Big(  \log_\bpd\big(
 e_\bpd +\tilde  g_m \big)\Big) +tr_\bpd\Big( \log_\bpd \big(e_\bpd - \Upsilon_n \big)
\Big)\Big]}
{
} \widetilde{G}_{\mathbf{g}}
\,
\ee
with
\be\label{g}
{\mathbf{g} }=
\Upsilon_n\bpd \left(e_\bpd  -\Upsilon_n\right)_\bpd^{-1} \,.
\ee
$\Upsilon_n$   (\ref{Upstil}) can be understood as
\be
\Upsilon_n=\Upsilon\big|_{g_{i>n}=0}\,,
\ee
where
 \be \label{Upstilb}
\Upsilon= \sum^\infty_{\stackrel{k=1}
{i_1<\ldots<i_{k }} } \tilde{g}_{i_1}\star \ldots\star
\tilde{g}_{i_k}  \, \qquad \mbox{with}\,\quad
\tilde{ g}_i=\sum_{k\ge 0}  \underbrace{{ g}_i
\bpd  \ldots
\bpd g_i }_k   \, .
\ee

Let
\be\label{UpsS}
 \S=\sum_{i=1}^\infty  \tilde{ g}_i\,.
\ee
Butterfly formulae follow from
(\ref{Fb}) by virtue of the following fundamental relations proven in Section
\ref{fundprove}
 \bee\label{multprogr}
{\mathbf{g} }=\Upsilon\bpd  \big(e_\bpd - \Upsilon \big)_\bpd^{-1}\,
&=&
\left(e_\oprodeq  -\S\right)_\oprodeq^{-1}\oprodeq \S
\,,\\ \label{multlog}
tr_\bpd\Big( \log_\bpd \big(e_\bpd - \Upsilon \big)
\Big) &=& tr_\oprodeq \Big(\log_\oprodeq \big(
 e_\oprodeq-\S \big)\Big) \,
 . \eee

By virtue of (\ref{multprogr}), (\ref{multlog}), Eqs.~(\ref{Fb}), (\ref{g})
 give the generating formula (\ref{oprodnmul}) with
\be \label{PTGBtil} {E}(\Gf(\mu)
)=\exp \Big (
    -tr_\oprodeq
             \big(
             ln_\oprodeq(
                        Id_\oprodeq-\Gf(\mu)
                         )
             \big)
       \Big )
\exp_{\ten}
\Big(
      \G(\Gf(\mu))
   \oprodeq\big(
          Id_\oprodeq-\G(\Gf(\mu))
           \big)^{-1}_\oprodeq
\Big)\,,
\ee%
  \be\label{Veqtil}
\Gf(\mu)= \sum_{j=1}^\infty  \tilde{g}_j(\gm^j)\,\q
\tilde{ g_i}(\mu^i)= \gm^i{ g}_i\oprodeq(e_\oprodeq -  \gm^i g_i)_\oprodeq^{-1}\,
\q\ee
\be
\G\big(\Gf(\mu)\big) \oprodeq \, \G\big(\Gf'(\mu)\big)=
\G\big(\Gf(\mu)\oprodeq\, \Gf'(\mu)\big)\,.
\ee

The remarkable property of equation (\ref{PTGBtil}) is
that multiple  $\oprodeq $ products
 of $g_i$ in (\ref{Veqtil}) cancel against those resulting from the
 butterfly product $\oprodeq$  in (\ref{PTGBtil})   in such a way that no
 term containing $g_j \oprodeq  g_j$  survives.
 This result is anticipated because
 the terms with the same $g_i$ are associated with some $\III^{2n}_{g_{i }}$
which is normal-ordered, and hence
 never undergo a Wick contraction giving rise to
 $\oprodeq$ products of $g_i$. The simplest way to see how this
  cancellation occurs is to use
   formula (\ref{resgMV}) which   shows that this  is indeed true for the product
 of a pair of multilinear currents. By induction   it remains
 true for the  product of any number of multiparticle currents.

   Effectively, this allows
 us to replace $\tilde g_i$ (\ref{Veqtil}) by $g_i$ and the product $\oprodeq$ by
 $\oprod$ in (\ref{PTGBtil}) to obtain formula (\ref{PTGB}).

\subsection{Proof of fundamental relations}
\label{fundprove}
The proof  of (\ref{multprogr}) and (\ref{multlog}) is based on
 the mutual associativity of the involved butterfly products.
 Let us first prove relation (\ref{multprogr}).

$\mathbf{I.}\quad$ Using that $\stackrel{-}{\oprod}=\big(\oprodeq-\bpd\big)$, Eq.~(\ref{multprogr}) can be
 equivalently rewritten as
 \be
 \label{multprogr1}
 \Upsilon =
\S+\S \stackrel{-}{\oprod}
\Upsilon \,. \ee
With the help of  (\ref{opordeqnu}) we obtain
\be\label{Sopordeqnu}\S
\stackrel{-}{\oprod}
 \!\!\! \sum_{\stackrel{1\le k\le n}  {i_1<\ldots<i_{k }\le n} }
\tilde{g}_{i_1}\bst\ldots\bst \tilde{g}_{i_k}=
\sum_{\stackrel{2\le k\le n}  {i_1<\ldots<i_{k }\le n} }
\tilde{g}_{i_1}\bst\ldots\bst \tilde{g}_{i_k} \,,\,\, \ee
which by virtue of  (\ref{Upstilb}) gives (\ref{multprogr1}),
 thus proving  (\ref{multprogr}).

 Note that the identity (\ref{multprogr1}) can be equivalently written
 in the form
 \be
 (e_{\stackrel{-}{\oprod}} - \S)\stackrel{-}{\oprod}( e_{\stackrel{-}{\oprod}}+
 \Upsilon ) = e_{\stackrel{-}{\oprod}} \,.
 \ee
{}From where it follows in particular that
$
\S\stackrel{-}{\oprod}  \Upsilon =  \Upsilon \stackrel{-}{\oprod}  \S\,.
$

$\mathbf{II.}\quad$
Let us first  prove the determinant relation (\ref{multlog})  directly by checking its
 variation with respect to $g_i$. Practically, it is convenient to proceed by
 induction first checking that the variation $\delta_1$
 with respect to $\delta g_1$ gives identity
 \bee \label{multlogd}
\gd_1
tr_\bpd\Big( \log_\bpd \big(e_\bpd - \Upsilon \big)
\Big) &=&  \gd_1 tr_\oprodeq \Big(\log_\oprodeq \big(
 e_\oprodeq-\S \big)\Big) \,
  . \eee
 After this is checked, one can set $g_1=0$ and prove
analogously that the variation with respect to $g_2$ gives identity, {\it etc}.

By virtue of (\ref{multprogr}), (\ref{multlogd}) is equivalent to
  \bee\label{multlogdd}
\gd_1
tr_\bpd\Big( \log_\bpd \big(e_\bpd - \Upsilon \big)
\Big) &=&-   tr_\oprodeq  \Big(\gd_1(\S)+\gd_1(\S) \oprodeq \Upsilon\bpd
  \big(e_\bpd - \Upsilon \big)_\bpd^{-1}\,  \Big)\,.
\eee
On the other hand,
\bee
\label{multlogd1}\gd_1 tr_\bpd\Big( \log_\bpd \big(e_\bpd - \Upsilon \big)
\Big) =
-tr_\bpd\Big(  \gd \tu_1\,
\bst\, (\Upsilon' +e_\bst)\bpd  \big(e_\bpd - \Upsilon \big)_\bpd^{-1}\,
 \Big)
 \,,\quad
 \eee
where
\be
\label{multlogd2}
\Upsilon'=
\left(\tilde{ g}_{2} +e_\bst\right)
\bst \ldots\bst \left(\tilde{ g}_n +e_\bst\right)- e_\bst  .
\ee

By virtue of  (\ref{troprodciknueq}) and  (\ref{multlogd1})
\bee\label{Upstilmn2}\gd_1 tr_\bpd\Big( \log_\bpd \big(e_\bpd - \Upsilon \big)
\Big) =-tr_\oprodeq\Big(
\gd\tu_1\bpd  \big(e_\bpd - \Upsilon \big)_\bpd^{-1}
+
\gd\tu_1 \bst  \Upsilon'
  \bpd  \big(e_\bpd - \Upsilon \big)_\bpd^{-1}
 \Big).\,\,
\eee
Since $ \big(e_\bpd - \Upsilon \big)_\bpd^{-1}\,
=e_\bpd+\Upsilon\bpd \big(e_\bpd - \Upsilon \big)_\bpd^{-1}\,   $,
\be\label{Upstilmn4}
\gd_1 tr_\bpd\Big( \log_\bpd \big(e_\bpd - \Upsilon \big)
\Big) =- tr_\oprodeq\Big( \gd\tu_1\,+
 \gd\tu_1\,\big(\bst\, \Upsilon'
 +  \bpd \Upsilon
\big)\bpd  \big(e_\bpd - \Upsilon \big)_\bpd^{-1}\,
 \Big).\ee
 By virtue of    (\ref{star+-b}), this proves (\ref{multlogdd}) using that
 $\delta_1 (\S)= \delta \tilde{g}_1$.

$\mathbf{III.}\quad$ Determinant relation (\ref{multlog})
is a particular case of a general relation not referring to the
detailed structure of the operators $\Upsilon$ and $\S$.
Namely, let $A$ be a biassociative
algebra with the mutually associative product laws $\bullet$ and $\diamond$
and respective traces obeying
\be
\label{trtr}
tr_\diamond (a\bullet b)= tr_\bullet (b\diamond a)\q \forall a,b \in A\,.
\ee
Let
\be
\diamond  = \bullet - \circ\,.
\ee
Then
\be
\label{log}
tr_\bullet (\log_\bullet (e_\bullet+ a\diamond (e_\diamond-a)^{-1}_\diamond) -
tr_\circ (\log_\circ (e_\circ+ a))+ tr_\diamond
(\log_\diamond  (e_\diamond - a))=0\,,
\ee
or, equivalently,
\be
\det{}_\bullet (e_\bullet+ a\diamond (e_\diamond -a)^{-1}_\diamond) \det{}_\diamond  (e_\diamond - a) =\det{}_\circ (e_\circ+ a)
\,.
\ee
This gives (\ref{multlog}) upon identification of
$\oprodeq, \bpd$ and $\stackrel{-}{\oprod}$ with  $\bullet$,
$\circ$ and $\diamond $, respectively, taking
into account that $\det{}_\diamond  (e_\diamond - a)=1$
because
\be
\tr_{\stackrel{-}{\oprod}}
\underbrace{(\S \stackrel{-}{\oprod}\ldots \stackrel{-}{\oprod} \S)}_n=0 \quad \mbox{at} \quad n>0\,.
\ee

To prove (\ref{log}) it is convenient to introduce the product
\be
\stackrel{t}\circ = \circ +t \diamond \q t\in [0,1]\,,
\ee
obeying
\be
\label{dprod}
 \f{\p}{\p t} \stackrel{t}\circ = \diamond\q
 \stackrel{0}\circ = \circ\,,\quad \stackrel{1}\circ = \bullet\,.
\ee

For any $a_0\in A$,
\be
\label{at}
a_t = a_0\diamond  (e_\diamond-t a_0)^{-1}_\diamond \,
\ee
obeys
\be
\label{pa}
\f{\p}{\p t} \,a_t = a_t \diamond  a_t\q a_t \stackrel{t}\circ (1+a_t)^{-1}_{\stackrel{t}\circ}=
a_{t'} \stackrel{{t'}}\circ (1+a_{t'})^{-1}_{\stackrel{{t'}}\circ}\qquad \forall t,t'\,.
\ee

{}From (\ref{trtr}), (\ref{dprod}) and (\ref{pa}) it follows that
\be
 \f{\p}{\p t}\,\, tr_{\stackrel{t}\circ} (\log_{\stackrel{t}\circ} (e_{\stackrel{t}\circ}+ a_t) ) =tr_\diamond  (a_t)\,.
\ee
With the help of  (\ref{at}), integration of this relation gives (\ref{log}).

\section{Space-time current operator algebra}
\label{Contact}

Once current operator algebra in the twistor space is known, it is
straightforward to extend it to space-time. Indeed,
  unfolded form of the current equation (\ref{rank2eq}) uniquely
determines dependence on space-time coordinates in terms of
twistor coordinates by virtue of (\ref{Xsdvigtoka})  or (\ref{DCYlow}).
 It is important
that the latter formulae hold for  general conserved currents, \ie not
necessarily built from free fields. This is because the unfolded current
equation holds independently of the origin of conformal conserved currents.

More generally, this implies that, in the framework of unfolded
formulation,  operator algebra is fully determined by the
associative operator algebra of functions of generalized twistor
variables of the system in question. Analysis and classification of
twistor algebras is much simpler than that of space-time operator
algebras. As shown in the previous
section, for free massless fields, this algebra is based
on the star-product HS algebra. More precisely, twistor operator algebra  is the
universal enveloping algebra of the HS algebra (for more detail see
 \cite{Vasiliev:2012tv}). This simple observation
clarifies in particular the difference between $2d$ and $3d$ conformal
theories pointed out
in \cite{Maldacena:2011jn,Maldacena:2012sf}:  $3d$
conformal HS algebras admit no deformation, while $2d$ conformal HS
algebras belong to a parametric family of algebras described in particular
 in \cite{Prokushkin:1998bq} (and references therein).

\subsection{$A$-currents}
\label{ACUR}
Space-time operator algebra  is  the algebra of currents
$\JNN_{{\gamma }}(\yy_1,\yy_2|X)$ (\ref{rhoJ}) at $\yy_1=\yy_2=0$ with parameters $\gga
={\gamma^{mn}_{ab}} (\A )$, that depend  on $\A_\pm$ (\ref{param12})
and obey  (\ref{:ggagga':}),
\be\label{Jgamma} \JNN_{{\gamma }} (X)= \JNN_{{\gamma }}
(Y_1,Y_2|X)\Big |_{Y_{1,2}=0} \,. \ee
It should be stressed that setting to zero $\yy$-variables does not imply
any loss of generality since the dependence of currents on twistor variables
is fully accounted by the freedom in the parameters $\gga$. To compute operator algebra
for all conformal currents and their descendants
 it suffices to consider  parameters
${\gamma^{mn}_{ab}} (\A_+(Y_1|X),\A_-(Y_2|X))$   that only depend on
the  operators $\A_{\pm \,C }$ (\ref{param12}) which represent
$Y$-derivatives. However, in most of our consideration, we  keep the
dependence on the $X$-dependent operators $\A_\pm^C$ since this does
not affect the analysis too much.

A useful trick is to interpret $\JNN_{{\gamma}} (Y_1,Y_2|X ) $ (\ref{rhoJ})
 as the current $\III^2_g$ (\ref{Jg}) with a function $g(\zz)=\gD(\yy_1,\yy_2 ,X ;{{\gamma }}|\zz )$ that
depends on  $\yy_i,$   $X$ and $\gamma$ as parameters
\be\label{evolintab}
  \JNN_{{\gamma}} (\yy_1, \yy_2|X )
   = \III^2_{\gD(\yy_1, \yy_2,X ;{{\gamma }} ) } \,.
  \ee
 To  find $\gD(\yy_1,\yy_2,X;{{\gamma }} )$ we observe that,  due to evolution
formula  (\ref{DCYlow}) along with  the following identity  for    ${C}_m^a( \yy|X)$ of definite parity
 \be \label{DCYlowna}
  \int\! d^M  \zz\,  \D_a(\zz - \yy|X)
C_m^a(\zz |0)= (-1)^{\ppi_m}
\! \int\! d^M  \zz\,  \D_a(\zz + \yy|X)
C_m^a(\zz |0)
  \ee
(no summation over $a$),  $\JNN_\gga$ (\ref{rhoJX12}) acquires the form
 \be \label{rhoJDab} \JNN_\gga(\yy_1,\yy_2|X)=
-\!\!\sum_{m,n,a,b}\!\! \gamma_{a\,b}^{m \,n}
\left( \rule{0pt}{14pt} \A_+ \left( \yy_1|X \right),
 \A_-\left( \yy_2|X \right)\right)
\!\int\! d\zz_{1, 2} \mathfrak{D}^{{\ppi_m} {\ppi_n}}_{a\,\,b}(\yy_1,\yy_2
,X |\zz_1,\zz_2)
  \TT^{ab}_{m\,n}( \zz_1, \zz_2|0)  \quad
\ee
with $a,b=+,-$,\,\, $ \TT^{ab}_{j\,k}$  (\ref{Tab})\,,
 \be\label{Dmn}
 \mathfrak{D}^{\ppi_m \ppi_n}_{a\,b}(\yy_1,\yy_2
,X|\zz_1,\zz_2)=(\kappa^a_1)^{\ppi_m}( \kappa^b_2)^{\ppi_n}
\D^{\ppi_m}_{a}(\zz_1, \kappa^a_1\yy_1|-X)
\D^{\ppi_n}_{b}(\zz_2,\kappa^b_2\yy_2|-X), \ee $\kappa_j^a$
(\ref{kappa}) and \be\label{Dm} \D^{\ppi_m}_{a}(\zz,  \yy |X
)=\D^{\ppi_m}_{a}(\yy,  \zz |X ) =\half\big(\D_{a}(\zz -\yy |X
)+(-1)^{\ppi_m}\D_a(\zz+ \yy |X )\big)\,. \ee By virtue of
(\ref{Dsol}), (\ref{lowindD}) and (\ref{intDDpmM}), for
non-degenerate $X$,
  \be\label{DE}
\D^{{\ppi_k}}_{-}( W\,, Y  |X) =\,  \D _{-}(X)
  \,\E^{\ppi_k}\big( W\,, Y|  X \big)\q
  \D^{\ppi_k}_{+}( iW\,,i Y |X) =\,  \D _{+
  }(X)
  \,\E^{\ppi_k}\big( W\,,Y|  X \big)
 \,,
\ee
  where
 \be \label{Dsollow} \D_\pm (X )
=-\f{i}{2^M\pi^{\f{M}{2}}} \exp \left(\pm\f{i\pi I_X}{4}\right)
\f{1}{\sqrt{\hhh|\det (X)|}} \,, \ee
  \be \label{Ecos}\E^0\big(W\,,Y\,|X  \big)=
 \exp\left( -  \f{ i}{4\hhh} X^{-1}_{AB}(WW+YY)^{AB}  \right)
\cos\left(   \f{ 1}{2\hhh} X^{-1}_{AB} W^A Y^B
  \right)\,,\ee
  and
 \be \label{Esin}\E^1\big(W\,,Y\,|X \big)=i
 \exp\left( -  \f{ i}{4\hhh} X^{-1}_{AB}(WW+YY)^{AB}  \right)
\sin\left(    \f{ 1}{2\hhh} X^{-1}_{AB} W^A Y^B
  \right)\,.\ee

$\mu$-symmetrization of (\ref{rhoJDab}) with $\mu$ (\ref{mug})
 gives
 \bee\label{rhoJDabmu} \JNN_\gga(\yy_1,\yy_2|X)=
-\sum_{a,b}\int d\zz_1\,d\zz_2\,\,
\gD{}^{m\,n}_{a\,b}(\yy_1,\yy_2
,X ;{{\gamma }} |\zz_1,\zz_2)
  \TT^{ab}_{m\,n}( \zz_1, \zz_2|0)\,,  \quad
\eee where \be\label{getasymab} \!\!\!\gD{}^{{m}
{n}}_{a\,b}(\yy_1,\yy_2 ,X ;{{\gamma }} |\zz_1,\zz_2) =
-\half(1+\mu)\big(\gamma_{ab}^{mn} (  \A_+{}( \yy_1|X )  ,
\A_-(\yy_2|X )   )
   \mathfrak{D}^{\ppi_m \ppi_n}_{a\,\,b}(\yy_1,\yy_2
,X |\zz_1,\zz_2)\big)\,,
  \ee
   that by construction
   satisfies
(\ref{Comrelg}) as well as (\ref{:gg':})
by virtue of (\ref{:ggagga':}).

Using butterfly formula (\ref{oprodnmul}), Eq.~(\ref{evolintab})
 determines the operator product of any number of multilinear space-time currents
$
 \JNN^{2k}_{\gga }=:\underbrace{\JNN_{\gga }\ldots \JNN_{\gga }}_k:\,
$
\be\label{oprodnmulX}
\JNN^{2k_1}_{\gga_1}(\yy^1_1,\yy^1_2|X^1)\ldots \JNN^{2k_n}_{\gga_n}(\yy^n_1,\yy^n_2|X^n)
= \left(\f{\p^{ k_{1}}}{ (\p
 \gm^{ {1} })^{k_{1}}}\ldots \f{\p^{ k_{n}}}{ (\p
 \gm^{ {n} })^{k_{n}}}
  {E}(\GmX(\gm)
)\right)\Big |_{\gm=0}\, ,   \ee
where
\be\label{VeqX}
 \GmX(\gm)=\sum_{j=1}^\infty  \gm^j g_j \q
 g_j=\gD (\yy^j_1,\yy^j_2,X^j;\gga_j)\,
\,,
\ee
\be \label{PTGBX} {E}(\GmX(\mu)
)=\exp \big [ -\stropr(ln_\oprod(Id_\oprod-\GmX(\mu))\big ]
\exp_{\ten}\big(\G(\GmX(\mu))\oprod (Id_\oprod-\G(\GmX(\mu))
)^{-1}_\oprod\big)\,
\ee
with $\G (g)= \III^2_{  g}$.
The product of  bilinear currents is obtained at
$k_1=\ldots=k_n=1$.

Operator product of
 space-time  multilinear currents results from (\ref{oprodnmulX})
at $\yy=0$
\be\label{prodmulti0}
\JNN^{2k_1}_{\gga_1}( X^1)\ldots\JNN^{2k_n}_{\gga_n}( X^n)=
\JNN^{2k_1}_{\gga_1}(\yy^1_1,\yy^1_2|X^1)\ldots\JNN^{2k_n}_{\gga_n}(\yy^n_1,\yy^n_2|X^n)
\Big|_{\yy=0}.\ee
In this case, formulae  are simplified further due to  simplification of
$
 \gD{}^{m\,n}_{a\,b}(\yy_1,\yy_2,X ;{{\gamma }} |\zz_1,\zz_2)\Big |_{\yy=0}
 $
  on the \rhs of (\ref{oprodnmulX}), that
  only depends on $\rho$-invariant  parameters $\gga^+{}{}_{a\,b}^{m\,n}$, \ie
\be\label{getasymab2new0}\gD{}^{0\,m\,n}_{a\,b}(X  ;{\gga} |\zz_1,\zz_2)
:=\gD{}^{m\,n}_{a\,b}(\yy_1,\yy_2,X ;{{\gamma }} |\zz_1,\zz_2)\Big |_{\yy=0}
=
\gD {}^{m\,n}_{a\,b}(\yy_1,\yy_2, X;{{\gga^+ }} |\zz_1,\zz_2)
 \big|_{\yy=0}
 \,.\quad   \ee
In fact, this is anticipated as a consequence of (\ref{rhoTab}).
 More formally,  we observe
  that, by virtue of (\ref{kappaprop}),
 \be\label{muD1rho}(\kappa^a_1)^{\ppi_m} \D^{\ppi_m}_{a}(\zz, \kappa^a_1\yy|X )=
(i)^{\ppi_m}   (\kappa^a_2)^{\ppi_m}
\D^{\ppi_m}_{a}(\zz,  i\kappa^a_2  \yy|X )\,.
\ee
Hence
\be\label{rhoD}
   \mathfrak{D}{}^{\ppi_m \ppi_n}_{a\,\,b}(\yy_1,\yy_2
,X|\zz_1,\zz_2) =
 i^{\ppi_{m}+\,\ppi_{n}}
 \mathfrak{D}{}^{\ppi_n \ppi_m}_{b\,\,a}(i\yy_2,i\yy_1
,X|\zz_2,\zz_1)\,
\ee
and
\be\label{murhoD}
\mu( \mathfrak{D}{}^{\ppi_m \ppi_n}_{a\,\,b}(\yy_1,\yy_2 ,X|\zz_1,\zz_2))
=i^{\ppi_{m\,,n}} \mathfrak{D}{}^{\ppi_n \ppi_m}_{b\,\,a}(i\yy_2,i\yy_1
,X |\zz_1,\zz_2)\,. \ee
Since,  by virtue of (\ref{param12i}),
$\gga^\pm$ (\ref{ggaoe}) satisfy
\be
   \gga^\pm{}_{a\,b}^{m\,n} (\A_+{}(   \yy_1|X)  , \A_-(\yy_2|X)   )
= \pm (-i)^{\ppi_m+\ppi_n}
  \gga^\pm{}{}_{b\,a}^{n\,m}(
\A_+{}( i  \yy_2|X)  , \A_-(i\yy_1|X)   )\,,
\ee
Eq.~(\ref{getasymab}) can be written in the form
\bee\label{getasymab2}\nn&&\ls\ls\ls\gD{}^{m\,n}_{a\,b}(\yy_1,\yy_2
,X ;{{\gamma }} |\zz_1,\zz_2) = -\half \Big\{
\Big( \gga^+{}{}_{a\,b}^{m\,n}+ \gga^-{}{}_{a\,b}^{m\,n} \Big) (
\A_+{}(   \yy_1|X )  , \A_-(\yy_2|X)   )
 \,\,\,   \mathfrak{D}{}^{\ppi_m \ppi_n}_{a\,\,b}(\yy_1,\yy_2
,X |\zz_1,\zz_2)
\\ &&\qquad\qquad\quad+
\Big( \gga^+{}{}_{a\,b}^{m\,n}- \gga^-{}{}_{a\,b}^{m\,n} \Big)(
\A_+{}( i  \yy_2|X )  , \A_-(i\yy_1|X )   )
 \,\,\,\mathfrak{D}{}^{\ppi_m \ppi_n}_{a\,\,b}(i\yy_2,i\yy_1
,X |\zz_1,\zz_2)\Big\}\,.   \eee
To prove (\ref{getasymab}) it remains to observe that $\gga^-$-dependent terms
cancel at $Y_{1,2}=0$.

Since, by virtue of (\ref {PAIRintop}), (\ref  {PAIRintdi2}),
 butterfly product $  \G\big(\gD\big) \oprod  \G\big(\gD'\big) $
(\ref{Gkrugloe}) deals  with
 integrations of   $\gD,\, \gD'$  over variables $\zz $, that does not involve
 variables $\yy$,
 Eq.~(\ref{oprodnmulX})
along with  (\ref{prodmulti0}), (\ref{getasymab2new0}) gives
\be\label{oprodnmulX0}
\JNN^{2k_1}_{\gga_1}(  X^1)\ldots\JNN^{2k_n}_{\gga_n}(  X^n)
= \left(\f{\p^{ k_{1}}}{ (\p
 \gm^{ {1} })^{k_{1}}}\ldots \f{\p^{ k_{n}}}{ (\p
 \gm^{ {n} })^{k_{n}}}
  {E}(\GmX^0(\gm)
)\right)\Big |_{\gm=0 } \,,   \ee
where
\be\label{VeqX0}
 \GmX^0(\gm)=\sum_{j=1}^\infty \gm^j g^0_j \q
 g^0_j=\gD^0 ( X^j;\gga_j)\,
\,,
\ee
\be\label{ggga000}  \gD^0{}^{m\,n}_{a\,b}( X;{{\gga }} |\zz_1,\zz_2) =-
  \gga^+{\,}_{a\,b}^{m\,n}  (
\A_+{}(   \yy_1|X)  , \A_-(\yy_2|X)   )
 \,\,\,   \mathfrak{D}{}^{\ppi_m \ppi_n}_{a\,\,b}(\yy_1,\yy_2
,X|\zz_1,\zz_2)\big|_{\yy=0}
 \,.\quad   \ee

According to (\ref{kmprod})
\be \label{but0}
   \gD{}^{0} (X^i  ;{\gga}_i  )\oprod
\gD{}^{0} (X^j  ;{\gga}_j  )  =
  \theta(j-i)\gD{}^{0 } (X^i  ;{\gga}_i  )\op
\gD{}^{0 } (X^j  ;{\gga}_j  ) +\theta(i-j) \gD{}^{0 } (X^i  ;{\gga}_i  )\opd
\gD{}^{0 } (X^j  ;{\gga}_j  ) \,,\quad\ee
where  Eqs.~(\ref{PAIRintop}), (\ref{PAIRintdi2}),
(\ref{Dmn}) and
 (\ref{ggga000}) along with the following consequence of
Eq.~(\ref{intDD})  \be\label{intDDk}
\int d^M p\, \D^{\ppi_k}_\mp(p, \yy| X ) \D^{\ppi_l}_{\pm}(p,   \yy'| X')= -i
\gd^{\ppi_k \ppi_l}    \D^{{\ppi_k} }_{\mp}( \yy ,\yy' |X- X')\,
   \ee
give
\bee \label{op0}
&&\big( \gD{}^{0\, } (X^q  ;{\gga}_q  )\op
\gD{}^{0\,} (X^p  ;{\gga}_p  )\big)^{m\,l}_{ad}(\zz_1,  \zz_2)
 \qquad\\ \nn =&& -2i\gd_{nk}   \Big(
 (\gga^{+}_{q})_{a\,-}^{m\,n}  (
\A_+{}(   \yy^q_1|X^q)  , \A_-(\yy^q_2|X^q)   )
\,\,(\gga^{+}_{p})_{+\,d}^{\,k\,\,l}  (
\A_+{}(   \yy^p_1|X^p)  , \A_-(\yy^p_2|X^p)   )
  \,\\ \nn&&
 \,\,\, \,\gd^{\ppi^2_q \ppi^1_p}\,  \mathfrak{D}{}^{{\ppi_m \ppi_l}}_{a\,\,d}(\yy^q_1,\yy^p_2
,X^q,X^p|\zz_1,  \zz_2)
     \,
{\D}{}^{\ppi^2_q}_{- }(\yy^q_2,\yy^p_1|  X^p-X^q
 )\Big)
\big|_{\yy=0}\, \,
,\quad   \eee
\bee \label{opd0}
&& \gD{}^{0\, } (X^q  ;{\gga}_q  )\opd
\gD{}^{0\, } (X^p  ;{\gga}_p  )\big)^{m\,l}_{ad}(\zz_1,  \zz_2)
 \qquad\\ \nn  =&&  -2i\gd_{nk}   \Big(
 (\gga^{+}_{q})_{a\,+}^{m\,n}  (
\A_+{}(   \yy^q_1|X^q)  , \A_-(\yy^q_2|X^q)   )\,\,
(\gga^{+}_{p})_{-\,d}^{\,k\,\,l}  (
\A_+{}(   \yy^p_1|X^p)  , \A_-(\yy^p_2|X^p)   )
  \,\\ \nn&&
 \,\,\, \gd^{\ppi^2_q \ppi^1_p} \,  \mathfrak{D}{}^{{\ppi_m \ppi_l}}_{a\,\,d}(\yy^q_1,\yy^p_2
,X^q,X^p|\zz_1,  \zz_2)
 \,
\,   {\D}{}^{\ppi^2_q}_{+ }(i\yy^q_2,i\yy^p_1|  X^p-X^q
 )\Big)
\big|_{\yy=0}\, \,
,\quad   \eee
where
 \be\label{Dmn2}
 \mathfrak{D}^{\ppi_m\ppi_n}_{a\,b}(\yy_1,\yy_2
,X_1,X_2|\zz_1,\zz_2)=(\kappa^a_1)^{\ppi_m}( \kappa^b_2)^{\ppi_n}
\D^{\ppi_m}_{a}(\zz_1, \kappa^a_1\yy_1|-X_1) \D^{\ppi_n}_{b}(\zz_2,\kappa^b_2\yy_2|-X_2)
\ee
and we  assume   that ${\gga}_j={\gga}_j{\,}^{m_jn_j}_{a_jb_j}$
has definite  parities  with respect to the first and
second arguments
\be\label{fixpar}
\ppi_j^1:=
\ppi_{m_j}  \q \ppi_j^2:=\ppi_{n_j} \qquad \forall \,m_j,\,n_j.
\ee
Recall, that additional factors of $i$ in ${\D}{}^{ \ppi_k}_\pm$ of (\ref{opd0})
result from Eqs.~(\ref{kappa}) and (\ref{Dmn}).

 To obtain explicit  formulae for   $g^0_1\oprod\ldots \oprod g^0_n$
and $tr_\oprod(g^0_1\oprod\ldots \oprod g^0_n)$\, with $g_j^0$ (\ref{VeqX0}),
we introduce butterfly product of parameters $\gga_i= \gga_i
(\A_+{}(   \yy^{i }_1|X^{i })  , \A_-(\yy^{i }_2|X^{i })   )
$ as follows
\bee \label{but0gga}
&&  (\gga_{i,j})^{ml}_{ad}= (\gga_i \oprod
 {\gga}_j )^{ml}_{ad} %\\ \nn &&
 =\gd_{nk}\big\{\theta(j-i )\tau^{bc}+
 \theta(i-j )\tau^{cb}\big\}
 (\gga_i)^{mn}_{ab}
(\gga_j)^{kl}_{cd}
 \eee
and, inductively, $\gga_{ j_1,...,j_p } =\gga_{ j_1}\oprod\ldots\oprod\gga_{j_p }$,
\bee \label{but0ggaind}
&& %(\gga_{i_1,...,i_k,j_1,...,j_p })^{ml}_{ad}=
(\gga_{i_1,...,i_k } \oprod
 \gga_{ j_1,...,j_p })^{ml}_{ad}%\\ \nn &&
 =\gd_{np}\big\{\theta(j_1-i_k )\tau^{bc}+
 \theta(i_k-j_1 )\tau^{cb}\big\}
 (\gga_{i_1,...,i_k })^{mn}_{ab}
(\gga_{ j_1,...,j_p })^{p\,l}_{cd}\,,
 \qquad\eee
 where  the arguments of $ \gga_j
(\A_+{}(   \yy^{j }_1|X^{j })  , \A_-(\yy^{j }_2|X^{j })   )
$ are omitted for simplicity.
The trace is defined  analogously
\be\label{trgga}
tr_{\oprod}(\gga_{ j_1,...,j_p })=
\gd_{mk}\big\{\theta(j_1-j_p )\tau^{ ba}+
 \theta(j_p-j_1 )\tau^{ab}\big\}
 (\gga_{ j_1,...,j_p })^{mk}_{ab}\,.
 \ee
This gives by virtue of (\ref{op0}),  (\ref{opd0})
\bee  \label{but00kgga}
\!&&\big(g^0_{j_1}\oprod\ldots \oprod g^0_{j_n}\big)^{m \,l }_{a \,d}
(\zz_1,  \zz_2|X)=
\big(\gD{}^{0 } (  X^{j_1}  ;{\gga}_{j_1}  )\oprod
\ldots \oprod
\gD{}^{0\, } (  X^{j_n}  ;{\gga}_{j_n}  )\big)^{m \,l}_{a \,d}
(\zz_1,  \zz_2)
\qquad\quad\\  \nn&=&
   -2^{ n-1 }
 \Big\{
\Big(  \gga^+_{j_1} \oprod    \ldots   \,\oprod
          \gga^+_{j_n} \Big)^{m \,l }_{a\, d}
 \,%\\ \nn&&
 \mathfrak{D}{}^{\ppi_m \ppi_l }_{\,a\,\,d}(\yy^{j_1}_1,\yy^{j_n}_2
,X^{j_1},X^{j_n}|\zz_1,  \zz_2 )  \,\prod_{k=1}^{n-1} \mathbb{E}_{j_k,\,j_{k+1}}
 \,\,\,
  \Big\}
\big|_{\yy=0}\, \,
 \quad   \eee
and, using again (\ref{DInert+}) and (\ref{intDDk}),
\bee     \label{but00kggatrb}
tr_\oprod (g^0_{j_1}\oprod\ldots \oprod g^0_{j_n})(X)
=
tr_\oprod (
\gD{}^{0 } (  X^{j_1}  ;{\gga}_{j_1} )\oprod
\ldots \oprod
\gD{}^{0\, } (  X^{j_n}  ;{\gga}_{j_n} )
)\qquad  \\ \nn=
(-1)^{\ppi^1_{j_1}}2^{n-1 }
\Big\{
  tr_{\oprod}\Big(  \gga^+_{j_1} \oprod    \ldots \,\oprod
          \gga^+_{j_n} \Big)\, \mathbb{E}_{j_n,\,j_{ 1}}
   \,  \prod_{k=1}^{n-1} \mathbb{E}_{j_k,\,j_{k+1}}
 \Big\}\big|_{\yy=0}\, ,\quad
\eee
where, according to (\ref{Dsollow}),
  (\ref{Ecos}) and
  (\ref{Esin}),
        \be\label{Dfrac} \mathbb{E}_{j,\,k}= \,\gd^{\ppi^2_j \ppi^1_k} \,
\f{ \E^{\ppi^2_{j}}_{j, k }}{\Sq^{j ,\,k}}\,,\ee
\be\label{ordersqrji} \Sq^{j,\,k}=
   {  (4\pi)^{\f{M}{2}}}\exp\left(  \mathrm{sign}(k-j) \f{i\pi I_{X^k-X^j} }{4}\right)
\sqrt {|\det(X^k-X^j)|}\,,
\ee
  \be \label{Ecosjk}\E^{0\, }_{j,\,k}=
 \exp\left( -  \f{ i}{4\hhh} (X^k-X^j)^{-1}_{AB}(\yy_2^j\yy_2^j+\yy_1^k\yy_1^k)^{AB}  \right)
\cos\left(   \f{ 1}{2\hhh} (X^k-X^j)^{-1}_{AB} \yy_2^j{}^A \yy_1^k{}^B
  \right)\,,\ee
 \be \label{Esinjk}\E^{ 1\,}_{j,\,k}=i
 \exp\left( -  \f{ i}{4\hhh} (X^k-X^j)^{-1}_{AB}(\yy_2^j\yy_2^j+\yy_1^k\yy_1^k)^{AB}  \right)
\sin\left(    \f{ 1}{2\hhh} (X^k-X^j)^{-1}_{AB} \yy_2^j{}^A \yy_1^k{}^B
  \right)\,.\ee
\subsection{$F$-currents}
\label{FCUR}

  $F$-currents   $\JJJ_\eta(\yy|X)$   are represented by $\JNN_\gga(\yy|X)$
   at $\gga_{ab}=\eta$ $\forall a,b$.
Space-time $F$-currents are $\JJJ_\eta(\yy|X)$   evaluated at $\yy
=0$. As explained in Section \ref{Quantization},
 the $\rho-$odd part $\eta ^-$ (\ref{etaoe})  does not contribute to
$\JJJ_\eta(0,0|X  )$ (\ref{quanJetaspacerho}). Hence,
 space-time $F$-currents are represented by $\III^2_{  g  }$
(\ref{Jg}) with  $g= g_{\{0,X;\eta^+\}} $.

Naively, this conclusion disagrees with the results of  Section
\ref{Charges}, where it was shown that, for $M=2$, nonzero
contribution to the charges gives $\eta^-$ (\ref{etaoe}), obeying
$\rho (\eta^-)=-\eta^-$. However, one should take into account that the
$dX$-dependent part of the closed form  (\ref{rhoWarpiab})
contains derivatives $\f{\p}{\p U^A}$ equivalent to $\B_A$. Hence,
the parameters of charges and currents differ by $M$ factors of
$\B_A$. As a result, their $\rho$-parities differ by a factor of
$(-1)^{\f{M}{2}}$, which is $-1$ at $M=2$. Note that the  symmetry
properties of parameters match for all $M$. However, in the general
case, the correspondence is   less trivial because, as demonstrated
in \cite{gelcur}, physical parameters contribute to the closed
form (\ref{rhoWarpiab}) with additional singular factors that
support  integration over spinor variables and compensate the
mismatch in $\rho$-parity. For example, in the $4d$ case of $M=4$,
the symmetry parameter contributes to the physical charge with the
singular factor bilinear in twistor variables, which just brings in
a factor of $-1$. In the twistor sector, necessary sign
factors result from the appropriate choice of integration contours
in the twistor space.

Operator product of $n$
free conformal $F-$  multilinear currents $\JJJ^{2k_j}_{\eta_j}$
with $\rho-$ invariant parameters $\eta_j$ is given by
\be\label{oprodnmulFX0}
\JJJ^{2k_1}_{\eta_1}(  X^1)\ldots\JJJ^{2k_n}_{\eta_n}(  X^n)
= \left(\f{\p^{ k_{1}}}{ (\p
 \gm^{ {1} })^{k_{1}}}\ldots \f{\p^{ k_{n}}}{ (\p
 \gm^{ {n} })^{k_{n}}}
  {E}(\GmX^0(\gm)
)\right)\Big |_{\gm=0 } \,,   \ee
with
\be\label{VeqX0F}
 \GmX^0(\gm)=\sum_{j=1}^\infty \gm^j g^0_j\q g^0_j=\gD^0 ( X^j;\eta_j)\,
\,,
\ee
\be\label{geta000}  \gD^0{}^{m\,l}_{a\,b}( X;{{\eta }} |\zz_1,\zz_2) =-
  \eta{\,}^{m\,l}  (
\A_+{}(   \yy_1|X)  , \A_-(\yy_2|X)   )
 \,\,\,   \mathfrak{D}{}^{\ppi_m\ppi_l}_{\,a\,\,b}(\yy_1,\yy_2,X|\zz_1,\zz_2)\big|_{\yy=0}
 \,.\quad   \ee

The formula to be used below for the derivation of OPE and
 correlators of  bilinear $F$-currents is obtained at
$k_1=\ldots=k_n=1$
\be\label{oprodnbiX0}
\JJJ_{\eta_1}(  X^1)\ldots\JJJ_{\eta_n}(  X^n)
= \left(\f{\p }{ \p \gm^{ {1} } }\ldots \f{\p }{  \p
 \gm^{ n }  }
  {E}(\GmX^0(\gm)
)\right)\Big |_{\gm=0}\,.    \ee

For     $F-$currents,
butterfly product  (\ref{but0gga}) amounts to the  matrix product
$\eta_j \cdot  {\eta}_k$ with respect
to color indices, \ie
\be \label{but0eta}
    (\eta_i \cdot  {\eta}_j )^{ml}
 =\gd_{nk}
 (\eta_i)^{mn}
(\eta_j)^{kl}
 \q
tr_{\cdot}(\eta_{ j_1}\cdot\eta_{j_2})=
\gd_{mk}
 (\eta_{ j_1}\cdot\eta_{j_2})^{mk} \,.
 \ee
 Assuming that ${\eta}_j={\eta}_j{\,}^{m_jn_j} $
has definite  parities $\ppi_j^1$ and $\ppi_j^2$  (\ref{fixpar}) with respect to the first and
second arguments and that $j_k\ne j_m$ for $k\ne m$, this gives
 \bee  \label{but00kF}
&&
\big(\gD{}^{0 \,} (  X^{j_1}  ;{\eta}_{j_1}  )\oprod
\ldots \oprod
\gD{}^{0\, } (  X^{j_n}  ;{\eta}_{j_n}  )\big)^{m\,\,l}_{a \,d}
(\zz_1,  \zz_2)\, \,\\  &=&
   -2^{ n-1 }
\Big\{
\nn
\big(  \eta_{j_1} \cdot   \ldots   \,\cdot          \eta_{j_n} \big)^{m\,\,l}
 \, %\\ \nn&&
 \mathfrak{D}{}^{\ppi_m\ppi_l}_{a \,d}(\yy^{j_1}_1,\yy^{j_n}_2
,X^{j_1},X^{j_n}|\zz_1,  \zz_2) \prod_{k=1}^{n-1} \mathbb{E}_{j_k,\,j_{k+1}}
  \Big\}
\big|_{\yy=0}\, \,
 ,   \\\label{but00kggatrbF}&&\stropr\big(g^0_{ {j_ 1}}\oprod\ldots \oprod g^0_{ j_{n}  }\big)=tr_\oprod (
\gD{}^{0\, } (  X^{j_1}  ;{\eta}_{j_1} )\oprod
\ldots \oprod
\gD{}^{0\, } (  X^{j_n}  ;{\eta}_{j_n} )
)   \\ \nn &=&
   (-1)^{\ppi^1_{j_1}}2^{n-1 }
\Big\{
  tr_{\cdot}\Big(  \eta_{j_1} \cdot   \ldots   \,\cdot          \eta_{j_n}  \Big)\, \mathbb{E}_{j_n,\,j_{ 1}}
      \prod_{k=1}^{n-1} \mathbb{E}_{j_k,\,j_{k+1}}
 \Big\}\big|_{\yy=0}\,.
 \eee
These formulae
make the computation of OPE and correlators of conserved currents
straightforward.

        \section{Correlators of free conformal currents}
  \label{Correlators}

\subsection{Any $M$ and $3d$   }
\label{Correlators3d}

From (\ref{oprodnbiX0}) it follows that the correlator  of $n$
free conformal $F-$ currents $\JJJ_{\eta_j}$ with $\rho-$ invariant parameters $\eta_j$
can be represented in terms of butterfly  algebra $A_\oprod$  as
\be
 \label{ncorr}\!\! \big\langle \JJJ_{\eta_1 }(X^1)\ldots\JJJ_{\eta_n}(X^n)\big\rangle
=
  \sum_{\S_n}\sum_r \f{1}{r!} \!\!\!\!\sum_{{k_1,\ldots , k_{ r}\ge 2}\atop {k_1+\ldots+k_{ r}=n}}
\!\!\!\!
  \f{1}{k_1\cdots k_r}\stropr\big( g^0_{  j^1_{ 1} , \ldots , j^1_{ k_1}}\big)
  \ldots
   \stropr\big( g^0_{  j^r_{ 1} , \ldots , j^r_{ k_r}}\big)
   \,,
\ee
   where $g^0 _{  j _{ 1} , \ldots , j_{ k }}=g^0_{j_{ 1}}\oprod\ldots \oprod g^0_{j_{k}  }$
   with
   $ g^0_j=\gD^0 ( X^j;\eta_j)$ (\ref{geta000}),
and
  summation over $\S_n$ implies symmetrization over indices
 $ j^1_{ 1} , \ldots , j^1_{ k_1} ,\ldots, j^r_{ 1} , \ldots ,j^r_{k_r}$.
 (Note that the terms where at least one $k_i=1$ do not contribute because
$ \stropr\big( g^0_{  j_{ k} } \big) =0
 $ in accordance  with the fact that, being normally ordered, fundamental currents
 have zero VEVs.)
Connected $n$-point functions  of free $F$-currents with $\rho$ invariant parameters  $ \eta_j $ are
\be\label{oprodn0Xirr} \big\langle\JJJ_{\eta_1 }(X^1
)\ldots\JJJ_{\eta_n}(X^n)\big\rangle_{con}  = \f{1}{n}  \sum_{\S_n}
   \stropr\big(g^0_{ { 1}}\oprod\ldots \oprod g^0_{ {n}  }\big)
  \, ,
\ee
where summation is over all permutations of $(1 , \ldots , n)$.

 Substitution of (\ref{but00kggatrbF}) into (\ref{ncorr}), (\ref{oprodn0Xirr}) gives
$n$-point functions of free $F$-currents.
These formulae hold for any $M\ge 2$.  At $M= 2$, they describe
$n-$point functions of $3d$ conformal currents. For any $M$, the
primary currents (\ref{primcurUV}) are associated with the
parameters
 \be\label{H0}
\eta_0\q \eta(\p_U)\q \eta(\p_V)\q \eta^{AB}\p_{U^A}\p_{V^B},\quad\mbox{with }\,\,\,
\eta^{AB}=-\eta^{BA}\,.
\ee

 \subsection{  $4d$  }
\label{Correlators4d} As discussed in  \cite{Vasiliev:2001zy,Mar},
usual  Minkowski space $\Mi$ is a subspace of the matrix space
$\M_M$ with appropriate $M$. Embedding of $4d$ Minkowski space-time
into $\M_4$  is most conveniently described in terms of
two-component indices $\ga,\gb =1, 2$ and $\da,\db =1', 2'$ in place
of the four-component indices $ A, B \ldots $ with the convention
that complex conjugation exchanges unprimed and primed indices. We
use notation with $ A= (\ga ,\da$) and $\yy^ A =(y^\ga  , \by^\da
)$, where $ \by^\da=\overline{y^\ga }$.
Two-component indices are  raised and
lowered according to \bee \nn A^\ga=\gvep^{\ga \gb}A_\gb\q
A_\gb=\gvep_{\ga \gb}A^\ga\q \gvep_{\ga\gb} = - \gvep_{\gb\ga}\q
\gvep_{12} = 1\,, \eee and analogously for primed indices.

In these terms,
 \bee\label{dotundot} X^{AB}=\left(
\beee{cc  } X^{\ga\gb}&X^{\da\gb}\\
X^{\ga\db}&\overline{X}^{\da\db} \eeee \right),\qquad\yy^{A }=\left(
\beee{c   } y^{\ga } \\
\by^{ \db} \eeee \right) \eee
with  $\overline{X^{\ga\db}}=
X^{\gb\da}$ and $\overline{X^{\ga\gb}}=\overline{X}^{\da\db}$. For
Minkowski coordinates, we  also  use notation $x^{\ga \db}$
instead of $X^{\ga \db}$. Minkowski time $t$ and space coordinates $x^i$ are
\bee
\label{4dfib} X^{\ga \db} =t\TTT^{\ga\db} +x^i \sigma_i^{\ga \db}\q
i=1,2,3 \,,
\eee
where $\TTT^{\ga\db}=\delta^{\ga\db}$ while
$\sigma_i^{\ga \db}$ are Hermitian traceless Pauli matrices.

As shown in \cite{Mar}, since $4d$ Minkowski space is identified with
\bee\label{antidiag} X^{AB}=\left(
\beee{cc  } 0&x^{\da\gb}\\
x^{\ga\db}&0 \eeee \right)\q \eee
the  inertia index of $X^{AB}$  equals to $\pm 4 $ or $0$.
As a result, $s{(X)}=- \det (x^{\gb\da}) $.

$4d$ correlators are given by Eqs.~(\ref{ncorr}), (\ref{but00kggatrbF})
with coordinates (\ref{antidiag}), $\yy=(y,\by)$,
$ \eta_{j }=\eta_{j }\big(\A_+(\yy{}^{j }_1|x^{j }),\A_+(\yy{}^{j }_2| x^{j })\big)\,,
$
 $\ppi^{1,2}_{j }$ being the parity of $\eta_{j }$ with respect to $\yy^{j }_{1,2}\,$,
  respectively, and
  \newcommand{\e}{e}
  \be\label{DfracM} \mathbb{E}_{j,\,k}= \,\gd^{\ppi^2_j \ppi^1_k} \,
\f{ \e^{\ppi^2_{j}}_{j, k }}{ { (4\pi)^{ 2}   }\det(x^{j }-x^{ {k}})}\,,
\ee
   \be\label{DGausspMc}
  \e^{0}_{j,\,k} =
\exp\left( -  \f{ i}{2\hhh}
(x^k-x^j)^{-1}_{\ga\db}
( y^k_1{}^\ga\by^k_1{}^\db + y^j_2{}^\ga\by^j_2{}^\db )
\right)
\cos\left(   \f{1}{2\hhh}(x^k-x^j)^{-1}_{\ga\db}
\big(y^j_2{}^\ga\by^k_1{}^\db +y^k_1{}^\ga\by^j_2{}^\db
  \big)
\right)\q
  \, \ee
   \be\label{DGausspMs}
  \e^{1}_{j,\,k} =i
\exp\left( -  \f{ i}{2\hhh}
(x^k-x^j)^{-1}_{\ga\db}
( y^k_1{}^\ga\by^k_1{}^\db + y^j_2{}^\ga\by^j_2{}^\db )
\right)
\sin\left(   \f{1}{2\hhh}(x^k-x^j)^{-1}_{\ga\db}
\big(y^j_2{}^\ga\by^k_1{}^\db +y^k_1{}^\ga\by^j_2{}^\db
  \big)
\right)\,.\qquad\qquad
  \ee

Connected $n$-point functions  of free $F$-currents
{ with $\rho-$ invariant parameters $\eta_j$} are \be\nn
\big\langle\JJJ_{\eta_{ { 1}} }(x^{ { 1}} )\ldots \JJJ_{\eta_{ { n}}}(x^{ {
n}})\big\rangle_{con} =\f{1}{n} \sum_{\S_n} \stropr\big( g^0_{  { 1} ,
\ldots , { n}}\big). \ee

To describe primary currents,  that belong to the cohomology group
$H^0(\gs^2_-{}_{Mnk})$ of $4d$ Minkowski current equation where
  \be \label{sigma-M}%
\gs_-^2{}_{Mnk} =\half d x^{{\ga}\pb}\left(
 \frac{\ptl}{\ptl v^{\ga}} \frac{\ptl}{\ptl \bu  ^{\pb}}
+  \frac{\ptl}{\ptl u^{\ga}} \frac{\ptl}{\ptl \bv^{\pb}}\right)
=  d x^{{\ga}\pb}\left(-
 \frac{\ptl}{\ptl y_1^{\ga}} \frac{\ptl}{\ptl \by_1  ^{\pb}}
+  \frac{\ptl}{\ptl y_2^{\ga}} \frac{\ptl}{\ptl \by_2^{\pb}}\right)
\q\ee%
 following \cite{param} it is useful to introduce ${\mathfrak{sl}_2}$
 generators
\be \label{ortalgy}%
\rho_{+}= y_1^\gga \frac{\ptl}{\ptl y_2^\gga}\,+\, \by_2^\pga
\frac{\ptl}{\ptl  \by_1^\pga}\q \rho_{-}= y_2^\gga \frac{\ptl}{\ptl y_1^\gga}\,
+\, \by_1^\pga \frac{\ptl}{\ptl  \by_2^\pga}\q
\rho_{0}= y_1^\gga \frac{\ptl}{\ptl y_1^\gga}\,+\, \by_2^\pga
\frac{\ptl}{\ptl  \by_2^\pga}- y_2^\gga \frac{\ptl}{\ptl y_2^\gga}\,
-\, \by_1^\pga \frac{\ptl}{\ptl  \by_1^\pga}\,,
\ee%
that commute  to $\gs_-^2{}_{Mnk}$ and act  on the space of bilinear
currents and, hence, on the space of parameters.

 Treating
 $
 \rho _{+}\,$  as a  positive grade operator,
one can find highest $ {\mathfrak{sl}_2}$-vectors in
$H^0(\gs^2_-{}_{Mnk})$ to reconstruct full $H^0(\gs^2_-{}_{Mnk})$ by the action of
${\rho}_{-}$. This gives the following set of parameters
\be\label{primMin}
 \eta\left(\frac{\ptl}{\ptl  y_1},\frac{\ptl}{\ptl  y_2}\right)
 \q \bar{\eta}\left(\frac{\ptl}{\ptl \by_1},\frac{\ptl}{\ptl \by_2}\right)
 \q\rho_{-}^k\left(\tilde{\eta}\left(\frac{\ptl}{\ptl y_2 }, \frac{\ptl}{\ptl \by_1}
 \right)\right)\quad \mbox{with }  \rho_{-}(\eta)=\left[ {\eta} \,,\,    \rho_{-} \right] \,\,\,( k=0,1,\ldots),   \ee
representing all primary currents.
Note that here
$\tilde{\eta}%\left(\frac{\ptl}{\ptl y_2 }, \frac{\ptl}{\ptl \by_1}\right)\,,
\    \rho_{-}$ is replaced by $\left[ \tilde{\eta} \,,\,    \rho_{-} \right]$
using that this does not affect the final result since $Y=0 $ in
(\ref{but00kggatrbF}).

For example, consider
\be\label{examppq}
\tilde{\eta}^{(p,q)}\left(\frac{\ptl}{\ptl y_2 }, \frac{\ptl}{\ptl \by_1 }\right)=
\eta^{^{\gga(p)\,;\,\pga(q)}} \frac{\ptl}{\ptl y_2 ^{\gga_1}}\ldots
\frac{\ptl}{\ptl y_2 ^{\gga_p}} \frac{\ptl}{\ptl \by_1^{\pga_1}}\ldots
\frac{\ptl}{\ptl \by_1^{\pga_q}}.
\ee
Let \be\label{helM} \mathfrak{h}_1=\half\big(y_1^\gga \frac{\ptl}{\ptl y_1^\gga}\,-\, \by_1^\pga
\frac{\ptl}{\ptl  \by_1^\pga}\big)\q
\mathfrak{h}_2=\half\big(y_2^\gga \frac{\ptl}{\ptl y_2^\gga}\,-\, \by_2^\pga
\frac{\ptl}{\ptl  \by_2^\pga}\big) \ee be rank-$1$ helicity operators in $4d$
Minkowski space.
Evidently,
$$\mathfrak{h}_1(\tilde{\eta})=\half q \,\tilde{\eta}\q
\mathfrak{h}_2(\tilde{\eta})= -\half p\, \tilde{\eta}\,.$$

Let $\tilde{\eta}^{(p,q)}_k=\f{1}{k!} \rho^k\left(\tilde{\eta}^{(p,q)} \right)$.
Clearly, $\tilde{\eta}^{(p,q)}_k=0$ for $k>(p+q)$.
Since
$\mathfrak{h}_1(\rho_{-} )=-\half \rho_{-}  $, $
\mathfrak{h}_2(\rho_{-})=  \half \rho_{-}$,
 %(\ref{helM})
$$\mathfrak{h}_1(\tilde{\eta}^{(p,q)}_k)=\half (q-k)\,\tilde{\eta}^{(p,q)}_k\q
 \mathfrak{h}_2(\tilde{\eta}^{(p,q)}_k)=-\half  (p-k)\, \tilde{\eta}^{(p,q)}_k\,.$$

In particular, for $q=p=2$, this gives
\bee\label{stresstensor4}
\tilde{\eta}^{(2,2)}_0&=& {\eta}^{\ga\gb\,;\pa\pb}
\frac{\ptl}{\ptl y_2 ^{\ga }}
\frac{\ptl}{\ptl y_2 ^{\gb}} \frac{\ptl}{\ptl \by_1^{\pa}}
\frac{\ptl}{\ptl \by_1^{\pb}}\q
\\ \nn%1
\tilde{\eta}^{(2,2)}_1&=& {\eta}^{\ga\gb\,;\pa\pb}
\left\{
\frac{\ptl}{\ptl y_1 ^{\ga }}
\frac{\ptl}{\ptl y_2 ^{\gb}} \frac{\ptl}{\ptl \by_1^{\pa}}
\frac{\ptl}{\ptl \by_1^{\pb}}+\frac{\ptl}{\ptl y_2 ^{\ga }}
\frac{\ptl}{\ptl y_2 ^{\gb}} \frac{\ptl}{\ptl \by_2^{\pa}}
\frac{\ptl}{\ptl \by_1^{\pb}}\right\}\q
\\ \nn%2
\tilde{\eta}^{(2,2)}_2&=& {\eta}^{\ga\gb\,;\pa\pb}
\left\{
\frac{\ptl}{\ptl y_1 ^{\ga }}
\frac{\ptl}{\ptl y_1 ^{\gb}} \frac{\ptl}{\ptl \by_1^{\pa}}
\frac{\ptl}{\ptl \by_1^{\pb}}+2\frac{\ptl}{\ptl y_1 ^{\ga }}
\frac{\ptl}{\ptl y_2 ^{\gb}} \frac{\ptl}{\ptl \by_1^{\pa}}
\frac{\ptl}{\ptl \by_2^{\pb}}+\frac{\ptl}{\ptl y_2 ^{\ga }}
\frac{\ptl}{\ptl y_2 ^{\gb}} \frac{\ptl}{\ptl \by_2^{\pa}}
\frac{\ptl}{\ptl \by_2^{\pb}}\right\}\q
\\ \nn%3
\tilde{\eta}^{(2,2)}_3&=& {\eta}^{\ga\gb\,;\pa\pb}
\left\{
\frac{\ptl}{\ptl y_1 ^{\ga }}
\frac{\ptl}{\ptl y_1 ^{\gb}} \frac{\ptl}{\ptl \by_1^{\pa}}
\frac{\ptl}{\ptl \by_2^{\pb}}+\frac{\ptl}{\ptl y_1 ^{\ga }}
\frac{\ptl}{\ptl y_2 ^{\gb}} \frac{\ptl}{\ptl \by_2^{\pa}}
\frac{\ptl}{\ptl \by_2^{\pb}}\right\}\q
\\ \nn%4
 \tilde{\eta}^{(2,2)}_4&=& {\eta}^{\ga\gb\,;\pa\pb}
\frac{\ptl}{\ptl y_1 ^{\ga }}
\frac{\ptl}{\ptl y_1 ^{\gb}} \frac{\ptl}{\ptl \by_2^{\pa}}
\frac{\ptl}{\ptl \by_2^{\pb}} \,.
\eee
The pairs
$\tilde{\eta}^{(2,2)}_0$, $\tilde{\eta}^{(2,2)}_4$
and $\tilde{\eta}^{(2,2)}_1$, $\tilde{\eta}^{(2,2)}_3$
generate the same currents modulo
exchange of the fundamental fields $C_1$ and $C_2$.
%that correspond to the well known three cases of different stress tensors provided that
%${\eta}^{\ga\gb\,;\pa\pb}=\overline{{\eta}^{\ga\gb\,;\pa\pb}}$.
Indeed, in this case
$\tilde{\eta}^{(2,2)}_0= \overline{\tilde{\eta}^{(2,2)}_4}\,,
\tilde{\eta}^{(2,2)}_1=\overline{\tilde{\eta}^{(2,2)}_3}\,,
\tilde{\eta}^{(2,2)}_2=\overline{\tilde{\eta}^{(2,2)}_2}$.
As a result, we are left with three independent structures associated with
$\tilde{\eta}^{(2,2)}_0$, $\tilde{\eta}^{(2,2)}_1$
and  $\tilde{\eta}^{(2,2)}_2$.
These correspond to three types of stress tensors
\bee\label{t1,1} T^{(1,1)}&=&\left(
\tilde{\eta}^{(2,2)}_0 C_1(y_1,\by_1)C_2(y_2,\by_2)+
\tilde{\eta}^{(2,2)}_4 C_1(y_1,\by_1)C_2(y_2,\by_2)\right)\Big|_{y=\by=0}
%&=&{\eta}^{\ga\gb\,;\pa\pb}T_{\ga\gb\,;\pa\pb}
\q\\ \label{th,h} T^{(\half,\half)}&=&\left(
\tilde{\eta}^{(2,2)}_1 C_1(y_1,\by_1)C_2(y_2,\by_2)+
\tilde{\eta}^{(2,2)}_3 C_1(y_1,\by_1)C_2(y_2,\by_2)\right)\Big|_{y=\by=0}
%&=&{\eta}^{\ga\gb\,;\pa\pb}T_{\ga\gb\,;\pa\pb}
\q
\\\label{t0,0} T^{(0,0)}&=&\left(
 \tilde{\eta}^{(2,2)}_2 C_1(y_1,\by_1)C_2(y_2,\by_2)\right)\Big|_{y=\by=0}
%&=&{\eta}^{\ga\gb\,;\pa\pb}T_{\ga\gb\,;\pa\pb}\q
\,,\eee
 constructed,
respectively, from free fields of spins $1$,  $\half$ and  $0$.
The respective three-point functions
reproduce three known  three-point functions of $4d$ stress tensors  (see, e.g., \cite{Stanev:2012nq,Zhiboedov:2012bm}).

Analogously,  our construction reproduces $s+1$ independent
structures for three-point functions of spin $s$ conserved currents
associated with free fields of spins $0,1/2,\ldots s/2$.

\section{Examples   }
In this section, obtained results are illustrated by the derivation of
$n$-point functions in the generalized space-times with arbitrary
$M$. At $M=2$, the resulting formulae give correlators of usual $3d$
conformal currents.
\label{Examples}
\subsection{Integer spins }
\label{Integer s}
 Consider bosonic $F-$currents.
 Let
 \be\label{paramE}
\eta_j^{mn}(\A_{+}(\yy_1|X) ,\A_{-}(\yy_2|X))= E {}^{mn}
\eta_j(\A_{+}(\yy_1|X) ,\A_{-}(\yy_2|X))+i\widetilde{E }{}^{mn}
\widetilde{\eta}_j(\A_{+}(\yy_1|X) ,\A_{-}(\yy_2|X)) \ee with
   \bee\label{Eunitar}
  E ^{km}=  \gd^{mk}\q \widetilde{E} ^{km}= -\widetilde{ E} {}^{mk}\q
\widetilde{E}{}^{mk}\widetilde{E}{}_{k}{}^{n}= -\gd{}^{mn}\q
    \delta{}^{m}_{m}= \NNN \,  \quad
  \eee
(indices are raised and lowered by the $O(\NNN)$ invariant metric
$\delta_{nm}$), and
$ \eta_j\,, $   $\widetilde{\eta}_j $ obeying
\bee \label{rhoE}\eta_j(\A_{+}(\yy_1|X) ,\A_{-}(\yy_2|X))=\eta_j(\A_{+}(i\yy_2|X) ,\A_{-}(i\yy_1|X))
\q\\ \nn \widetilde\eta_j(\A_{+}(\yy_1|X) ,\A_{-}(\yy_2|X))=-\widetilde\eta_j(\A_{+}(i\yy_2|X) ,\A_{-}(i\yy_1|X))\,
\qquad
\eee
so that, for boson currents, $\eta_j$ and $\widetilde{\eta}_j$ (\ref{paramE}) are $\rho-$invariant.
Note  that, by (\ref{rhoE}), parameters ($\widetilde \eta_j$)$\eta_j$ are
(anti)symmetric in color indices, being associated with
currents of (odd)even spins.

Substitution of (\ref{paramE}) into  (\ref{but00kggatrbF}) gives
connected  $n$-point functions (\ref{oprodn0Xirr})
   \bee  \label{nnointciklB}
\ls\big\langle\JJJ_{\eta_{ { 1}} }(X^{ { 1}} )\ldots \JJJ_{\eta_{ {n}}}(X^{
{ n}})\big\rangle_{con}^b=
      2^{n-1}
\eta_{   (n)}    \big(\A)     \sum_{\S_n}
\f{\Big(\exp i  R_{   (n)}
\cos   K_{1,  {2}} \cdots
 \cos   K_{{n-1}  ,  {n}}
 \cos   K_{ {n }  ,  {1}}
\Big) (\yy  )}
 {\Sq^{{ 1}, {2 }}\ldots\Sq^{{n-1}, {n }} \Sq^{ {n }  ,  {1}} }
  \Big|_{\yy= 0}
 \,\,\quad
   \eee
for  boson-boson  $F-$currents and
  \bee \label{ppointciklFe}
 \ls\big\langle\JJJ_{\eta_{ { 1}} }(X^{ { 1}} )\ldots
 \JJJ_{\eta_{ {n}}}(X^{ { n}})\big\rangle_{con}^f=
     -i^n 2^{n-1}
\eta_{   (n)}  \big( \A\big)     \sum_{\S_{n}}
                    \f{\Big(\exp i  R_{   (n)}
\sin   K_{1, {2}} \cdots
 \sin   K_{{n-1}  ,  {n}}
 \sin   K_{ {{n} }  ,  {1}}
\Big) (\yy)} {\Sq^{{ 1}, {2 }}\ldots\Sq^{{n-1}, {n }} \Sq^{ {n }  ,
{1}}}
        \Big|_{\yy= 0}
 \,\,\quad
   \eee
    for fermion-fermion $F-$currents. Here
\be
\label{defQPijk}
  K_{i,\,j} =   \f{1}{2\hhh}
 (X^j-X^i)^{-1}_{AB} \yy_1^j{}^A  \yy_2^i{}^ B\,,
 \ee
 \be\label{defRijk}  R_{i,\,j }=\f{1}{4\hhh}
(X^i-X^j)^{-1}_{AB}
 (\yy_2^i{}^A  \yy_2^i{}^B +\yy_1^j{}^A  \yy_1^j{}^B)\,, \qquad   R_{  (n)}
=    R_{ 1,  {2} }+\ldots+ R_{  {n-1},   {n }}+R_{ {n}, {1 }  }
\,
\ee
and       \bee\label{etaprodY}
\eta_{{ (n)}}(\A) =   \f{1}{n}tr_{\cdot}\Big(  (E\eta_{j_1}+i\widetilde{E}\widetilde{\eta}_{j_1})
\cdot\ldots\cdot(E\eta_{j_n}+i\widetilde{E}\widetilde{\eta}_{j_n})
\Big)(\A)\qquad
\\ \nn= \f{\NNN}{2n}  \left(   \prod_{j=1}^n
\big(\eta_{j}(a_1^{j},a_2^{j}) + \widetilde{\eta}_{j}(a_1^{j},
a_2^{j})\big)+
        \prod_{j=1}^n \big(\eta_{j}(a_1^{{j}} ,a_2^{j})- \widetilde{\eta}_{j}(a_1^{j},a_2^{j})
       \big)\right)
       \eee
       with
       \be
        a_1^{j}=\A_+(\yy_1^j|X^j)\q a_2^{j}\,=\,\A_-(\yy_2^j|X^j).
        \ee
Note that (\ref{etaprodY}) is a consequence of
 $$tr_{\cdot}(\underbrace{\widetilde{E}\cdot\ldots\cdot\widetilde{E}}_{2m })=(-1)^{m}\NNN\q
 tr_{\cdot}(\underbrace{\widetilde{E}\cdot\ldots\cdot\widetilde{E}}_{2m+1})=0.$$
From more general perspective, Eq.~(\ref{etaprodY}) expresses
$\rho-$invariance of correlators.

Substitution of (\ref{nnointciklB}) and (\ref{ppointciklFe}) into (\ref{ncorr}) gives
$n$-point functions for   $F-$currents of integer spins including
all disconnected contributions. Note that  formulae
(\ref{nnointciklB}) and (\ref{ppointciklFe})
are already projected to parameters $\eta_j$, $\tilde{\eta}_j$ (\ref{paramE})
of definite parities with respect to the first and second arguments. This implies
that the condition that parameters $\eta$ should have definite parities can be
discarded. We will use this property in the sequel in practical applications of
(\ref{nnointciklB}) and (\ref{ppointciklFe}), only
requiring $\eta_j$, $\tilde{\eta}_j$ to
be $\rho-$invariant, \ie to obey (\ref{rhoE}).

\subsection{Half-integer spins }
\label{Exampleshalf}

Consider  $F$-supercurrents.
To
distinguish between boson and fermion color indices we denote them
by Latin and Greek letters, respectively (the latter should not be
confused with spinor indices). Hence, a $F-$supercurrent acquires the
form
\be
\JJJ_\eta(\yy_1,\yy_2|X)=\half\left(\eta^{m\nu}\T_{m\,\nu}(\yy_1,\yy_2|X)
+\eta^{\nu\, m}\T_{ \nu\,m}(\yy_1,\yy_2|X)\right)\,,
\ee
where  parameters are required to be $\rho-$ invariant, \ie
\be\label{rhoEO}
\eta^{m\,\nu}(\A_{+}(\yy_1|X) ,\A_{-}(\yy_2|X))
=i \eta^{\nu \,m}(i\A_{-}(\yy_2|X) ,i\A_{+}(\yy_1|X)).
\ee

Let
 \be\label{paramEO}
\eta_j^{m\gn}(\A_{+}(\yy_1|X) ,\A_{-}(\yy_2|X))= E {}^{m\gn}
\eta_j(\A_{+}(\yy_1|X) ,\A_{-}(\yy_2|X))+i\widetilde{E }{}^{m\gn}
\widetilde{\eta}_j(\A_{+}(\yy_1|X) ,\A_{-}(\yy_2|X)) \ee with
   \bee\label{EunitarO}
  E ^{m\gn}= E ^{\gn m}  \q \widetilde{E} ^{m\gn}= -\widetilde{ E} {}^{\gn m }\q
\widetilde{E}{}^{\gm k}\widetilde{E}{}_{k}{}^{\gn}= -\gd{}^{\gm \gn}\q
    E{}^{\gm k}E{}_{k\gm}= \NNN \,.
  \eee
From (\ref{rhoEO}) it follows
 \bee\label{paramE=}&&
\eta_j(\A_{+}(\yy_1|X) ,\A_{-}(\yy_2|X))
=i\,  \eta_j(i\A_{-}(\yy_2|X),i\A_{+}(\yy_1|X)  )\q\\ \nn &&
\widetilde{\eta}_j(\A_{+}(\yy_1|X) ,\A_{-}(\yy_2|X))
=-i\,
 \widetilde{\eta}_j(i\A_{-}(\yy_2|X),i\A_{+}(\yy_1|X)  )\,.
 \eee

 For supercurrents,  (\ref{but00kggatrbF}) is different from zero at
$n=2m$. Taking into account that parameters $\eta_j$ (\ref{:etaeta':})
are Grassmann odd, analogously to the case of bosonic currents, the substitution of
(\ref{paramEO}) into (\ref{but00kggatrbF}) gives by virtue of (\ref{Ecos}),
  (\ref{Esin}),    (\ref{rhoEO})
 and (\ref{EunitarO})
  \bee \label{ppointsicl2}&&\ls
\big\langle\JJJ_{\eta_{ { 1}} }(X^{ { 1}} )\ldots \JJJ_{\eta_{ { 2m}}}(X^{
{ 2m}})\big\rangle_{con} =   i^m 2^{ -1}  \eta_{{  (2m)}}(\A )
 \sum_{\S_{2m} } \f{ (-1)^{\pi_{\S_{2m}}}}
 {\Sq^{{ 1}, {2 }}\ldots\Sq^{{2m-1}, {2m }}
 \Sq^{{2m }, 1 }}\,
           \\ \nn&&\ls
  \Big(-
 \cos   K_{1,  {2}} \sin   K_{2,  {3}}  \ldots
 \,
 \cdots
  \cos   K_{{2p-1 }  ,  {2p }}\sin K_{{2p  }  ,  {2p+1 }}
  \cdots
   \sin K_{{2m }  ,  {1}}
  \quad\qquad\\ \nn
  & &\ls\!+
  \sin   K_{1,  {2}}\cos   K_{2,  {3}} \ldots
    \sin   K_{{2p-1 }  ,  {2p }}\cos  K_{{2p  }  ,  {2p+1 }}
 \cdots
   \, \cos K_{{2m  }  ,  {1}}
\Big) (\yy  ) \exp \left[i   R_{(2m)}(\yy  )\right]
  \Big|_{\yy= 0}
   \eee
with
$\Sq^{ij}$ (\ref{ordersqrji}),
$K$ (\ref{defQPijk}),  $R$   (\ref{defRijk})  and
$\eta_{{  (2p)}}(\A )$  (\ref{etaprodY})
 ($ {\pi_{\S_{2m} }}$ is the  parity of a permutation  of $\S_{2m}$).
Substitution of
(\ref{ppointsicl2})   into (\ref{ncorr}) gives $n$-point functions
for   $F-$supercurrents of half-integer spins, including disconnected
contributions.

\subsection{Correlators of primary  currents of integer spins }
\label{3deven}
\subsubsection{General case}
  As mentioned in Section \ref{Correlators3d},
primary  currents in the matrix space are associated with the parameters
(\ref{H0}).  The primaries associated with antisymmetric tensor $\eta^{AB}$
we call {\it special}. Since, at $M=2$, $\eta^{\ga\gb}\sim \epsilon^{\ga\gb}$,
 in that case there is a single spin zero special current
 that has  dimension $\Delta=2$.

In the variables $U,$ $V$ (\ref{UVYY}), the functions $
K_{i,\,j }$ (\ref{defQPijk}) and $ R_{i,\,j }$ (\ref{defRijk}) have the form
 \bee\label{defQPijkUV}
  K_{i,\,j}  &=&   \f{1}{2\hhh}
 (X^j-X^i)^{-1}_{AB} (U^i{}^A+V^i{}^A)  (V^j {}^{B}-U^j {}^{B} )
\,, \qquad \\ \label{defQPijkR} R_{i,\,j }  &=&\f{1}{4\hhh}
(X^i-X^j)^{-1}_{AB}
 \big((V^i {}^{A}+U^i {}^{A} )   (V^i {}^{B}+U^i {}^{B} )
 +(V^j {}^{A}-U^j {}^{A} )   (V^j {}^{B}-U^j {}^{B} ) \big)
    \, . \eee
Setting for example in (\ref{paramE})
\be\label{paramprim}%
\eta_j=\eta_j(\p_{U^j})\q\widetilde{\eta}_j=\widetilde{\eta}_j(\p_{U^j})\,,
\ee   satisfying (\ref{rhoE}), we obtain that the nonzero contribution to
correlators comes from
 \bee\label{defQPijkp}
P_{i,\,j}=K_{i,\,j} |_{V=0}   &=& -  \f{1}{2\hhh}
 (X^i-X^j)^{-1}_{AB}  U^i{}^A U^j {}^B\,,\\\label{defRijkp}
    Q_{(p)}=  R_{(p)} |_{V=0}&=&
     Q_{1,\, {2}  ,\,\, {3}}+\ldots+ Q_{{p-2},\, {p-1}  ,\,\, {p}}+
   Q_{{p-1},\, {p }  ,\,\, {1}}+Q_{{p},\, {1 }  ,\,\, {2}}   \,
 ,
 \eee
 where \be \label{defQPijkU0}  Q_{i,j,k}
=\f{1}{4\hhh}\left((X^i-X^j)_{AB}^{-1}+(X^j-X^k)_{AB}^{-1}\right)
  U^j{}^{A }   U^j{}^{B}   \,.
\ee

Substitution  of (\ref{paramprim}) into (\ref{nnointciklB}),
(\ref{ppointciklFe}) gives
    \be \label{ppointciklBp}
\big\langle\JJJ_{\eta_{ { 1}} }(X^{ { 1}} )\ldots \JJJ_{\eta_{n}}(X^{ {
n}})\big\rangle_{con}^{b}=
   2^{n-1}
\eta_{(n)}
    \big(  \p_{U }   \big)\sum_{\S_n}
%\qquad\qquad \\ \nn
           \f{\Big(\cos   Q_{   (n)}
\cos   P_{1,\, {2}} \cdots
 \cos   P_{{n-1}  ,\, n} \cos   P_{ {n }  ,\, {1}}
\Big) (U  )} { \,\Sq^{{ 1}, {2 }}\ldots\Sq^{{n-1}, {n }} \Sq^{n, 1} }
  \Big|_{U= 0}
 \,,
   \ee
  % for $n$ boson-boson primary $F-$currents,
       \be \label{ppointciklFep}
\big\langle\JJJ_{\eta_{ { 1}} }(X^{ { 1}} )\ldots \JJJ_{\eta_{ { 2m}}}(X^{
{ 2m}})\big\rangle_{con}^{f}=
     (-1)^{m+1 } 2^{2m-1}
\eta_{   (2m)} \big( {\p_U }  \big)
\sum_{\S_{2m}}\f{\Big(\cos  Q_{
(2m)} \sin   P_{1,\, {2}} \cdots  \sin   P_{ {{2m } }  ,  {1}}\Big)
(U)} { \,\Sq^{{ 1}, {2 }}\ldots\Sq^{{2m-1}, {2m }} \Sq^{{2m}, {1 }} }
         \Big|_{U= 0}
 \,,
   \ee
   \bee \label{ppointciklFop}&&
\big\langle\JJJ_{\eta_{ { 1}} }(X^{ { 1}} )\ldots
\JJJ_{\eta_{2m+1}}(X^{ { 2m+1}})\big\rangle_{con}^{f}=
   \qquad\qquad \\ \nn&&
     (-1)^{m  }  2^{2m}
\eta_{   (2m+1)} \big( {\p_U }  \big)\sum_{\S_{2m+1} }
          \f{\Big(\sin  Q_{ (2m+1)}
\sin   P_{1,\, {2}} \cdots \sin   P_{ {{2m } }  ,\,\, {2m+1}} \sin   P_{ {{2m+1} }  ,\,\, {1}}\Big)
(U)} { \Sq^{{ 1}, {2 }}\ldots\Sq^{{2m}, {2m +1}} \Sq^{{2m +1}, {1}} }
  \Big|_{U= 0}
 \,.
   \eee
 %  for      fermion-fermion primary $F-$currents.
   (Recall that
    $\eta_{{  (n)}}(\A)$ is given in (\ref{etaprodY}).)
 In the derivation  of (\ref{ppointciklBp})-(\ref{ppointciklFop}) one
should take into account antisymmetry of $ R_{i,\,j }\big|_{V=0}$
and $K_{i,\,j}\big|_{V=0}$ in $i,j$ and full symmetrization of the
\rhs's of (\ref{nnointciklB}) and (\ref{ppointciklFe}), which effectively
produces the factors of $\cos Q_{\ldots}$ or $\sin Q_{\ldots}$ via
appropriate (anti)symmetrization of $\exp iR_{(n)}$ in
Eqs.~(\ref{nnointciklB}) and (\ref{ppointciklFe}).

As anticipated, Eqs.~(\ref{ppointciklBp})-(\ref{ppointciklFop}) are
in accordance with the results of
\cite{Giombi:2011rz,Didenko:2012tv}.

       \subsubsection{Two-point function and unitarity}
 For  primary currents with  parameters
 $\eta^{mn}_j$ of the form  (\ref{paramE}) with  $\eta_j$ and $\widetilde{\eta}_j $
 (\ref{paramprim}),
  taking into account that $Q_{(2)}=0$,
  we obtain from (\ref{ppointciklBp}) and (\ref{ppointciklFep})
  in the boson-boson  and fermion-fermion cases, respectively,
 \bee\label{2pointB}
 \big\langle\JJJ_{\eta_1 }(X^1 )\JJJ_{\eta_2}(X^2) \big\rangle^{b}=
 -  \f{ 2  \NNN \exp
\left(\f{  i\pi I_{{X^1} -{X^2} }}{2}\right) }{  (4\pi)^{ {M}}
   {\hhh|\det ({X^2}-{X^1})|}}
 \big(\eta_1{}\eta_2 + \widetilde{\eta_1}\widetilde{\eta}_2 \big)\big(\p_{U^1} \,,  \p_{U^2} \big)
 \cos^2\left(P(U^1,U^2) \right)\Big|_{U^1=U^2= 0} , \qquad\eee
\bee\label{2pointF} \big\langle\JJJ_{\eta_1 }(X^1 )\JJJ_{\eta_2}(X^2) \big\rangle^{f}=
    \f{    2\NNN      \exp
\left(\f{  i\pi I_{{X^1} -{X^2} }}{2}\right)}{(4\pi)^{ {M}}
      \hhh|\det ({X^2} -{X^1})|}
 \big(\eta_1{}\eta_2 + \widetilde{\eta_1}\widetilde{\eta}_2\big) \big(
\p_{U^1} \,,  \p_{U^2}\big)
 \sin^2\left(P(U^1,U^2) \right)\Big|_{U^1=U^2= 0},\qquad
\eee where
\be\nn
P (U^1,U^2) = -  \f{1}{2\hhh}
 (X^1 -X^2)^{-1}_{AB}  U^1 {}^A U^2 {}^B\,.
  \ee

Since  at $M=2$,  $I_{ X^1-X^2  }=\{-2,\,\,0,\,\,2\}$, as
anticipated,
 $$
  \big\langle\JJJ_{\eta_1 }(X^1 )\JJJ_{\eta_2}(X^2) \big\rangle=\big\langle\JJJ_{\eta_2 }(X^2 )
  \JJJ_{\eta_1 }(X^1) \big\rangle\quad \forall\,\, X^1,\, X^2\,; \quad M=2.
  $$%
Since the overall sign in front of the two-point function is
different for $I_{X^1-X^2}=0$ and $I_{X^1-X^2}\ne0$, for $M=2$ Eqs.~(\ref{2pointB}),
(\ref{2pointF})  can be rewritten in the form
\be\label{2pointB2}
 \big\langle\JJJ_{\eta_1 }(X^1 )\JJJ_{\eta_2}(X^2) \big\rangle^{b}=
   -\f{ 2  \NNN (4\pi)^{ -2} }{
    \hhh \det ({X^2}-{X^1}) }
 \big(\eta_1{}\eta_2 + \widetilde{\eta_1}\widetilde{\eta}_2 \big)\big(\p_{U^1},\p_{U^2} \big)
  \cos^2\left(P(U^1,U^2) \right) \Big|_{U^1=U^2= 0}\,,
 \ee%
\be\label{2pointF2} \big\langle\JJJ_{\eta_1 }(X^1 )\JJJ_{\eta_2}(X^2) \big\rangle^{f}=
   \f{    2\NNN        (4\pi)^{-2}}{
      \hhh \det ({X^2} -{X^1}) }
 \big(\eta_1{}\eta_2 + \widetilde{\eta_1}\widetilde{\eta}_2\big) \big(
\p_{U^1} ,  \p_{U^2}\big)\,
 \sin^2\,\left(P(U^1,U^2) \right)\Big|_{U^1=U^2= 0}\,.\qquad
\ee%

To find correlators of special primaries,
we set $\eta_1(\p_{\yy_1^1},\p_{\yy_2^1}) = \eta^{AB}\p_{U^A} \p_{V^B}
$, $\eta_2(\p_{\yy_1^2},\p_{\yy_2^2}) =\eta'{}^{AB}\p_{U'{}^A}
\p_{V'{}^B}$  in (\ref{paramE}) with $\eta^{AB} =-\eta^{BA}$,
$\eta'{}^{AB} =-\eta'{}^{BA}$\, ($\tilde{\eta}=0$). From  (\ref{ppointciklFe}), we
obtain
 \be\label{spec2}
\big\langle\JJJ_{\eta }(X )\JJJ_{\eta'}(X') \big\rangle=
  - \f{    \NNN      \exp
\left(\f{  i\pi I_{{X} -{X}' }}{2}\right)({X}'-{X})^{-1}_{AB}({X}'-{X})^{-1}_{CD}
 \eta{}^{AC}\eta{}'{}^{BD}  }{(4\pi)^{ {M}}
      \hhh|\det ({X}' -{X})|}
  \, .\ee
  At $M=2$
the special primaries describe
  $\Delta=2$ scalar currents
  with
  $
  \eta^{\ga\gb}=\eta^{\prime\ga\gb}= \epsilon^{\ga\gb}
  $.
  Since
$({X}'-{X})^{-1}_{\ga\gb}({X}'-{X})^{-1}_{\gga\delta}
 \epsilon{}^{\ga\gga}\epsilon{}^{\gb\delta}   =  2 \det{}^{-1} ({X}'-{X})\,,
 $
this gives\be
\big\langle\JJJ_{\eta }(X )\JJJ_{\eta'}(X') \big\rangle=
    \f{    \NNN    }{
   {8\pi^2\hhh \det^2 ({X}'-{X})}}
  \, .\ee

As anticipated, the signs of two-point functions respect unitarity.
Indeed, suppose that a state
\be
|A\rangle = \int_\Sigma \phi^{nm}(X) J_{nm}(X)|0\rangle\,
\ee
is represented by an integral over
a space-like surface $\Sigma$ with some measure $\phi^{nm}(X)$.
(More precisely, $\Sigma$ is some local Cauchy bundle introduced
in \cite{Mar}). In particular, at $M=2$,
for any  space-like $\Sigma$,
\be
I_{{X-X'}}=0\quad \forall X,X'\in \Sigma\,.
\ee
In this case, the two-point functions (\ref{2pointB2})
and (\ref{2pointF2})  turn out to be positive-definite
for the currents of even spins and negative-definite for the currents of
odd spins.\footnote{Integration over a general (\ie not necessarily
space-like) surface
will not lead to an integral of some expression of a definite sign
since in this case $I_{{X-X'}}$ will take all possible values for various
$X,X'\in \Sigma$. Nevertheless, the final result should be positive
because the space of states resulting from the integration over space-like
surfaces, as well as over twistor space, spans the full Hilbert space of states $H$, so
that any other integration prescription will result in a vector from $H$.}
This  just matches the property (\ref{rdag}) that the $\rho$-invariant currents of
(odd)even spins are (anti)Hermitian, implying  that $\langle A|A\rangle >0$ for all spins.
(Here it is important that the parameters $\eta^{nm}_i$ (\ref{paramE})
are Hermitian for real $\eta$ and $\widetilde \eta$).

      \subsubsection{Three-point function}
  \label{3-pointnew}

For primary parameters (\ref{paramE}) of the form
 $$
\eta^{mn}_j=\eta^{mn}_j(\p_{U^j})\q\widetilde{\eta}^{mn}_j=\widetilde{\eta}^{mn}_j(\p_{U^j})\,,
 $$
Eqs.~(\ref{ppointciklBp}) and (\ref{ppointciklFop}) give
  for  boson-boson and fermion-fermion currents, respectively,
 \bee      \label{3pointB}
&&\langle \JJJ_{\eta_1 }(X^1 )\JJJ_{\eta_2}(X^2)\JJJ_{\eta_3}(X^3)\rangle{}^b=
     \f{   \NNN}{\pi^{ 3M/2}}
  \exp\left( i\f{\pi}{4}\left({I_{X^1-X^2}+I_{X^2-X^3}+I_{X^1-X^3}}\right)\right)
\qquad\qquad
 \\ \nn &&
 {{\big|\det (X^1-X^2)\,\det (X^1-X^3)\,\det(X^2-X^3)\big|}} ^{-\half} \,\,\qquad\qquad
  \\ \nn &&
  \Big( \eta_1(\p_{U^1}) \big(  \eta_2 (\p_{U^2}) \eta_3(\p_{U^3})
\,+\,  \widetilde{\eta}_2(\p_{U^2})  \widetilde{\eta }_3(\p_{U^3})\big)
+ \widetilde{\eta}_1(\p_{U^1})\big(
\widetilde{\eta }_2(\p_{U^2})   \eta_3(\p_{U^3}) \,+\,  \eta_2(\p_{U^2})
\widetilde{\eta}_3 (\p_{U^3}) \big)\Big)  \qquad
\\\nn&&
 \cos\Big(Q_{1,2,3}+Q_{2,3,1}+Q_{3,1,2}\Big)
    \cos \left( P_{1,2}\right)\cos \left( P_{2,3}\right)\cos \left( P_{3,1}\right)
    \big( {U^1},U^2 ,{U^3}  \big)\Big|_{U^1= U^2  =U^3 =0}
 \,,\,\quad
   \eee
\bee      \label{3pointF}&&
 \langle \JJJ_{\eta_1 }(X^1
)\JJJ_{\eta_2}(X^2)\JJJ_{\eta_3}(X^3)\rangle{}^f=
 -\f{  \NNN}{\pi^{ 3M/2}}
  \exp\left( i\f{\pi}{4}\left({I_{X^1-X^2}+I_{X^2-X^3}+I_{X^1-X^3}}\right)\right)
\qquad\qquad  \\ \nn&&
 {{\big|\det (X^1-X^2)\,\det (X^1-X^3)\,\det(X^2-X^3)\big|}} ^{-\half} \,\,\qquad
 \,\,\qquad \\ \nn&&
 \Big( \eta_1(\p_{U^1}) \big(  \eta_2 (\p_{U^2}) \eta_3(\p_{U^3})
\,+\,  \widetilde{\eta}_2(\p_{U^2})  \widetilde{\eta }_3(\p_{U^3})\big)
+ \widetilde{\eta}_1(\p_{U^1})\big(
\widetilde{\eta }_2(\p_{U^2})   \eta_3(\p_{U^3}) \,+\,  \eta_2(\p_{U^2})
\widetilde{\eta}_3 (\p_{U^3}) \big)\Big) \qquad\\
\nn&&\sin \Big(Q_{1,2,3}+Q_{2,3,1}+Q_{3,1,2}\Big)
    \sin \left(P_{1,2}\right)\sin \left( P_{2,3}\right)\sin \left( P_{3,1}\right)\,
    \big( {U^1},U^2 ,{U^3}  \big)\Big|_{U^1= U^2  =U^3 =0} \,\,\quad  \eee
with  \be\nn
P_{i,\,j} = -  \f{1}{2\hhh}
 (X^i-X^j)^{-1}_{AB}  U^i{}^A U^j {}^B\q
     Q_{i,j,k}
=\f{1}{4\hhh}\left((X^i-X^j)_{AB}^{-1}+(X^j-X^k)_{AB}^{-1}\right)
  U^j{}^{A }   U^j{}^{B}   \,.
 \ee

For $M=2$, the factor of $  \exp\left( i\f{\pi}{4}\left({I_{X^1-X^2}+I_{X^2-X^3}+I_{X^1-X^3}}\right)\right)$
 is sensitive to the order of \\$\JJJ_{\eta_1 }(X^1 ),\JJJ_{\eta_2}(X^2),\JJJ_{\eta_3}(X^3) \, $
 in the correlator. However, as anticipated, for space-like separation of all three points with
 $ I_{{X^1 -X^2}}=I_{{X^1 -X^3}}=I_{{X^2 -X^3}}=0$
 the sign factor is $1$ independently  of the order of
currents.

Consider one special and two regular primary  currents setting in (\ref{paramE})
  $\widetilde{\eta}=0$, $\eta_1(\p_{\yy_{1,2}^1}) = \eta_1^{AB}\p_{U^1{}^A}
  \p_{V^1{}^B}
$ with some antisymmetric
  $\eta_1^{AB} $,
   $\eta_2(\p_{\yy_{1,2}^2}) =\eta_2(\p_{U^2})$  and
$ \eta_3(\p_{\yy_{1,2}^3} ) =\eta_3 (\p_{U^3})$.
  From   (\ref{ppointciklFe})   it follows
 \bee\nn &&%
  \big\langle\JJJ_{\eta_1}(X^1
)\JJJ_{\eta_2}(X^2)\JJJ_{\eta_3}(X^3)\big\rangle{}=
 \f{ -4i \NNN}{(4\pi)^{ 3M/2}}
  \exp\left( i\f{\pi}{4}\left({I_{X^1-X^2}+I_{X^2-X^3}+I_{X^1-X^3}}\right)\right)
\qquad\qquad  \\ \nn&&
 \left({{\big|\det (X^1-X^2) \det(X^1-X^3)\det (X^2-X^3)\big|}}\right)^{-\half} \,\,
   \eta_1^{AB}   \eta_2(\p_{U^2})   \eta_3(\p_{U^3})
   \qquad\\&&
\nn
    ({X}^1-{X}^2)^{-1}_{AC}({X}^3-{X}^1)^{-1}_{BD}\,U^2{}^C U^3{}^D
    \Big(\cos \big( Q_{2,3,1}+ Q_{1,2,3} \big)\sin P_{2,3} \Big) \big(  U^2 ,{U^3}  \big)\Big|_{  U^2  =U^3 =0}\,.\quad   \eee

For two special and   one  regular currents we set in (\ref{paramE})
 $\widetilde{\eta}=0$, $\eta_j(\p_{\yy_{1,2}^j} ) = \eta_j^{AB}\p_{U^j{}^A} \p_{V^j{}^B}
$, with some antisymmetric
  $\eta_j^{AB} $   with $j=1,2$ and
$ \eta_3(\p_{\yy_{1,2}^3} ) =\eta_3 (\p_{U^3})$.
  From   (\ref{ppointciklFe})   it follows
 \bee   \nn && \big\langle\JJJ_{\eta_1 }(X^1
)\JJJ_{\eta_2}(X^2)\JJJ_{\eta_3}(X^3)\big\rangle{}=
 \f{4\NNN}{(4\pi)^{ 3M/2}}
  \exp\left( i\f{\pi}{4}\left({I_{X^1-X^2}+I_{X^2-X^3}+I_{X^1-X^3}}\right)\right)
\qquad\qquad  \\ \nn&&
 \left( \big|\det (X^1-X^2) \det(X^1-X^3) \det (X^2-X^3)\big| \right)^{-\half} \,\,
  \eta_1^{AD}   \eta_2^{CB}   \eta_3(\p_{U^3})
  \qquad\\&&
\nn({X}^1-{X}^2)^{-1}_{AC}({X}^2-{X}^3)^{-1}_{BF}({X}^3-{X}^1)^{-1}_{ED}\,U^3{}^E
U^3{}^F
   \sin \Big( Q_{2,3,1} \Big)
  \big( {U^3}  \big)\Big|_{ U^3 =0}\,.\quad  \qquad \eee

    \section{Conclusion}
\label{conc}

In this paper,  operator algebra of $3d$ conserved currents is
reconstructed in terms of certain associative algebra $\alM$ of
distributions $\gD(\yy)$ in the twistor space. Remarkably,
description of OPE of the infinite tower of HS currents turns out to
be much simpler than for a finite set. The reason is that, as long
as distributions $\gD(\yy)$ are kept arbitrary, their products at
different $\yy$ are well defined even at coinciding space-time
points $X$. On the other hand, to describe OPE of currents of
particular spins, it is necessary to consider
 appropriate derivatives of $\gD(\yy)$ at $Y=0$ leading to
 (derivatives of) $\delta(\yy)$ at $\yy=0$ in the operator product.
 To regularize such expressions,
 one has to consider currents $J(Y|X)$ at different $X$.

 Once algebra $\alM$ is known, it is easy to reconstruct the
 dependence on space-time coordinates $X$, which is
 completely determined by the unfolded equations in terms of $\yy$-dependence.
 Practically, the map is given by the generalized $\D$ function which
 can be called twistor-to-boundary $\D$-function  (propagator).
 This step is insensitive to  particular realization of
 conserved currents since the conservation condition for conformal
 currents completely determines twistor-to-boundary $\D$-function.
 As a result,  space-time operator algebra is determined by the twistor
 algebra $\alM$ which encodes full information on the dynamical origin
 of the system.

 For currents built from free massless fields,
 after an appropriate half-Fourier transform, algebra $\alM$
 coincides  with the universal enveloping algebra of the conformal HS algebra
   \cite{Vasiliev:2012tv}. This conclusion
 fits the analysis of \cite{Colombo:2012jx,Didenko:2012tv}
 on the bulk side, where the computation  performed in terms of HS
 algebra was essentially based on HS symmetry, allowing the authors of \cite{Didenko:2012tv}
 to find all connected $n$-point functions up to overall coefficients.
 Similarly, the computation of this paper, allowing to determine $n$-point functions
 with exact relative coefficients, is controlled by the symmetry
 associated with  $\alM$. Since $\alM$ relates operator products involving
 different numbers of currents, it was called  multiparticle algebra in
 \cite{Vasiliev:2012tv}. (Note that it has the meaning of a multiparticle
 symmetry from the bulk point of view where boundary currents correspond to
  elementary fields.) The final result turns out to be remarkably simple
  both for the operator algebra and for
  $n$-point functions, being formulated in terms of
  butterfly product constructed from the star product of the HS algebra.
In particular, the generating function of $n$-point functions has
the suggestive form of certain determinant with respect to butterfly
product.

Obtained results  may have different applications. A
particularly interesting direction is to analyze deformations of
operator algebras to be associated with deformations of the
underlying free dynamical system, \ie interactions. For example,
$AdS_4/CFT_3$ HS algebras (see \cite{kon} and references therein)
contain no free continuous parameters. Hence, the corresponding $3d$
operator algebras are rigid in agreement with the conclusion of
Maldacena and Zhiboedov \cite{Maldacena:2011jn}. On the other hand,
$AdS_3/CFT_2$ HS algebras contain a free parameter $\nu$
\cite{Prokushkin:1998bq}. Hence, in agreement with the analysis of
$HS$ $AdS_3/CFT_2$ correspondence (see \cite{Gaberdiel:2012uj} and
references therein), the boundary  algebras are not unique.

On the other hand, as explained in more detail in
\cite{Vasiliev:2012tv},  $\alM$ is a promising candidate for the
symmetry underlying a string-like extension of HS theory, which we
call multiparticle theory.
Hopefully, further study along these lines may shed light on the
structure of HS theory with mixed-symmetry fields and, eventually,
on yet unknown HS multiparticle  version of String Theory. It is
tempting to speculate that the origin of butterfly formulae for the operator
algebra and $n$-point functions should receive  natural explanation
in terms of the multiparticle theory.

Amplitudes can be  associated with $n$-point functions
 carrying stripped indices of creation and
annihilation operators. Interaction deformation of such $n$-point
functions, computed in this paper for free currents, may help to
clarify deeper structures underlying tremendous progress in the
analysis of multiparticle processes of field theory. Parallelism
with sophisticated field-theoretic methods due to twistor
description is obvious (see, e.g., \cite{ArkaniHamed:2012nw} and
references therein).

 Wightman functions of currents $\langle J(X^1)J(X^2)\ldots J(X^n)\rangle $
evaluated in  this paper  should be distinguished from  chronological
functions. The latter can be easily obtained from the former at
least away from singularities via insertion of appropriate
step-functions in time. On the other hand, Wightman functions
contain   additional factors,
 that depend on the causal relations between coordinates
 $X^i$  and originate from  the accurate definition of the involved $\D$-functions.
 Both in our scheme and in that of \cite{Colombo:2012jx,Didenko:2012tv}
 $\D$-functions and Green functions result from
 Gaussian integrals in the twistor space. In our approach,   signs of
 $\D$-functions are determined by evaluation of Gaussian integrals in the
 complexified  Fock-Siegel space \cite{gelcur} which is
 a twistor extension of the usual Siegel space \cite{Siegel,Mumford}. It would be
 interesting to extend this technics to the bulk computations of
 \cite{Colombo:2012jx,Didenko:2012tv} where evaluation of Gaussians
 was so far a bit formal, hence leaving undetermined sign factors resulting
 from the square root of the determinant.

Being based on unfolded dynamics, our construction can be easily
extended to a larger space where the symmetry, originally
interpreted as conformal at the boundary, acts. This is achieved via
extension of the unfolded current conservation equations, that still
have the form of flatness conditions (\ref{D0}). Most natural option
is to go to the  bulk $AdS$ space. This is achieved via extension of
$\D$-functions to the bulk by solving unfolded equations
(\ref{dgydgyh+}) with the initial data (\ref{Ddelta}), that gives a
twistor-to-bulk extension of the rank-two $\D$-function. In fact, up
to details distinguishing between $\D$-functions and Green
functions, the respective twistor-to-bulk $\D$-functions
(propagators) that indeed respect the initial data (\ref{Ddelta}) in
the twistor space are well known  (see, e.g.,
\cite{Giombi:2012ms,Didenko:2012tv} and references therein). In
accordance with the general analysis of \cite{Vasiliev:2012vf}, this
simple observation uplifts the whole setting from the boundary to
the bulk in a very straightforward way, making boundary and bulk
computations literally equivalent. In particular, our computation
reproduces the results of Didenko and Skvortsov
\cite{Didenko:2012tv} on $3d$ connected $n$-point functions,
determining all relative coefficients for different $n$ and
extending them to supercurrents. This approach works equally well in
higher dimensions allowing us, in particular, to evaluate $n$-point
current correlators in four dimensions in Section
\ref{Correlators4d}.

As argued in \cite{Vasiliev:2012vf}, equivalence of the boundary and
bulk pictures should hold beyond the level of free fields as well.
However, apart from two specific HS models corresponding to free
boundary theories, explicit relation is more involved requiring
solution of nonlinear unfolded equations in the both of dual
pictures. To make the $AdS_4/CFT_3$ correspondence complete at the
nonlinear quantum level, one should construct a generating
functional  invariant under (appropriately deformed) multiparticle
symmetries. As shown in
 \cite{Vasiliev:2005zu}, in unfolded dynamics approach, gauge invariance of such a functional
 implies that it is represented by an integral of a closed form, which is independent of
 local variations of the integration surface. (This is analogous to the
 charge $Q$ (\ref{rhoQ}) that can be equally well evaluated  via integration
  over twistor space or  space-time.) Hence, the corresponding generating
 functional can be evaluated via integration  over $(i)$ twistor space,
 $(ii)$ partially
 twistor space and partially  boundary surface, or $(iii)$ over bulk, giving the same
 result. Such an equivalence (duality) may seem obscure
 unless manifest equivalence of different formulations via unfolded dynamics is
 accounted.

\section* {Acknowledgments}

The authors are grateful to  V.~Didenko, E.~Feigin, E.~Skvortsov, I.Tipunin, I.~Tyutin and,
especially, M.~Soloviev and B.~Voronov for useful discussions  and to  M.~Gumin for
comments on the original version of this paper.  We thank the organizers
and participants of spring 2012 ESI workshop on Higher Spin Gravity,
where this work was initiated, for creation of friendly and
productive atmosphere and stimulating conversations. Also we thank
 the Galileo Galilei Institute for Theoretical Physics for the hospitality
and the INFN for partial support during the final stage of this work.
This research
was supported in part by RFBR Grant No 11-02-00814-a.

\addtocounter{section}{1} \addcontentsline{toc}{section}{\,\,\,\,\,\,References}

\section* {$\rule{0pt}{1pt}$}

\end{document}